\documentclass[reprint,superscriptaddress,preprintnumbers,amsmath,amssymb,aps,prd,tightenlines,longbibliography,nofootinbib]{revtex4-2}

\usepackage{graphicx}
\usepackage[T1]{fontenc}
\usepackage[dvipsnames]{xcolor}
\usepackage{amsmath,amssymb,bm}
\usepackage{mathtools}
\usepackage[colorlinks=true
,urlcolor=blue
,anchorcolor=blue
,citecolor=blue
,filecolor=blue
,linkcolor=red
,menucolor=blue
,linktocpage=true
,pdfproducer=medialab
,pdfa=true
]{hyperref}
\usepackage{ulem}

\newcommand{\be}{\begin{equation}}
\newcommand{\ee}{\end{equation}}
\newcommand{\bea}{\begin{equation}\begin{aligned}}
\newcommand{\eea}{\end{aligned}\end{equation}}
\newcommand{\SNR}{{\rm SNR}}

\begin{document}

%=============================================================================

\title{Crowdsourcing Gravitational Waves from Superradiant Axions}

\author{Sebastian A.R. Ellis}
\email{sebastian.ellis@kcl.ac.uk}
\affiliation{King’s College London, Strand, London, WC2R 2LS, United Kingdom}

\author{Orion Ning}
\email{orion.ning@berkeley.edu}
\affiliation{The Leinweber Institute for Theoretical Physics, University of California, Berkeley, CA 94720, USA}
\affiliation{Theoretical Physics Group, Lawrence Berkeley National Laboratory, Berkeley, CA 94720, USA}

\author{Nicholas L. Rodd}
\email{nrodd@lbl.gov}
\affiliation{The Leinweber Institute for Theoretical Physics, University of California, Berkeley, CA 94720, USA}
\affiliation{Theoretical Physics Group, Lawrence Berkeley National Laboratory, Berkeley, CA 94720, USA}

\author{Jan Sch\"utte-Engel}
\email{janschue@berkeley.edu}
\affiliation{Department of Physics, University of California, Berkeley, CA 94720, USA}
\affiliation{RIKEN iTHEMS, Wako, Saitama 351-0198, Japan}

\begin{abstract}
Black hole superradiance is a powerful probe of ultralight axions.
If nature contains a boson with a mass of order $10^{-12}\,$eV, \textit{mere vacuum fluctuations} will lead to its efficient production around spinning stellar mass black holes, forming a gravitational atom that both drains the black hole spin and decays to produce near-monochromatic gravitational waves.
Existing superradiance constraints derive primarily from spin measurements of a handful of identified black holes.
Here we instead present a detailed study of the population level effect: gravitational waves arising from both the 100 million black holes in the Milky Way and the stochastic signal from axion clouds throughout the universe.
We study the impact of a broad range of systematic uncertainties on the black hole properties and compute the projected axion sensitivity for LIGO, as well as the future instruments Einstein Telescope, Cosmic Explorer, and a high-frequency Magnetic Weber Bar.
We demonstrate that LIGO can robustly probe axion masses from roughly $10^{-13}\,$eV to $4 \times 10^{-12}\,$eV.
If the black hole population extends to masses slightly below $5\,M_{\odot}$ -- as hinted for by LIGO inspiral observations -- LIGO would approach $10^{-11}\,$eV.
Under that same assumption we show that a future high-frequency detector could push considerably higher, potentially beyond $10^{-10}\,$eV in the most optimistic scenarios, reaching towards the lowest masses within the projected sensitivity of axion dark matter searches.
\end{abstract}

\maketitle

%%%%%%%%%%%%%%%%%%%%%%%%%%%%%%%%%
\section{Introduction}
%%%%%%%%%%%%%%%%%%%%%%%%%%%%%%%%%

Over the past decade the axion has emerged as the most sought after extension of the Standard Model.
The axion could solve the Strong CP problem~\cite{Peccei:1977hh,Peccei:1977ur,Weinberg:1977ma,Wilczek:1977pj}, provide the particle description of dark matter~\cite{Abbott:1982af,Preskill:1982cy,Dine:1982ah}, and remain as a relic of the compactification of string theory at high energies~\cite{Svrcek:2006yi,Arvanitaki:2009fg}.
These deep motivations have driven the development of an enormous range of discovery channels.

In this work we focus on arguably the most startling axion discovery channel: gravitational waves (GWs).
The mere existence of an ultralight pseudoscalar axion in the spectrum of nature could trigger the superradiance instability of a rotating black hole (BH)~\cite{Arvanitaki:2010sy}, opening a broad path to axion detection \textit{even if it is not dark matter}.
This instability exists as the wave analog of the Penrose process~\cite{Penrose:1969pc}; whilst a BH without spin is stable under perturbations~\cite{Regge:1957td}, certain perturbations can extract angular momentum from a spinning BH.
These initial perturbations can grow resonantly if the field that perturbs the Kerr metric~\cite{Kerr:1963ud} is massive~\cite{Damour1976,Ternov:1978gq,Detweiler:1980uk_1,Furuhashi:2004jk,Dolan:2007mj}; see also Ref.~\cite{Brito:2015oca} for a review.
The end result is the development of a large axion cloud around the BH -- in the form of a gravitational atom -- that has such a large axion occupation number that it begins to deplete its energy through axion to graviton annihilation, $aa \to g$~\cite{Arvanitaki:2010sy}.
In summary, the existence of an axion could be established by the observation of GWs emerging from BHs in the Milky Way and universe beyond.

The process of axion superradiance is both well established and largely understood.
Indeed, the observation of individual BHs with large spins has been used to establish strong constraints on axions with masses in the rough range $6 \times 10^{-14}\,\textrm{eV} \lesssim \mu \lesssim 6 \times 10^{-12}\,\textrm{eV}$, where the axion Compton wavelength is comparable to the BH horizon, see Refs.~\cite{Arvanitaki:2010sy,Arvanitaki:2014wva,Mehta:2020kwu,Baryakhtar:2020gao,Unal:2020jiy,Ng:2020ruv,Hoof:2024quk,Witte:2024drg,Caputo:2025oap,Aswathi:2025nxa}.\footnote{Note that axions can further generate a distinct GW signature by perturbing the dynamics of binary BH mergers~\cite{Yang:2017lpm,Choudhary:2020pxy,Zhang:2022rex,Bamber:2022pbs,Aurrekoetxea:2023jwk,Kim:2025wwj,Roy:2025qaa}.}
Nevertheless, we only know of roughly twenty of the expected 100 million BHs in the Milky Way, all of which could contribute to a signal observable at the LIGO-Kagra-Virgo observatory or one of the many planned future GW instruments that look to improve the sensitivity and broaden the frequency range of existing devices.

\begin{figure*}
\centering
\includegraphics[width=0.49\textwidth]{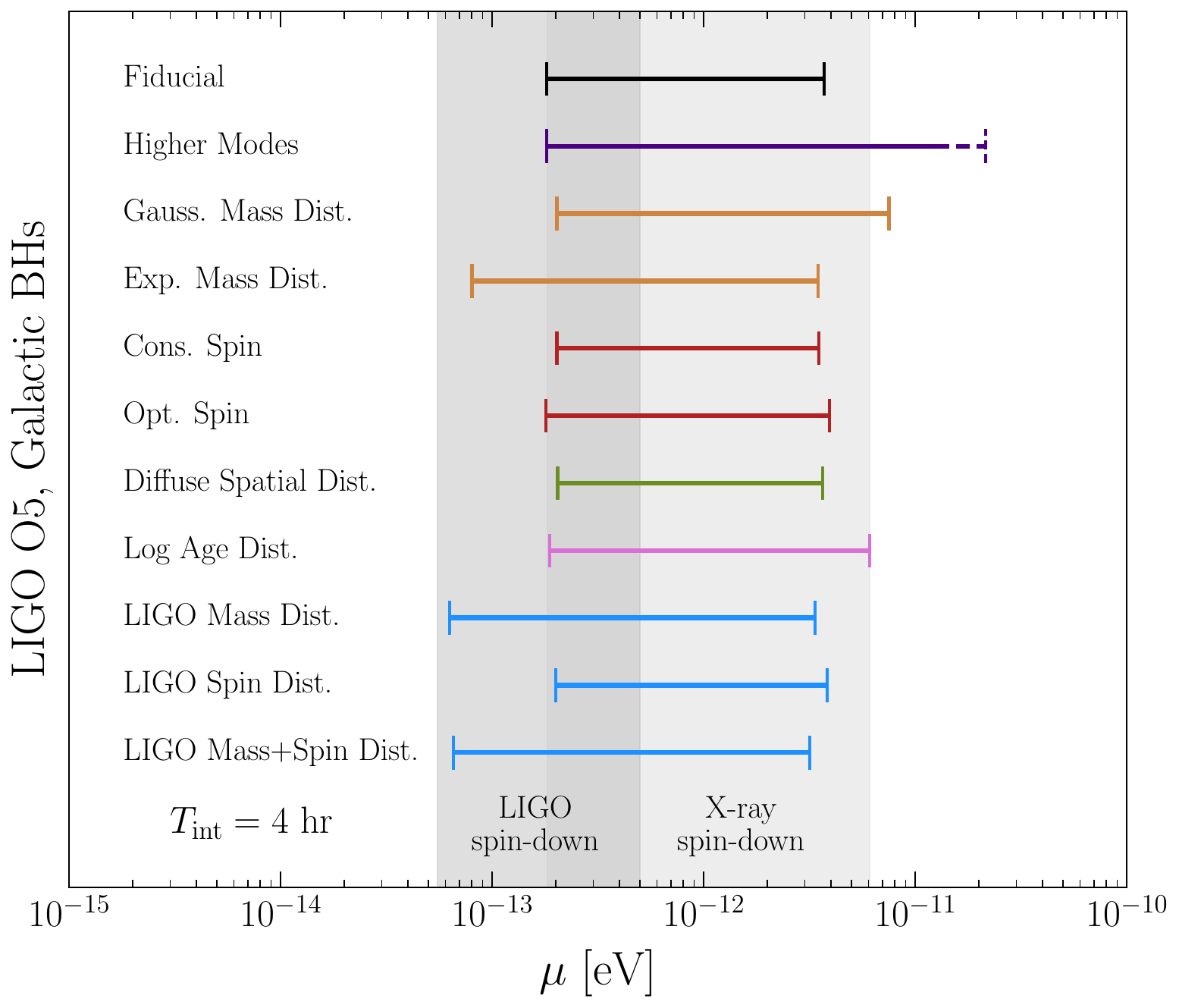}
\includegraphics[width=0.49\textwidth]{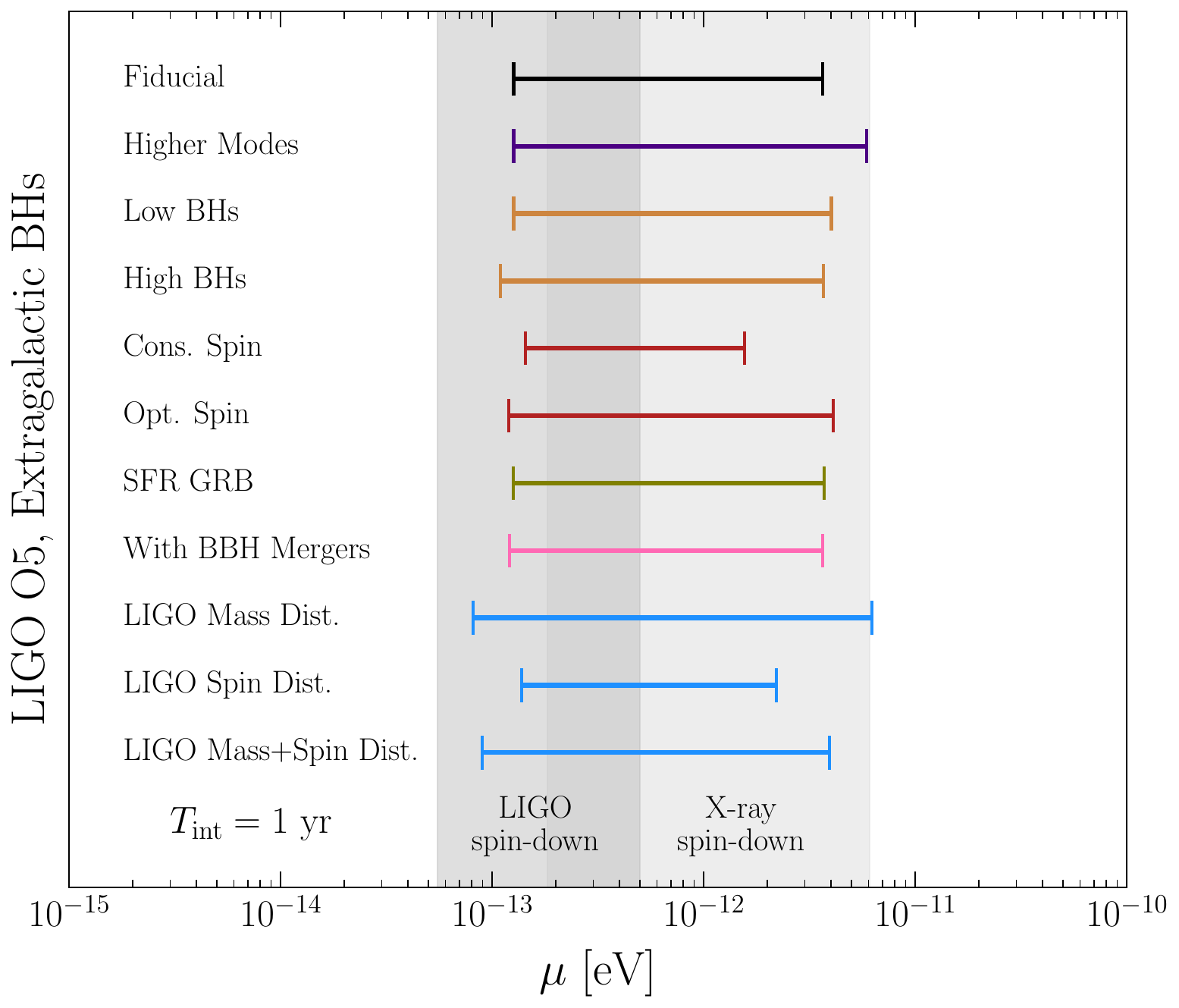}
\vspace{-0.2cm}
\caption{The projected sensitivity of GW searches for axion superradiance as a function of the axion mass $\mu$, for galactic (left) and extragalactic (right) analyses.
Results are shown for LIGO O5 and for a wide range of systematic variations to the underlying treatment of the BH population parameters and also the theoretical treatment of superradiance.
Broadly, GW searches are well placed to complement and potentially extend existing analyses focusing on the spin-down of individual BHs with spins inferred from X-ray and LIGO observations (gray regions), which are subject to systematic uncertainties orthogonal to those of a population analysis.
The dashed sensitivity on the left would be obtained only if LIGO sensitivity was extended to higher frequencies as discussed in the text.}
\label{fig:fig_1}
\end{figure*}

The GW signal arising from the full BH population has received less attention, although there have been a number of key developments, see e.g. Refs.~\cite{Arvanitaki:2014wva,Brito:2017wnc,Brito:2017zvb,Tsukada:2018mbp,Tsukada:2020lgt,Zhu:2020tht,Yuan:2021ebu,Sprague:2024lgq}.\footnote{Galactic GW superradiance searches have been performed, see e.g. Refs.~\cite{Arvanitaki:2014wva,Palomba:2019vxe,LIGOScientific:2021rnv,LIGOScientific:2025csr}, although the results are presented in the context of individual BHs rather than as constraints derived on a population, as is our focus.
LIGO has also performed model independent continuous wave searches~\cite{LIGOScientific:2025bkz,LIGOScientific:2025ouy} and searches for vector superradiance with an electromagnetic counterpart~\cite{Mirasola:2025car}.}
The goal of the present work is to present a comprehensive study of the gravitational wave signal crowd-sourced from the full BH population, with a particular focus on quantifying the associated systematics and understanding the sensitivity of existing and future instruments.
We primarily focus on the sensitivity that could already be obtained by LIGO~\cite{LIGOScientific:2016aoc,aLIGO_O5} -- revisiting the analyses of Refs.~\cite{Arvanitaki:2014wva,Zhu:2020tht} -- although we also consider potential successors in the form of Cosmic Explorer (CE)~\cite{Evans:2023euw} and Einstein Telescope (ET)~\cite{Hild:2010id}.
When superradiance is generated by a boson of mass $\mu$, the GW emission occurs at a linear frequency $\sim$$\mu/\pi$.
This suggests that high-frequency detectors could have a unique sensitivity to heavier axions, and to explore this we study the sensitivity of a Magnetic Weber Bar deployed at DMRadio (MWB-DMR)~\cite{Domcke:2024mfu}, which is a leading high frequency proposal (cf. Refs.~\cite{Berlin:2023grv,Carney:2024zzk} for examples of other promising proposals).
On this last point our results are highly complementary to the recent study of Ref.~\cite{Sprague:2024lgq} which considered the sensitivity of a Levitated Sensor Detector; for a review of various high frequency detectors, see Ref.~\cite{Aggarwal:2025noe}.

Our analysis is divided amongst the two distinct ways that a gravitational signature could appear in a detector.
First, the gravitational waves from axion superradiance can be long-lasting and coherent~\cite{Yoshino:2013ofa,Isi:2018pzk,Zhu:2020tht,Baryakhtar:2020gao,Yuan:2021ebu} (cf. the rapidly evolving GW frequency from binary BH mergers during the inspiral phase).
As a result, BHs within our own galaxy could generate a forest of narrowband coherent signatures that can be individually identified in a detector.
Second, whilst this coherent emission is expected of BHs throughout the universe, the extragalactic signal manifests as an incoherent sum of the redshifted contribution from axion clouds produced throughout cosmic history~\cite{Brito:2017wnc,Brito:2017zvb}.
The search strategy for such an incoherent signature is distinct, and therefore we treat this as a different channel through which the axion could be detected.

Our results for both cases for the LIGO detector are shown in Fig.~\ref{fig:fig_1}.
Sensitivities are shown purely as a function of the axion mass.
In general, superradiance constraints can be established also as a function of the axion decay constant $f_a$.
As we review later in this work, the relevance of $f_a$ arises through an axion quartic term that scales as $(\mu/f_a)^2$ such that as $f_a$ decreases there is a growing self-interaction between the axions in the superradiant cloud. 
For sufficiently small values of $f_a$, the self-interactions perturb the growth and decay of the cloud, eventually shutting off the GW emission entirely.
The impact of self-interactions has been comprehensively studied for the lowest modes of the gravitational atom~\cite{Baryakhtar:2020gao,Witte:2024drg}, but not more generally.
As the effects of self-interactions vary with BH properties, we have chosen not to include them in the population analysis presented here.
Nevertheless, by default we restrict our attention to the most well understood lowest superradiating mode, for which we estimate that our bounds are generally sensitive to the QCD axion parameter space.
We discuss this point in greater detail later in this work, although note already that computing our sensitivity as a function of $f_a$ is one of the most important extensions of the results we present.

Another key ingredient in our population-level study is the underlying properties of the BH population; in particular, we require a model for the distribution of BH spin, mass, age, and positions.
This is an area where there have been enormous developments in the past decade thanks largely to the GW window onto BHs opened by LIGO.
A large number of BHs have already been discovered, with observations now suggesting the stellar mass BH window could extend below $5\,M_{\odot}$~\cite{LIGOScientific:2020zkf,LIGOScientific:2024elc} and potentially up to hundreds of solar masses~\cite{LIGOScientific:2025rsn}.
With this in mind, Fig.~\ref{fig:fig_1} displays results for a wide range of variations to the underlying BH population parameters.

Figure~\ref{fig:fig_1} demonstrates that population-level searches can probe a wide range of axion masses.
For instance, they can relatively robustly push down to masses $\mu \simeq 10^{-13}\,$eV.
There is far greater uncertainty regarding the higher mass sensitivity, which as we quantify hinges both on the existence of lighter BHs and the ability to account for the contribution of higher modes in a theoretically controlled way.
The dashed part of the curve in the figure assumes that LIGO or a comparable interferometer can extend its sensitivity to higher frequencies as discussed in Refs.~\cite{Miao:2017qot,Martynov:2019gvu,Schnabel:2024hem,Jungkind:2025oqm}, although as detailed in those references this would be a significantly non-trivial extension of LIGO and is a primary driver for our consideration of dedicated high-frequency instruments.
Specifically, for LIGO and other observatories examined in this work, we assume the extrapolation is quadratic in the noise-equivalent power spectral density (PSD).

In addition to the gravitational signature that is our focus, there is another straightforward consequence of axion superradiance that has been studied extensively: the spin-down of a BH.
This has led to multiple groups using measurements of BH mass and spins to derive exclusion limits on the axion parameter space, which are shown in Fig.~\ref{fig:fig_1}.
As the figure highlights, GW searches can be highly complementary to the spin-down observations, although they are associated with their own unique set of systematics.
To briefly expand on this final point, there are two primary methods for inferring BH spins that have been used to derive the constraints displayed.
The first is X-ray observations of galactic BHs, with sensitivity primarily driven by Cygnus X-1.
We show results from Refs.~\cite{Baryakhtar:2020gao, Witte:2024drg} derived including the $m=1$ and $2$ superradiant modes, which constrain axions with masses $2 \times 10^{-13}\,\textrm{eV} \lesssim \mu \lesssim 6 \times 10^{-12}\,\textrm{eV}$, although we note that Ref.~\cite{Witte:2024drg} also presented constraints extending to $10^{-11}\,$eV when including modes with $m=3,4$.
The primary systematic in these analyses comes from the X-ray modeling used to infer the BH spin.
For Cygnus X-1, the measurement can vary from $\chi > 0.9985$, to $\chi \sim 0.04$ in potentially extreme models; for a detailed discussion see, for example, Ref.~\cite{Zdziarski:2024zfg}.
The second technique for inferring BH spins is to extract it directly from the small imprint the spins leave in the gravitational waveform measured in instruments such as LIGO.
This has been performed on the recent GW231123 event which corresponded to the observed merger of two $\gtrsim$$100\,M_{\odot}$ BHs~\cite{LIGOScientific:2025rsn}.
In Refs.~\cite{Caputo:2025oap,Aswathi:2025nxa} this event was used to constrain axions in the approximate range $[0.6,5]\times 10^{-13}\,$eV (note Ref.~\cite{Aswathi:2025nxa} also considered GW190517~\cite{LIGOScientific:2021usb} although that does not extend the constrained range).
Such methods have their own challenges, indeed the measured spins for BHs in GW231123 is quite uncertain and there might be further systematic uncertainties from the waveform models. On top of that the mere presence of a binary companion can disrupt the superradiant clouds~\cite{Zhu:2024bqs}.
In summary, in spite of enormous progress in recent years, BH properties remain challenging to reliably constrain, and that uncertainty manifests as various systematics arising in each method used to search for axion superradiance.
The advantage of a population approach is simply that it accesses a different set of systematics -- those relevant for the full BH population -- as opposed to the systematics on individual objects or events.

The remainder of this work is dedicated to quantifying the above discussion and extending the results to additional detectors.
We organize the discussion as follows.
In Sec.~\ref{sec:superradiance} we present a self-contained review of the phenomenon of superradiance, outlining both the GW signal we consider and areas where theoretical uncertainties remain regarding the effect.
In Secs.~\ref{sec:gal} and~\ref{sec:extragal} we compute the signatures of galactic and extragalactic superradiance, providing the full details of how the results in Fig.~\ref{fig:fig_1} were generated and providing results for other detectors.
In Sec.~\ref{sec:highmasses} we provide a discussion dedicated to how high in axion mass experiments might be able to probe with superradiance.
We outline our conclusions in Sec.~\ref{sec:conclusion} and leave a number of technical details and supplemental results to the appendices.
Throughout we set $G = \hbar = c = 1$ and work with the mostly minus metric.

%%%%%%%%%%%%%%%%%%%%%%%%%%%%%%%%%
\section{Axion Superradiance}
\label{sec:superradiance}
%%%%%%%%%%%%%%%%%%%%%%%%%%%%%%%%%

Our method for sourcing GWs from axions is BH superradiance.
The principles of axion superradiance are well established; comprehensive discussions can be found in, for example, Refs.~\cite{Dolan:2007mj,Arvanitaki:2010sy,Brito:2017zvb,Baumann:2019eav,Zhu:2020tht,Sprague:2024lgq}.
In this section we present a review of the phenomena, drawing extensively on the above references in the process.
Particular attention is placed on where the calculation is on firm footing and where it is not; as emphasized already, our fiducial results in Secs.~\ref{sec:gal} and \ref{sec:extragal} rely only on reliable calculations, whereas in Sec.~\ref{sec:highmasses} where we explore the most optimistic reach in axion mass, less controlled approximations are considered.
Several technical details of our calculations are deferred to App.~\ref{app:superradiance_extras}.

Superradiance is the wave analog of the Penrose process: a wave scattering off a BH can gain energy extracted from the BH mass and spin.
For a massive axion that can form a bound state around the BH this process can occur continuously leading to an exponential growth of the bound state, cut-off only once energy can no longer be extracted or due to the onset of a nonlinearity, which can be driven by axion self-interactions.
Once formed, the axion cloud then dissipates through inverse decays, $aa \to g$, generating GWs of angular frequency $\omega \simeq 2 \mu$.

In what follows we provide a more quantitative description of the physics, breaking our discussion into three key steps in the evolution: 1. The condition for superradiance to occur; 2. Exponential growth of an axion bound state; and 3. Cloud dissipation through GWs.
Throughout, we entirely neglect the effect of axion self-interactions.
As discussed in the final subsection, we expect these effects to be unimportant for a QCD axion for our fiducial results, although as we look to push to higher masses this restriction represents an important theoretical uncertainty on our findings.

%%%%%%%%%%%%%%%%%%%%%%%%%%%%%%%%%
\subsection{The Superradiance Death Line}
%%%%%%%%%%%%%%%%%%%%%%%%%%%%%%%%%

We first establish the threshold that determines whether a BH can undergo superradiance.
The starting point of the analysis is the Klein-Gordon equation for a scalar field, $\Phi$, in curved space,
\be
\big( g^{\mu \nu} \partial_{\mu} \partial_{\nu} + \mu^2 \big) \Phi = 0.
\label{eq:KG-Kerr}
\ee
We take $g^{\mu \nu}$ to be the unperturbed Kerr metric described with Boyer--Lindquist coordinates, $\{t,r,\theta,\phi\}$, and denote the initial BH mass and spin as $M$ and $J$.
The unperturbed metric is appropriate as large-scale backreaction from the axion cloud is insignificant: the cloud can grow at most to $\sim$10\% of the BH mass.
Further, the timescale for the cloud's evolution is far longer than the natural timescale of the BH, its rotational period $\simeq$\,$M$, such that we can adiabatically evolve the BH mass and spin.
To analyze the problem, it is convenient to introduce three dimensionless variables, the dimensionless spin $\chi = J/M^2 \in [0,1]$, the outer horizon $\bar{r}_+ = 1 + \sqrt{1-\chi^2}$  (the dimensionful horizon is $M \bar{r}_+$), and the axion-BH gravitational coupling $\alpha = \mu M$.
The gravitational coupling is particularly important: the strongest GW signal occurs for $\alpha \sim 1$.
This already anchors the scale of the axion masses superradiance can be sensitive to: taking $M = 10 M_{\odot}$ as a characteristic BH mass scale, $\alpha = 0.5$ for $\mu \simeq 6.7 \times 10^{-12}\,$eV (cf. Figs.~\ref{fig:fig_1} and \ref{fig:Tg_Th}).

In Boyer--Lindquist coordinates, the solution to Eq.~\eqref{eq:KG-Kerr} is separable and takes the form~\cite{Brill:1972xj,Carter:1968ks},
\be
\Phi = e^{-i\omega t} e^{i m \phi}\, R(r)\,S(\theta).
\label{eq:ansatz_BL}
\ee
The stationarity and axisymmetry of the Kerr geometry imply two Killing vectors and hence two conserved quantities for each solution, here labeled $\omega \in \mathbb{C}$ and $m \in \mathbb{Z}$ for the frequency and magnetic quantum number.
$\Phi$ has been further factorized into a radial and spheroidal functions $R$ and $S$.
Solutions for $\Phi$ can be obtained in various limits of $\alpha$.
For a weak gravitational coupling, $\alpha \ll 1$, a perturbative solution can be obtained~\cite{Baumann:2019eav}.
When $\alpha \gg 1$ solutions can be determined with the WKB method~\cite{Zouros:1979iw,Arvanitaki:2010sy}.
For $\alpha \sim 1$, one is left with numerics~\cite{May:2024npn,Siemonsen:2022yyf,Brito:2014wla,DellaRocca:2025xwz}.
In general, the solutions are qualitatively hydrogen-like -- a cloud of axions around the central BH with a Bohr radius $\sim M (n/\alpha)^2$ -- and can be labeled by integers $n = 0,1,\ldots$ and $\ell = 0,1,\ldots,n-1$, with $m \in [-\ell,\ell]$.

Given a solution for $\Phi$, we first determine whether it can undergo superradiance.
To form a diagnostic, we decompose the frequency into its real and imaginary parts, $\omega = \omega_R+ i\omega_I$.
Equation~\eqref{eq:ansatz_BL} then scales as $\Phi \propto e^{\omega_I t}$, implying that if $\omega_I > 0$ the mode can grow exponentially, the hallmark of superradiance.
Solutions for $\Phi$ reveal that this can occur when
\be
\omega_R < \frac{m\chi}{2 M \bar{r}_+}.
\label{eq:SR_condition}
\ee
This condition acts as the superradiance death line.
Only a BH with mass, spin, and modes that satisfy the condition are potentially observable and such a BH then slowly evolves until falling below the death line.

Although we have not derived it, the origin of Eq.~\eqref{eq:SR_condition} can be understood heuristically.
The frame-dragging angular velocity of the outer horizon is $\Omega_H = \chi/2M \bar{r}_+$.
The phase of the axion wave has an angular velocity of $\omega/m$, so that for $\omega/m < \Omega_H$ an observer being dragged along just outside the horizon sees the wave as carrying negative energy.
As in the Penrose process, under these conditions, an outgoing wave can now emerge amplified by energy extracted from the BH.
This simple criterion exactly matches Eq.~\eqref{eq:SR_condition}.

Before moving to the dynamics, let us further analyze the death line, with a particular view to understanding how large an axion mass superradiance can probe; cf. Fig.~\ref{fig:fig_1}.
Firstly, $\chi/\bar{r}_+ \in [0,1]$ is a monotonic function of $\chi$ and therefore superradiance is most easily achieved for an extremal BH, with $\chi=1$.
For such a BH, Eq.~\eqref{eq:SR_condition} simplifies to $\omega_R < m/2M$, although for a general spin the constraint is approximately $\omega_R \lesssim m \chi/2M$.
Next, for $\alpha \ll 1$, one finds $\omega_R \simeq \mu$ to ${\cal O}(\alpha^2)$ corrections.
In this limit, superradiance can only occur when $\mu < m/2M \lesssim m \times 7 \times 10^{-12}\,\textrm{eV}\, (10\,M_{\odot}/M)$.
This constraint reveals that probing higher axion masses requires either smaller $M$ or larger $m$: lighter BHs or higher superradiance modes.
Both paths come with complications.
The minimum mass of BHs in our universe is an open question.
The behavior of the higher mode bound states depends on non-linear axion interactions that we return to later in this section~\cite{Arvanitaki:2010sy,Baryakhtar:2020gao,Witte:2024drg}.
Further, for fixed $M$ we have to increase $m$ to reach larger axion masses $\mu$.
However, in this limit the growth time increases dramatically which is also evident from Fig.~\ref{fig:Tg_Th}.\footnote{As shown in Ref.~\cite{Zouros:1979iw}, even with a fixed $M$, superradiance can occur for arbitrarily large $\mu$ (corresponding to the $\alpha \gg 1$ regime).
Yet the development of the cloud becomes exponentially slow, growing as $\tau_c \simeq 10^7 e^{3.68 \alpha} M$, cf. Eq.~\eqref{eq:tauc_sa}.
(Here we use the corrected result from Ref.~\cite{Arvanitaki:2010sy}.)
The exponential suppression in the growth rate arises as follows.
To satisfy the superradiance condition $m$ and hence $\ell$ must grow with $\mu$.
The larger $\ell$ generates a centrifugal barrier the wave must tunnel through to reach the horizon so that superradiance can occur, which leads to an exponential suppression in the rate.
Therefore, independent of nonlinearities, higher modes do not open access to arbitrarily large axion masses.}

%%%%%%%%%%%%%%%%%%%%%%%%%%%%%%%%%
\subsection{Exponential Cloud Growth}
%%%%%%%%%%%%%%%%%%%%%%%%%%%%%%%%%

\begin{figure}
\centering
\includegraphics[width=0.45\textwidth]{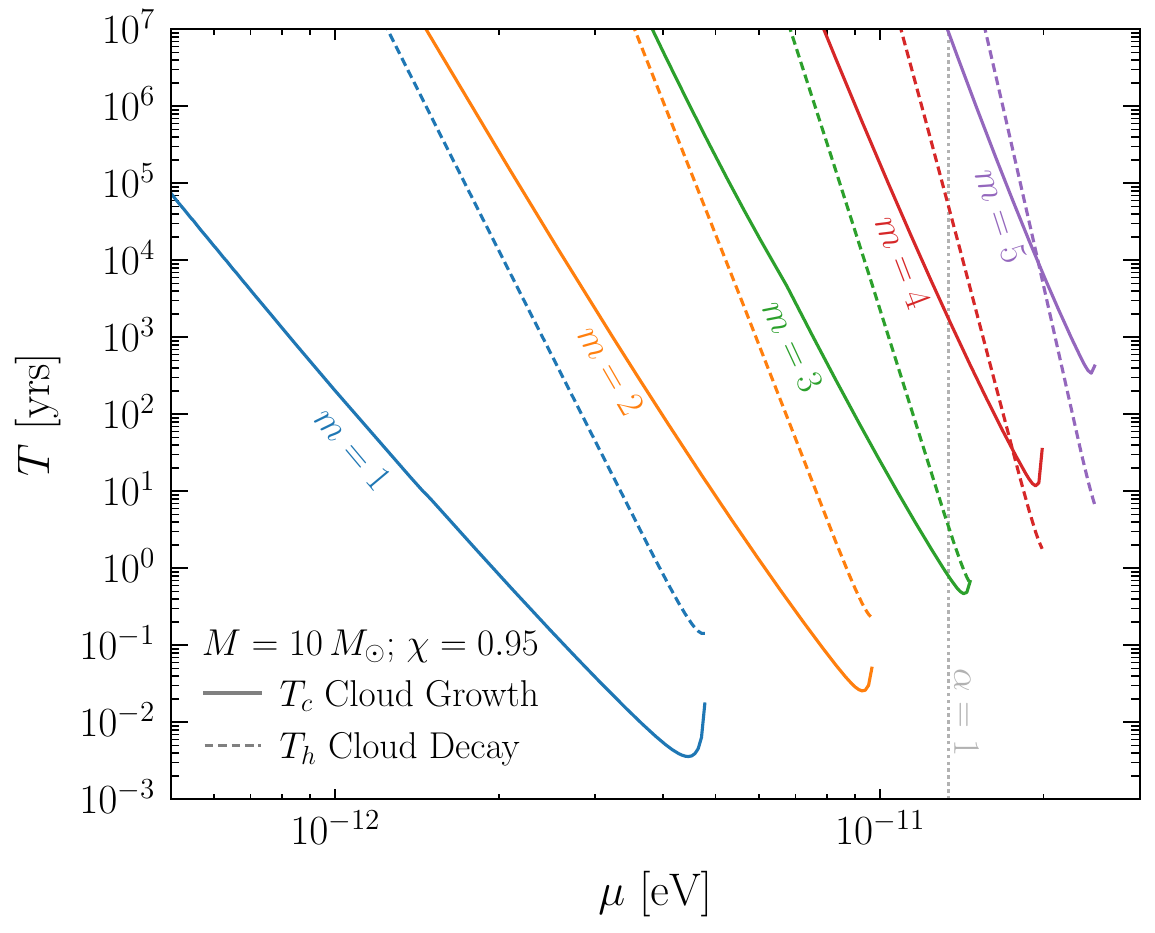}
\vspace{-0.4cm}
\caption{The two key timescales for axion superradiance: cloud growth, $T_c$, and cloud decay, $T_h$.
We show the timescales as a function of the axion mass, $\mu$, for the first five modes associated with a BH with an initial mass and spin of $10\,M_{\odot}$ and $0.95$.
From this example we see that lower modes grow faster, higher modes probe larger $\mu$, and eventually $T_h < T_c$ which shuts off the cloud growth and the signal's observability.}
\label{fig:Tg_Th}
\end{figure}

When the superradiance condition is satisfied, an axion cloud begins to grow exponentially around the BH, and ignoring nonlinearities, continues to do so until the cloud has extracted so much angular momentum and mass that the mode and BH fall beneath the death line.
Here we outline the cloud growth dynamics.
The characteristic timescale of the growth is dictated by $\tau_c = (2\omega_I)^{-1}$; the factor of two arises as we are interested not in the growth of the scalar field, but of its energy density $\rho \simeq \mu^2 \Phi^2$.
As detailed in App.~\ref{app:superradiance_extras}, in general we compute $\omega_R$ and $\omega_I$ using a combination of numerical methods when $\alpha$ is large and perturbative results when it is small, smoothly matching the two.
To provide intuition, if we take $\alpha \ll 1$ and $\chi=1$ the result simplifies to~\cite{Baumann:2019eav}
\be
\tau_c \simeq \frac{M}{2C_{n\ell} m^{2\ell+1}}\, \alpha^{-4\ell-5}.
\label{eq:tauc_sa}
\ee
Here $C_{n\ell}$, given in App.~\ref{app:superradiance_extras}, is a function that tends to decrease with increasing $n$ and $\ell$, with a maximum value $C_{21} = 1/48$ (cf. $C_{32} \simeq 5 \times 10^{-6}$).
Note that the $m=0$ mode never fulfills the superradiance condition, cf. Eq.~\eqref{eq:SR_condition}.
Accordingly, the lowest mode satisfying the superradiance condition grows the fastest, starting with $(n\ell m)=(211)$, for which $\tau_c \simeq 24 M \alpha^{-9}$.
Equation~\eqref{eq:tauc_sa} encodes additional lessons.
First, for our cases of interest $|\omega_R| \gg |\omega_I|$.
Second, the cloud evolution is far slower than the natural timescale of the BH, $\tau_c \gg \Omega_H^{-1}$, justifying an adiabatic evolution of the BH parameters.
Lastly, although by default we focus on the 211 mode, when considering clouds with larger $m$, the fastest growing mode always has $\ell=m$; as discussed shortly, the most relevant mode also has $n-1=m$.

The e-folding timescale for the growth of the axion cloud, $\tau_c$, is not the total time it takes the cloud to grow. 
Superradiance allows vacuum fluctuations of the axion with energy $\sim$\,$\mu$ to develop into a cloud with a mass of order the original BH, a process that requires many e-folds.
The dynamics can be tracked with a simple pair of differential equations for the BH mass and spin~\cite{Brito:2014wla,Brito:2017zvb},
\be
\dot{M} = -2 \omega_I M_c = - \dot{M}_c,\hspace{0.5cm}
\dot{J} = - 2 m \omega_I M_c/\omega_R,
\label{eq:MJ_DE}
\ee
with $M_c$ the mass of the axion cloud.
The first equation formalizes that the cloud grows exponentially at a rate $2 \omega_I$ and does so at the expense of the BH mass.
The second equation specifies the evolution of the BH spin, and quantifies that each axion extracts an angular momentum $m$ from the BH, which gets weighted by the rate of axions being produced, $\dot{M}_c/\omega_R$.
The simplicity of Eq.~\eqref{eq:MJ_DE} belies three assumptions.
Firstly, it is assumed that dissipation of the axion cloud to GWs is negligible on the timescale of the cloud's growth.
This assumption is quantifiable: it is broadly valid for the low $m$ modes, but not in general (see Fig.~\ref{fig:Tg_Th}).
The second, in the spirit of the adiabatic approximation, is that $\omega$ remains constant throughout the cloud's evolution.
In general this is a good approximation, although it can lead to corrections to the timescales for $\alpha \sim 1$.
Finally, we assume that the evolution of each mode of the cloud can be tracked independently, ignoring any potential mixing.
As studied in Refs.~\cite{Baryakhtar:2020gao,Collaviti:2024mvh,Witte:2024drg}, this need not be true.
We return to the final assumption later in this section.

Equation~\eqref{eq:MJ_DE} can be solved as follows.
The equations combine to $\dot{J} = (m/\omega_R) \dot{M}$, or $\Delta J = (m/\omega_R) \Delta M$.
If the final BH and cloud masses are $M_f$ and $M_c^f$, then $\Delta M = M_f - M = - M_c^f$.
To determine $\Delta J$, if $M^2 \chi$ is the initial spin, the evolution continues until the BH falls below the death line of Eq.~\eqref{eq:SR_condition}, which can be determined by solving for the spin at which the inequality is saturated: $J_f = 4m \omega_R M_f^3/(m^2 + 4 \omega_R^2 M_f^2)$.
Accordingly, from $\Delta J$ we infer $\Delta M$ and hence the final cloud mass (here $M$ and $\chi$ are the initial BH mass and spin),
\be
M_c^f \simeq \frac{\alpha \chi}{m} M.
\label{eq:Mcf}
\ee
which again assumes $\alpha \ll 1$, with $\alpha$ evaluated at the initial mass, the full expression is provided in App.~\ref{app:superradiance_extras}.
With this result, we can compute the total growth time for the cloud,
\be
T_c \simeq \tau_c\,\ln (M_c^f/\mu).
\label{eq:Tc}
\ee
The result without assuming small $\alpha$ reveals that $M_c^f \lesssim 0.1 M$, so that $\ln (M_c^f/\mu) \leq \ln (0.1 M/\mu)$, which for $M = 10\,M_{\odot}$ and $\alpha = 1$ is around 177, a large number of e-folds.
Note further that in the $\alpha \ll 1$ limit, the final spin is given by $\chi_f \simeq 4\alpha/m$: for small $\alpha$ the axion cannot extract significant mass from the BH, but as the superradiance condition remains valid for longer it can extract a larger fraction of its spin.

We can now construct a picture of the cloud growth dynamics.
In principle, all allowed modes start growing simultaneously. 
Yet small changes in $\tau_c$ have an enormous impact on the rate of growth as it feeds directly into the number of e-folds. 
For example, denoting the characteristic growth of a general mode as $\tau_c^{n\ell m}$, the two lowest $\ell=m=1$ modes have similar timescales for small $\alpha$, $\tau_c^{311} \simeq 2.8\,\tau_c^{211}$.
As the modes have the same $m$, both have their growth shut off simultaneously, at the time given in Eq.~\eqref{eq:Tc}, evaluated for the $211$ mode.
At that time, if we take $M_c^f = 0.1 M$, $M=10 M_{\odot}$, and $\alpha=1$, the $311$ mode has undergone roughly 63 rather than 177 e-folds, so that the mass of the cloud is $\sim$$10^{50}$ smaller.
Accordingly, we generally expect only the mode with the smallest $\tau_c$ to be relevant.
For $m=1,2,3$ that fastest growing mode has $n=m+1$.
Starting at $m=4$ this condition breaks, as $\tau_c^{644} < \tau_c^{544}$~\cite{Arvanitaki:2010sy}.
However, as the leading order correction to the frequency is $\omega_R \simeq \mu(1-\alpha^2/2n^2)$, clouds with larger $n$ have a larger $\omega_R$ and therefore as they extract the BH spin fall below the superradiance death line faster; cf. Eq.~\eqref{eq:SR_condition}.
As noted in Refs.~\cite{Siemonsen:2019ebd,Yuan:2021ebu} this implies that for GW emission only the mode with $n=m+1$ is relevant, as even if a higher $n$ mode formed faster, as the lower $n$ mode continues to grow past their death line, their $\omega_I$ is driven negative so that their cloud is reabsorbed by the BH before there is appreciable gravitational emission.

Details aside, the 211 mode undoubtedly grows the fastest, developing far before any other mode becomes relevant.
Modes with higher $m$ matter solely as they can continue growing after lower modes have been shut off.
The amount of growth the $322$ mode undergoes while the $211$ fully develops is negligible and most of its evolution occurs with the mass and spin having already been depleted from the $211$ formation.
As such, we approximate the modes as evolving one after another.
By default we consider only the 211 mode in our analyses, however, when we do consider higher modes we start with the lowest allowed mode and the initial BH parameters, track the properties of its growth, and then move immediately onto the next mode.\footnote{In principle, the growth of a higher $m+1$ mode can cut off the GW emission from the $m$ mode~\cite{Zhu:2020tht}.
Nevertheless, we find that across the parameter range we consider the dissipation is always faster than the growth of the next mode.}
To quantify this intuition, a plot of $T_c$ for the first five modes and a $10\, M_{\odot}$ near-extremal BH is shown in Fig.~\ref{fig:Tg_Th}.

%%%%%%%%%%%%%%%%%%%%%%%%%%%%%%%%%
\subsection{Cloud Dissipation through Gravitational Waves}
%%%%%%%%%%%%%%%%%%%%%%%%%%%%%%%%%

With the axion cloud formed, the next stage in the BH life cycle is the decay of the cloud to gravitational radiation.
The interaction that allows this is unavoidable: the gravitational coupling of the axion.
This mediates for the process $aa \to g$ that is kinematically allowed as the axions are bound.
Although the perturbative rate for this process is miniscule, with two axions in the initial state, the rate is enhanced by the square of the enormous number of axions in the cloud.

The key step to quantifying the decay is the power emitted into gravitational radiation as computed in Ref.~\cite{Yoshino:2013ofa},\footnote{Equation~\eqref{eq:dotEh} was computed assuming the background metric is flat rather than Kerr, however this only impacts the numerical coefficient.
For the 211 mode, the result has also been computed assuming a Schwarzschild background, finding a larger prefactor of $\simeq\! 0.025 = 16 D_{21}$~\cite{Brito:2014wla}.
A full numerical calculation for this mode finds the result is in between these two values~\cite{Siemonsen:2022yyf}.
As the only analytical result available with higher modes included is the one performed in flat space, here we choose to use that result for all modes, although we emphasize that this implies our results correspondingly underestimate the GW strain and overestimate the timescale for the cloud to decay.}
\be 
\dot{E}_h = D_{n\ell} \left( \frac{M_c}{M} \right)^2 \alpha^{4\ell+10},
\label{eq:dotEh}
\ee
where we provide the form for the numerical coefficient $D_{n\ell}$ in App.~\ref{app:superradiance_extras} ($D_{21} = 1/640$).
Here $M_c$ and $M$ are the cloud and BH mass, with the latter being the BH mass after the buildup of the cloud ($M_f$ in the notation of the previous subsection, which is also the mass that should be used to evaluate $\alpha$ here).
That $\dot{E}_h \propto M_c^2$ is a direct consequence of the two axions in the initial state.
Although Eq.~\eqref{eq:dotEh} is perturbative in $\alpha$ and computed assuming a flat background metric, Ref.~\cite{Yoshino:2013ofa} also performed a numerical computation of the energy loss in the correct Kerr background.
Those results displayed the expected power-law scaling with $\alpha$, with good agreement at larger spins, giving further confidence in this approach.
Nevertheless, correcting Eq.~\eqref{eq:dotEh} to a numerically determined result in the Kerr background would place our calculation here on firmer footing.

We can determine the timescale for the cloud to decay directly from Eq.~\eqref{eq:dotEh}. 
The gravitational radiation is exactly how the axion cloud dissipates, so that $\dot{M}_c = - \dot{E}_h$.
As a result, we can compute the evolution of the cloud as a function of time after it starts decaying as
\be
M_c(t) = \frac{M_c^f}{1+t/T_h},
\label{eq:Mct}
\ee
with $M_c^f$ the maximum mass of the cloud formed before the gravitational emission, and the corresponding decay timescale given by
\be
T_h = \frac{M^2}{D_{n\ell}\,M_c^f} \,\alpha^{-4\ell-10}.
\label{eq:Th}
\ee
Results for this are shown in Fig.~\ref{fig:Tg_Th}.

Arising from $aa \to g$, the dissipated gravitational radiation is emitted at a frequency $2\omega_R$.
As the cloud loses mass, there is a positive frequency drift~\cite{Arvanitaki:2014wva,Zhu:2020tht,Baryakhtar:2020gao}; this can be relevant as the signal can drift out of the detector's observational window, effectively shortening the signal duration.
For the lowest mode we focus on in our fiducial analysis, the drift is generally small over our analysis timeframe, however it should be accounted for in a full analysis.
Further, the radiation emitted from the a cloud with quantum numbers $n\ell m$ will carry a magnetic quantum number of $2m$.\footnote{This implies only the gravitational emission from the lowest $m=1$ cloud is quadrupolar~\cite{Yoshino:2013ofa}.}
For a BH at a distance $r$, the strain of the dominant mode incident on our detector can be computed as~\cite{Isi:2018pzk,Sprague:2024lgq}
\be
h = \frac{2 \sqrt{\pi \dot{E}_h}}{\omega_R\,r},
\label{eq:h}
\ee
which is fully specified using the previous equations in this subsection; in particular, the time dependence of the strain is specified by using Eq.~\eqref{eq:Mct} in Eq.~\eqref{eq:dotEh}.

%%%%%%%%%%%%%%%%%%%%%%%%%%%%%%%%%
\subsection{Axion Self-Interactions}
%%%%%%%%%%%%%%%%%%%%%%%%%%%%%%%%%

The discussion so far has treated the axion as a free scalar field other than for the inclusion of the irreducible gravitational coupling.
That the perturbative gravitational coupling of an ultralight field plays an important role in the dynamics indicates that the enormous occupation numbers in the cloud can render relevant otherwise minute interactions.
Indeed, a coupling of the axion to dark-sector states can cause the cloud growth to saturate before it becomes sufficiently large to produce a gravitational signal~\cite{Fukuda:2019ewf,Mathur:2020aqv}.
More generically, we should expect that the same potential that provides the axion its mass will generate a quartic term ${\cal L} \supset \lambda \varphi^4/4!$ and thereby give rise to axion self-interactions that can play an important role in the dynamics of superradiance, see for instance Refs.~\cite{Arvanitaki:2010sy,Baryakhtar:2020gao,Omiya:2022gwu,Collaviti:2024mvh,Takahashi:2024fyq,Witte:2024drg}.

The relevance of the quartic is determined by the size of the axion decay constant, $f_a$.
For a generic cosine potential, $V = - \mu^2 f_a^2 \cos(\varphi/f_a)$, we have attractive self-interactions and $\lambda = \mu^2/f_a^2$.
For the QCD axion, the mass and decay constant are related, and one has~\cite{GrillidiCortona:2015jxo}
\be
\lambda_\textrm{QCD} = z_q(1-3z_q) \frac{m_{\pi}^2 f_{\pi}^2}{f_a^4} = (1-3z_q) \frac{\mu^2}{f_a^2},
\ee
where $m_{\pi}$ and $f_{\pi}$ are the pion mass and decay constant, whereas $z_q = m_u m_d/(m_u+m_d)^2$ is specified by the light quark masses.
Numerically, the quartic is tiny, $\lambda_\textrm{QCD} \simeq 5 \times 10^{-82} (M_\textrm{pl}/f_a)^4$, where $M_\textrm{pl} \simeq 1.2\times 10^{19}$ GeV, although the enormous field amplitude superradiance generates can render this coupling important.

A comprehensive analysis of how self-interactions modify the superradiant evolution of the two lowest modes was provided in Ref.~\cite{Baryakhtar:2020gao}.
In the story we have outlined so far where self-interactions are neglected, the 211 cloud grows until it drives the BH below the corresponding death line, after which the cloud is dissipated through gravitational radiation, and then on a longer timescale the 322 cloud begins to grow.
Self-interactions introduce two additional processes that can modify this picture with interactions amongst the modes.
First, they can source a non-linear pumping that accelerates the growth of the 322 mode at the expense of the 211 cloud.
In detail, two 211 axions can convert to a 322 axion and one with $m=0$ that falls into the BH.
As the size of the axion quartic is slowly increased, this interaction leads to the 322 cloud forming earlier than it would otherwise.
Second, self-interactions can source an energy loss channel whereby two 322 axions convert to a 211 as well as an axion emitted non-relativistically out of the system.
Therefore, as the quartic increases further, the two processes can drive an equilibrium between the 211 and 322 modes, as well as sourcing a new detection channel for the clouds through non-relativistic axion emission, which can be searched for as BH scalar sirens~\cite{Gavilan-Martin:2026zzw}.

Having identified the relevant interactions, it is then a detailed question to determine how, for a given value of $\chi$, $M$, and $\mu$, the cloud dynamics vary as a function of $f_a$.
Restricting attention to the fastest growing $m=1$ mode, the analysis in Ref.~\cite{Baryakhtar:2020gao} suggests that for a QCD axion coupling, self-interactions do not appreciably impact the dynamics unless one considers large values of $\alpha$.
However, as $\lambda$ is increased further, or if one moves beyond the lowest mode, self-interactions cannot be neglected.

With the above in mind, we make two important simplifying assumptions throughout this work.
First, by default we only consider the growth and decay of the 211 superradiant mode.
As part of our systematic studies and especially in Sec.~\ref{sec:highmasses} we consider the impact of including higher modes, but the assumption is always that self-interactions remain so small that they can be neglected.
Second, all our results on the sensitivity to superradiance are presented solely as a function of $\mu$.
For a given $\mu$, as we increase $\lambda$ (or decrease $f_a$), self-interactions will begin to modify the dynamics and eventually perturb the cloud growth so significantly that there is no appreciable GW emission to detect.
Even restricting our attention to the first two modes studied in Ref.~\cite{Baryakhtar:2020gao}, as the impact of self-interactions depends upon the BH mass and spin, which vary across the large population we consider, determining their impact is non-trivial.
A logically straightforward albeit numerically demanding procedure would be to take a Monte Carlo approach: for each $\mu$, rerun our galactic and extragalactic analysis for a given value of $\lambda$, account for the impact of the self-interactions on each BH, and then by repeating this for many values of $\lambda$ determine the corresponding 95\% sensitivity limit.
We have not pursued this here although it would be an important step to further quantifying the results we present.
Moving to even higher modes with large self-interactions would require further theoretical work, although aspects of this have been studied in Ref.~\cite{Witte:2024drg}.

%%%%%%%%%%%%%%%%%%%%%%%%%%%%%%%%%
\section{Galactic Superradiance}
\label{sec:gal}
%%%%%%%%%%%%%%%%%%%%%%%%%%%%%%%%%

Having reviewed the phenomenon of axion superradiance, we now turn to our forecast predictions for the associated GW signal that a population of BHs can generate.
We begin by considering the local neighborhood of galactic BHs, which could generate a detectable forest of near-monochromatic GW signals.
In this section we quantify the sensitivity to this forest, outlining first our modeling of the Milky Way BH population before turning to a computation of the sensitivity detectors have to the resulting GWs.

%%%%%%%%%%%%%%%%%%%%%%%%%%%%%%%%%
\subsection{The Black Hole Population of the Milky Way}
%%%%%%%%%%%%%%%%%%%%%%%%%%%%%%%%%

Before we can understand the galactic signature of BH superradiance, we must understand the properties of the BHs themselves.
In particular, we need to adopt a model to describe the full population and its distribution in mass, age, distance, and spin.
We emphasize that although there are expected to be around 100 million BHs in our galaxy~\cite{Timmes:1995kp}, only a handful of these have been discovered.
As such, any model will be associated with systematic uncertainties, and a key aspect of our analysis is in studying how our findings vary as a function of the various assumed distributions.

The full set of models we consider is detailed in Tab.~\ref{tab:configs_GA} and we expand on each below.
Our fiducial model largely follows Ref.~\cite{Sprague:2024lgq}, which studied the detectability of a galactic signal at a Levitated Sensor Detector.
Note that we do not consider the impact accretion disks may have around stellar-mass BHs, as they are expected to have no significant impact~\cite{Arvanitaki:2014wva}.

The first BH property to consider is the distribution of masses.
From observations of X-ray binaries, microlensing, astrometry, and core-collapse supernovae~\cite{Farr:2010tu, Ozel:2010su, Miller:2014aaa, Shao:2020tin, OGLE:2022gdj, Chakrabarti:2022eyq, El-Badry:2022zih, Fryer:2011cx,Gaia:2024ggk}, the distribution of the known BHs appears consistent with a Salpeter-like power-law distribution, $\xi(M) \propto M^{-2.35}$, extending over masses $5 \, M_{\odot} \leq M \leq 20 \, M_{\odot}$.
We adopt this as our default mass model.
Nevertheless, variations are certainly allowed.
We explore two alternative mass distributions suggested by fits in Ref.~\cite{Farr:2010tu}.
The first is a gaussian mass distribution, in which BH masses follow a normal distribution $p(M) = \mathcal{N}(9.2 \, M_{\odot},\, 3.3 \, M_{\odot})$ truncated to ensure $M \geq 1.4\,M_{\odot}$, and secondly an exponential mass distribution, for which the BH masses follow a decaying exponential with a minimum mass.
For the latter, which is parameterized as $p(M) = \exp[-(M - M_{\rm min})/M_0] / M_0$ for $M \geq M_{\rm min}$, we take $M_{\rm min} = 5.3 \, M_{\odot}$ and $M_0 = 4.7 \, M_{\odot}$.
Additionally, we also consider the binary black holes (BBH) primary mass distribution inferred from the LIGO-Virgo GWTC-3 catalogue, which has two distinct peaks at masses around 10 and 35\,$M_{\odot}$~\cite{KAGRA:2021duu}.
See also Ref.~\cite{LIGOScientific:2025pvj} for updated BBH distributions (GWTC-4).
This distribution has recently been demonstrated to be consistent with two populations of BHs: a primary population following a Salpeter-like power-law distribution and a secondary population from hierarchical mergers~\cite{Bouhaddouti:2026jgc}.
The evidence for a primary population following a Salpeter-like power-law suggests the distribution of BHs seen by LIGO is consistent with our fiducial model, which neglects the merger history of BHs.
Note that the discovery of compact objects with masses in the BH-neutron star mass gap $\sim\!2.5$--$5.0$\,$M_{\odot}$~\cite{LIGOScientific:2020zkf,LIGOScientific:2024elc}, motivates the possibility of BHs lighter than $5\,M_{\odot}$, which are present in both our Gaussian and LIGO-informed mass distributions (and which have been considered in other works, for example taking a Salpeter distribution extended down to $3\,M_{\odot}$ as in Ref.~\cite{Zhu:2020tht}).

\begin{figure}
\centering
\includegraphics[width=0.49\textwidth]{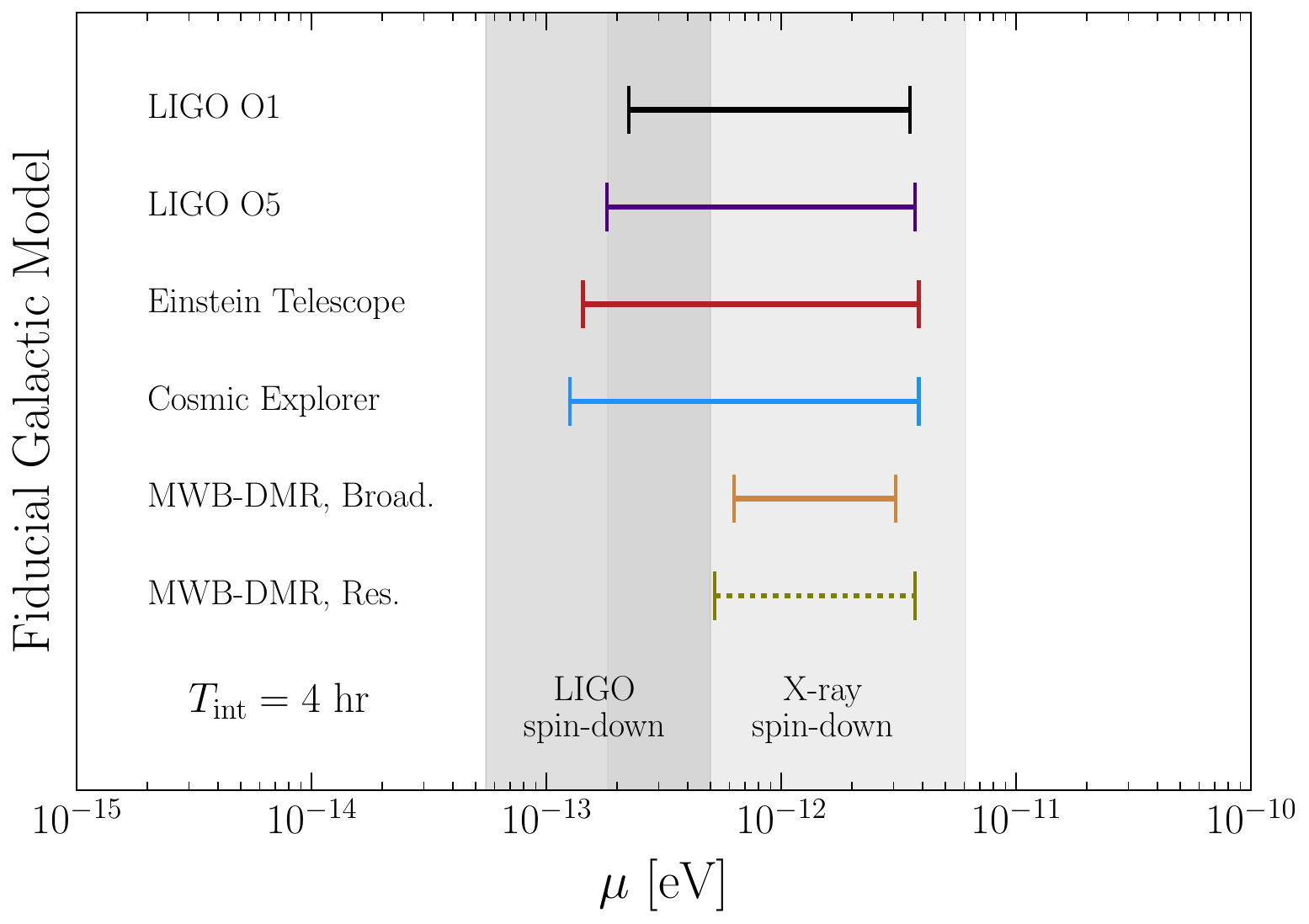}
\vspace{-0.6cm}
\caption{Projected sensitivity to an axion of mass $\mu$ under our fiducial astrophysical galactic model, assuming a coherent integration time of $T_{\rm int} = 4$\,hr and a total observation time $T_{\rm obs} = 1$\,yr.
We show results for each the observatories considered in this work: LIGO O1, LIGO O5, Einstein Telescope, Cosmic Explorer, and MWBs. 
For LIGO O5, the impact of modifying our fiducial model for the BH population is shown in Fig.~\ref{fig:fig_1}, with the details of the variations specified in Tab.~\ref{tab:configs_GA}.
Other details are as in Fig.~\ref{fig:fig_1}.}
\label{fig:galactic}
\end{figure}

The second key BH property we need to model is the distribution of their spins.
Given the paucity of measurements of isolated BHs, any model for the spins is associated with considerable uncertainty.
X-ray binary observations provide marginal evidence for a uniform distribution of spins~\cite{Reynolds:2020jwt}, and therefore we take $\chi \in U[0,1]$ as our fiducial model.
As a simple prescription for exploring the impact of this, we consider a conservative and optimistic variant where we take the spins as drawn from $U[0,0.5]$ and $U[0.5,1]$, respectively.
We also consider a spin distribution motivated by LIGO-Virgo GWTC-3 BBH observations, which suggests that the distribution of spin magnitudes peaks between $\chi \sim 0.1$--$0.2$, though still has reasonable support up to $\chi\sim 1$.\footnote{We note that if an axion exists, that also suggests that the inferred spin distributions seen by LIGO would likely be biased by spin-down coming from axion clouds, though we do not treat this possibility in further detail.}

The remaining two BH parameters are the birthrate and distance.
The importance of the latter is clear from Eq.~\eqref{eq:h}: closer BHs generate a brighter signal.
The age distribution is also critical.
The reason for this can be inferred from the discussion in Sec.~\ref{sec:superradiance}: the loudest signals arise for the largest $\alpha$, which also corresponds to the superradiant clouds that grow and decay the fastest---cf. Fig.~\ref{fig:Tg_Th} and recall we consider only the $m=1$ mode in our fiducial analysis.
Therefore, for a detector to catch the brightest signals, the BH generating the cloud must be relatively young as measured on the timescale of the Milky Way.

In order to model these two parameters, we divide the population into three components: a thin disk, thick disk, and bulge.
From a combination of white dwarf luminosity function and stellar stream measurements in these three regions~\cite{2017ApJ...837..162K, Helmi:2020otr} one can roughly estimate the ages of other stellar remnants like BHs. 
Complementing each other, broadly the white dwarf luminosity functions help to constrain ages across dense stellar environments, whereas stellar streams allow dating of accreted populations farther out in the halo.
Both form an idea of stellar ages distributed over the Galaxy, from which BH ages can then be inferred.
Given these findings,  Ref.~\cite{Sprague:2024lgq} implemented a log-flat birth rate for the BH ages between $10^3$--$8\!\times\! 10^9$\,yrs, $10^3$--$10^{10}$\,yrs, and $10^9$--$1.3\!\times\! 10^{10}$\,yrs for the thin disk, thick disk, and bulge, respectively.
This distribution skews the BH ages to be systematically younger, which although possible would be optimistic from the perspective of superradiance.
As such, we depart from Ref.~\cite{Sprague:2024lgq} on the age distribution, although we implement this model as a systematic.
Instead, for our fiducial age model, we use a linear uniform distribution over the Milky Way star formation history (SFH) as inferred in Ref.~\cite{2019ApJ...887..148F}, which reflects an older BH population whose birth rate tracks the stellar formation rate.

For the spatial distribution, we return to following Ref.~\cite{Sprague:2024lgq} and implement a model for the mass density profile in each of the disk and bulge regions.
From Refs.~\cite{McMillan:2011wd, SDSS:2005kst,Bissantz:2001wx} the disks (we assume the same form for both thin and thick disks) and bulge BHs are taken to follow the following distributions
\be
p_D(\rho, z) \propto e^{-|z|/z_{\rm d}-\rho/R_{\rm d}},
\hspace{0.4cm}
p_B(\bar{r}) \propto \frac{e^{-(\bar{r}/r_{\rm cut})^2}}{(1+\bar{r}/r_0)^{\alpha_{\rm b}}}.
\label{eq:gal_mass_dens}
\ee
Here $(\rho,z)$ are cylindrical coordinates in the galactic frame and $\bar{r}^2 = \rho^2 + (z/q)^2$
The parameters $z_{\rm d}$, $R_{\rm d}$, $r_\textrm{cut}$, $r_0$, $\alpha_{\rm b}$, and $q$ are given in App.~\ref{app:ext_GA}, where we also show a plot of the resulting distribution for distances from the Earth (note that from this point forward in the main text, $z$ refers to redshift).
The BHs are placed in the thin disk, thick disk, and bulge in proportion 31\,:\,12\,:\,7, which corresponds to the approximate mass ratio between the regions.
To test the impact of this choice, we consider the effects of a more diffuse distribution of BHs in space inspired by natal kicks modeled out of Ref.~\cite{Sweeney:2022fxx}, for which we simply double our scale radii and scale heights in our population sampling.

As a final remark, we note that although we consider a wide range of systematics in Tab.~\ref{tab:configs_GA}, there are further variations that have been considered in the literature, see e.g. Refs.~\cite{Olejak:2019pln,Gavilan-Martin:2026zzw}.
Nevertheless, our choices are sufficiently broad that in these cases it is clear how these alternative choices would impact our results from the correlation between the systematic variations and the corresponding results in Fig.~\ref{fig:fig_1}.

\begin{table*}[htb!]
\centering
\renewcommand{\arraystretch}{1.2}
\begin{tabular}{|p{4cm}|p{10cm}|}
\hline
\multicolumn{2}{|c|}{\textbf{Galactic BHs Systematic Configurations}} \\
\hline
\textbf{Name} & \textbf{Description} \\
\hline
Fiducial & Our fiducial analysis is defined by the following choices:
\newline \begin{minipage}[t]{\linewidth}
\begin{itemize}
    \item A Salpeter-like BH mass distributed $\xi(M) \propto M^{-2.35}$ restricted to masses in 5\,$M_{\odot}\,{-}\,20\,M_{\odot}$~\cite{Sprague:2024lgq, Farr:2010tu}
    \item A uniform initial spin distribution $\chi \in U[0,1]$
    \item A BH age distribution that follows the Milky Way SFH as in Ref.~\cite{2019ApJ...887..148F}
    \item Separate BH spatial distributions following the mass density profiles for the thin disk, thick disk, and bulge; see Eq.~\eqref{eq:gal_mass_dens}~\cite{Sprague:2024lgq,McMillan:2011wd, SDSS:2005kst,Bissantz:2001wx}
    \item Include only the lowest 211 superradiant mode
    \vspace{0.1cm}
\end{itemize}
\end{minipage} \\
\hline
Higher Modes & Inclusion of higher superradiant modes \\
\hline
Gauss. Mass Dist. & Change the BH mass distribution to a Gaussian centered at $\mu = 9.2\,M_{\odot}$ and $\sigma = 3.3$ $M_{\odot}$ following the fit in Ref.~\cite{Farr:2010tu}, although restricted to $M \geq 1.4\,M_{\odot}$ \\
\hline
Exp. Mass Dist. & Change the BH mass distribution to a delayed exponential with $M_{\rm min} = 5.3$ $M_{\odot}$ and $M_0 = 4.7$ $M_{\odot}$ following the fit in Ref.~\cite{Farr:2010tu} \\
\hline
Cons. Spin & Conservative spin: initial spin distribution $\chi \in U[0,0.5]$ \\
\hline
Opt. spin & Optimistic spin: initial spin distribution $\chi \in U[0.5,1]$ \\
\hline
Diffuse Spatial Dist. & Same BH spatial distribution but double scale radii and heights, loosely inspired by natal kicks in Ref.~\cite{Sweeney:2022fxx} \\
\hline
Log Age Dist. & Change the BH age distribution to log-uniform between $10^3$--$8 \!\times\! 10^9$\,yrs for the thin disk, $10^3$--$10^{10}$\,yrs for the thick disk, and $10^9$--$1.3 \!\times\! 10^{10}$\,yrs for the bulge~\cite{Sprague:2024lgq} \\
\hline
LIGO Mass Dist. & Change the BH mass distribution to the primary mass distribution observed in the LIGO GWTC-3 catalog~\cite{KAGRA:2021duu} \\
\hline
LIGO Spin Dist. & Change the BH spin distribution to the inferred component spin distribution observed in the LIGO GWTC-3 catalog~\cite{KAGRA:2021duu}  \\
\hline
LIGO Mass+Spin Dist.  & Combines both the LIGO mass and spin distributions \\
\hline
\end{tabular}
\caption{The systematic configurations we consider for the galactic superradiance search; see Fig.~\ref{fig:fig_1}.}
\label{tab:configs_GA}
\end{table*}

%%%%%%%%%%%%%%%%%%%%%%%%%%%%%%%%%
\subsection{Suburban Signals and Systematics}
%%%%%%%%%%%%%%%%%%%%%%%%%%%%%%%%%

For a given astrophysical configuration described above, we simulate an ensemble of $10^8$ BHs, each with a randomly drawn mass, age, location, and spin.
Then, for a fixed axion mass $\mu$, these parameters are used to determine what GW strain -- if any -- that each BH generates at the location of the Earth using the results in Sec.~\ref{sec:superradiance}.
In more detail, for a given BH we first consider the superradiance condition in Eq.~\eqref{eq:SR_condition}.
This dictates which modes of the axion superradiance cloud are allowed to grow, although again for our fiducial model we only consider the $m=1$ mode.
Having identified the relevant modes, we then consider whether the age of the BH permits a complete cloud buildup, $T_c$, as well as an emission timescale $T_h$ such that we would observe the GW today for a BH which emitted the signal a time $r/c$ ago, with $r$ the BH distance.
BHs that pass this criteria then emit a time-dependent GW strain $h$ as given in Eq.~\eqref{eq:h}, which we assume is emitted at a monochromatic linear frequency $f=\omega_R/\pi$.
When considering higher modes, if the $m=1$ cloud cannot be observed today, we trace through the sequential growth and decay each of the $n-1=\ell=m$ modes of the cloud, determining whether any would be presently observable.
As noted in Sec.~\ref{sec:superradiance}, we discard modes for which $T_c > T_h$, where the dissipation is faster than the build up time.

In this manner, we construct an ensemble of strains, $h$, from our extant BHs which continue to radiate superradiance-induced GWs, making up the resolved BH population signal in our galactic axion search.
We show illustrations of the distribution of resulting strains for various scalar masses $\mu$ in App.~\ref{app:ext_GA}.
Additionally, we show examples of the BH properties responsible for the strongest and weakest (with ${\rm SNR} \geq 5$, see the discussion below) individual strains in our fiducial galactic BH ensemble in Tabs.~\ref{tab:GA_BH_list_12} and~\ref{tab:GA_BH_list_11}.
Unsurprisingly, those results demonstrate that the brightest signals tend to be associated with young, nearby, large spin BHs.
However, we point out that these values are not anomalous outliers in our overall distributions of mass, spin, distance, and age, for which we illustrate visual examples in App.~\ref{app:ext_GA}.
In those examples, one can see that the strongest BH values are well within their respective overall distributions such that there would be numerous other signals with similar properties that would reliably show up to surpass the fiducial SNR threshold we later impose for axion sensitivity.
This is to say that the number of strong SNR signals are robust to large outlier effects that may be present in any single BH population simulation.

Given a distribution of strains, our prescription for determining whether that axion mass $\mu$ is observable is as follows.
For each source, the optimal signal-to-noise ratio obtained after an observation campaign lasting $T_{\rm obs}$ is
\be
\SNR^{\rm opt} = \frac{h \sqrt{T_{\rm obs}}}{\sqrt{S_n(f)}},
\ee
set by the strain $h$ as observed at the detector, and the noise-equivalent PSD of the specific detector evaluated at the signal frequency $f$, $S_n(f)$.
This SNR assumes that for the full observation time the signal can be treated as coherent and monochromatic.
Independent of any frequency drift intrinsic to the superradiance GW emission, the Earth's rotation introduces a frequency shift.
If the source location is known, this shift can be accounted for, but as we are performing a blind search accounting for this is computationally expensive~\cite{BradyCoherent:2000,Frasca_2005}. 
Many methods to reduce computational cost while optimizing for sensitivity have been proposed, as described in the review of Ref.~\cite{Riles:2022wwz}.
Here we assume a generic semi-coherent method as has been proposed in the literature, where a full campaign time $T_{\rm obs}$ is divided into $N$ coherently-analyzed subsets of duration $T_{\rm int}$ each~\cite{Astone:2010zz,Astone:2014,Piccinni:2018akm,Dergachev:2019wqa}; this approach has also been applied for galactic superradiance searches, see Ref.~\cite{Zhu:2020tht}.
The SNR is then given by
\be
\SNR = \frac{h \sqrt{T_{\rm int}}\, N^{1/4}}{\sqrt{S_n(f)}}.
\label{eq:Tint-N}
\ee
For our fiducial projections, we consider $T_{\rm int} = 4\,\text{hr}$ and $T_{\rm obs} = 1\,\text{yr}$, so that $N \simeq 2\times10^3$.
The value of $T_{\rm int}$ was chosen as a benchmark value demonstrated for all-sky semi-coherent searches with LIGO data in Ref.~\cite{Dergachev:2019wqa}. 
Additionally, for the signals driving our sensitivity, we show explicitly in App.~\ref{app:superradiance_extras} that the frequency drift of the superradiance signal over the benchmark time is negligible.
We note that increasing $T_{\rm obs}$ only marginally improves the sensitivity.
Beyond the weak $\SNR \propto T_{\rm obs}^{1/4}$ scaling, as shown in App.~\ref{app:ext_GA} -- see, in particular, Fig.~\ref{fig:GA_stat} -- the $\SNR$ varies sharply with $\mu$, which further reduces the sensitivity to the exact threshold.

For the detectors we study, we determine the appropriate noise-equivalent PSD from Refs.~\cite{LIGOScientific:2016aoc,aLIGO_O5} for LIGO, Ref.~\cite{Evans:2023euw} for CE, Ref.~\cite{Hild:2010id} for ET, and Ref.~\cite{Domcke:2024mfu} for the MWB.
For the MWB, when operating resonantly, the instrument can obtain a significantly improved sensitivity at the cost of bandwidth.
Therefore, one has to devise a scan strategy that determines how the total observation time is distributed across various resonant frequency values.
As studied in App.~\ref{app:ext_GA}, the range of frequencies giving rise to the brightest signals can be relatively narrow, a fact which would need to be exploited to develop an optimal resonant strategy.
We have not attempted to develop such a strategy here, and instead to understand what the MWB could achieve, make the extremely optimistic assumption that a strategy can be developed as if the resonant sensitivity of the MWB can be deployed in a broadband manner; for this reason we dot all curves associated with the resonant MWB, indicating they should be taken as maximally optimistic.

The above procedure converts the distribution of BHs to an ensemble of SNRs.
From this, we analyze the quantity $\lambda$, defined as $\lambda = \sum (\SNR \geq 5)$ or the number of signals with an SNR greater than the threshold five.
To compute confidence intervals bracketing purported axion masses, we evaluate $\lambda$ over a range of axion masses $\mu$ and examine where $\lambda \geq 3$, which comes from the mean of a Poisson distribution for which a null 0-count observation is observed in 5\% of repeated trials.
Consequently, masses where $\lambda \geq 3$ constitute our projected 95\% confidence interval sensitivity to the axion mass; a similar procedure was adopted in Ref.~\cite{Fedderke:2024wpy}.
We emphasize that this is an approximate prescription for evaluating sensitivity; to highlight this, a signal hypothesis that predicts a large number of $\SNR = 4.9$ events would surely be testable, but not under our prescription.
Further, when looking for potentially many narrowband features in the frequency domain there will be a look elsewhere effect that must be accounted for in quantifying the significance of any detection, see also the discussion in Ref.~\cite{Arvanitaki:2014wva}.
Lastly, as noted in Ref.~\cite{Zhu:2020tht}, there is an additional effect we have not included: beyond the brightest signals there can be many potentially overlapping sources which can generate an incoherent signal in the detectors.
An optimal analysis that accounts for each of these effects is not straightforward to write down, however, as these scenarios can be simulated, this could be a problem where machine learning techniques such as neural likelihood estimation~\cite{Cranmer:2019eaq,Papamakarios:2019fms} provide a path to an optimal analysis.

With the above qualifications noted, the results of our analysis prescription for the fiducial galactic model are shown in Fig.~\ref{fig:galactic}.
The results suggest that LIGO is already well-placed to perform a superradiance search with a complementary sensitivity to existing searches based on individual BHs with X-ray~\cite{Baryakhtar:2020gao,Witte:2024drg} and GW waveform inferred~\cite{Caputo:2025oap,Aswathi:2025nxa} spin measurements, although future instruments like ET and CE could extend that range.
In addition, a MWB could provide another way to conduct this search, with the advantage of being designed to reach higher frequencies without the need for extrapolated sensitivity (see also Fig.~\ref{fig:GA_observatory_sys} in the appendix).
However, for our fiducial analysis the high-frequency detectors are not particular competitive.
Their reach is dampened by the fact that in regions where a MWB and observatories like LIGO have overlapping frequency ranges, the other observatories tend to have higher sensitivity; whereas the MWBs can extend to higher frequencies than the other observatories, at these high frequencies (high axion masses) there will also be far fewer BHs that can undergo superradiance for most systematic configurations we consider.
Thus, generally, a higher frequency sensitivity is only manifestly advantageous for particular systematic configurations where there are sufficient BHs able to superradiate at large axion masses.
We explore this point further in Sec.~\ref{sec:highmasses}.

Our projections are broadly consistent with earlier Galactic population-level analyses we set out to revisit.
Ref.~\cite{Arvanitaki:2014wva} estimated up to $\sim$$10^4$ detectable GW annihilation signals at aLIGO for roughly $10^{-13} \,\rm{eV} \lesssim \mu \lesssim 10^{-11} \,\rm{eV} $ allowing for higher modes, which is consistent with our higher mode result.
Meanwhile, Ref.~\cite{Zhu:2020tht}, which restricts to the lowest mode, found an ensemble signal exceeding existing all-sky continuous wave upper limits (O1, O2) for $2 \times 10^{-13} \,\rm{eV} \lesssim \mu \lesssim 2.5 \times 10^{-12}\,\rm{eV}$.
Our fiducial results, especially for our projections for O1, are comparable, but the slightly broader projected exclusion range we find likely comes from the choice of adopting the idealized semi-coherent SNR of Eq.~\eqref{eq:Tint-N} rather than quoting upper limits of existing pipelines, as is done in Ref.~\cite{Zhu:2020tht}. 
Another possibility is that Ref.~\cite{Zhu:2020tht} adopts a more complicated natal velocity model for BHs, which would make their spatial distribution more diffuse, though otherwise the astrophysical models adopted in Ref.~\cite{Zhu:2020tht} are roughly explored within our systematic tests.

As emphasized repeatedly, our fiducial results carry systematic uncertainties.
Therefore, in Fig.~\ref{fig:fig_1} we demonstrate how the LIGO O5 sensitivities vary for the full set of variations listed in Tab.~\ref{tab:configs_GA}.
From these results, we see that the three most important variations are the BH age distribution, mass distribution, and the possibility of observing signals from higher modes.
In the case of the final two, LIGO has the possibility of probing axions with masses of $\sim$$10^{-11}\,$eV.
We explore this possibility further in Sec.~\ref{sec:highmasses}, and note that the impact of systematic variations on other instruments is shown in App.~\ref{app:ext_GA}.

%%%%%%%%%%%%%%%%%%%%%%%%%%%%%%%%%
\section{Extragalactic Superradiance}
\label{sec:extragal}
%%%%%%%%%%%%%%%%%%%%%%%%%%%%%%%%%

\begin{figure}
\centering
\includegraphics[width=0.49\textwidth]{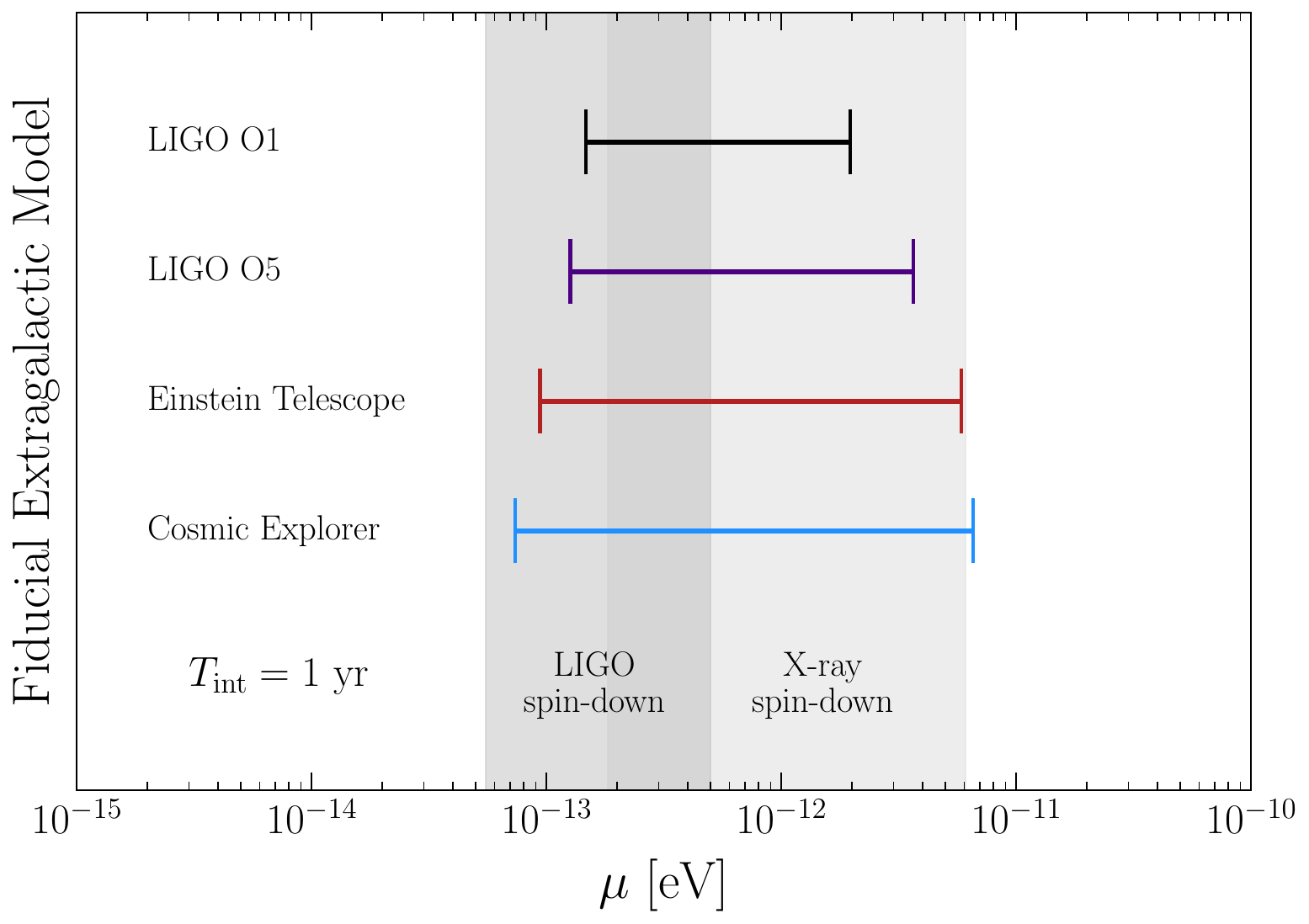}
\vspace{-0.6cm}
\caption{Projected sensitivity to the axion mass $\mu$ under our fiducial extragalactic model for four observatories considered in this work: LIGO O1, LIGO O5, Einstein Telescope and Cosmic Explorer.
Higher frequency detectors like the MWB, which are not competitive for the lower frequency extragalactic searches, are not shown.
In all cases we show the sensitivity to a stochastic signal after one year of integration.
The definition of our fiducial model for the BH parameters is provided in Tab.~\ref{tab:configs_EG}; for the impact of variations in these choices for LIGO O5 and other details, see Fig.~\ref{fig:fig_1}.}
\label{fig:extragalactic}
\end{figure}

The BHs within our own Milky Way represent only a tiny fraction of the BHs expected in the observable universe.
In this section we turn to the superradiance signal arising from all BHs in the cosmos.
While for the galactic BH population we focused on resolvable GW sources, turning to extragalactic BHs we focus on the stochastic GW background that arises from the incoherent superposition of many individual signals, as extragalactic BHs are generally too distant to be resolved individually.
As previously, our goal is to understand the range of axion masses such searches can probe as a function of GW detectors, as shown in Fig.~\ref{fig:extragalactic}, and as we vary our systematic assumptions regarding the distribution of BHs and our treatment of superradiance, see Fig.~\ref{fig:fig_1}.
In the remainder of this section we first outline how to model the extragalactic BH population, detailing in particular the associated uncertainties, before describing how a given model can be combined with the superradiance calculations in Sec.~\ref{sec:superradiance} to obtain the quoted sensitivities.

%%%%%%%%%%%%%%%%%%%%%%%%%%%%%%%%%
\subsection{The Cosmological Black Hole Distribution}
%%%%%%%%%%%%%%%%%%%%%%%%%%%%%%%%%

In this section we describe our modeling of the extragalactic BH population.
Broadly, we require a model for the rate of BH formation, $R_{\rm BH}$, which must depend on the stellar mass distribution, star formation rate, metallicity, star lifetime, and the stellar-BH mass relation, all of which in general are functions of redshift.
We describe each of these ingredients below, focusing primarily on the isolated BH population which serves as the fiducial model in this work.
(As a systematic check, we demonstrate that the addition of a population of BBHs has a negligible impact on our sensitivity, with details in App.~\ref{app:ext_EG} and results in Fig.~\ref{fig:fig_1}.)

Unlike the galactic distribution discussed in Sec.~\ref{sec:gal}, here all key quantities are inherently parameterized in terms of redshift and are generally specified in terms of cosmic densities, for instance per Mpc$^3$.
Additionally, we emphasize that characterizing the entire cosmological BH population introduces a substantially different class of systematic uncertainties.
While the galactic BH distribution is phenomenologically chosen such that it is in line with the (admittedly limited) set of known Milky Way BHs, the extragalactic BH distribution -- excluding mergers -- is primarily inferred from knowledge of the cosmological stellar population.

Accordingly, the starting point are the stars, whose mass distribution is characterized by the initial mass function (IMF) $\xi(M_{\star}) \equiv dN_{\star}/dM_{\star}$.
The IMF describes the number of stars $dN_{\star}$ in a mass interval $(M_{\star},\, M_{\star} + dM_{\star})$.
In a cosmological context, if we interpret $N_{\star}$ as the number of stars per volume element $dV$ and per time interval $(t,\, t+\Delta t)$ -- equivalently redshift interval $(z, \, z+dz)$ -- then we can adjust the normalization of the IMF with time by fixing this as follows
\be
\int_{M^\textrm{min}_\star}^{M_\star^\textrm{max}}\! dM_{\star}\, \xi(M_{\star}) = N_{\star}.
\ee
As in Refs.~\cite{Brito:2017wnc, Brito:2017zvb, Tsukada:2018mbp}, we adopt the conventional Salpeter IMF $\xi(M_{\star}) \sim M_{\star}^{-2.35}$~\cite{Salpeter:1955it}.
As described, we allow the IMF normalization to evolve, however for simplicity we assume that the scaling of the IMF does not change over time, although in general this is not the case.
One example would be the redshift dependence of metallicity indirectly altering stellar populations in the early universe versus later in time.

To obtain the star formation over cosmic time we require a model for the star formation rate (SFR), defined as the mass that forms stars per time and volume at a given cosmic epoch, or $\psi(z) = dM_{\star}/dtdV$.
We choose to parameterize the SFR as follows,
\be
\psi(z) = \nu \frac{a e^{b(z-z_m)}}{a-b+be^{a(z-z_m)}},
\label{eq:SFR}
\ee
where $a$, $b$, $z_m$, and $\nu$ are fitted coefficients.
For our fiducial scenario we take the model from Ref.~\cite{Vangioni:2014axa}, which fixes $\nu = 0.178 \, M_{\odot}/{\rm yr}/{\rm Mpc^3}$, $z_m = 1.72$, $a = 2.37$, and $b = 1.8$.
The specific values are inferred from the galaxy luminosity function derived from observations of star-forming galaxies.
However, there are other choices one could make for these coefficients, for instance those derived from gamma-ray burst (GRB) rates, which has the advantage of being able to trace star formation in faint distant galaxies.
We explore alternate forms of the SFR as part of our systematics analysis.

We combine the IMF and SFR to obtain the fully differential star formation rate at a given time or redshift as,
\be
\frac{dN_{\star}}{dt dV dM_{\star}}(t,M_\star) = \frac{\psi(z)\,\xi(M_{\star})}{\int_{M^\textrm{min}_\star}^{M_\star^\textrm{max}}\!dM_{\star}\, M_{\star} \,\xi(M_{\star})}.
\label{eq:dNdtdVdMs_main}
\ee
Note that the IMF normalization cancels out in this expression; this occurs as the normalization is controlled by the SFR.
To fully specify this expression we have to give a value to the minimum and maximum stellar masses.
For the lower bound, we take as $M^\textrm{min}_\star=0.1\, M_\odot$ following Refs.~\cite{Salpeter:1955it, Brito:2017wnc, Brito:2017zvb, Tsukada:2018mbp}.
The integral is not particularly sensitive to the upper bound, though for concreteness we set $M_\star^\textrm{max}=100 \, M_\odot$ as a rough estimate of the upper bound of observed stellar masses.

Finally, with the full star formation rate specified, we return to the problem of interest: the BH formation rate, $R_{\rm BH}$.
To construct this, we need a model for the stellar lifetimes, $\tau(M_\star)$, that captures how long the stars live before collapsing and enforces a mass-dependent delay between the stellar and BH birth rates; for this we adopt the model of Ref.~\cite{Schaerer:2001jc}.
Further, we need a relation between the stellar and remnant masses, which captures the mapping $M_\star \to M$ from the stellar to BH masses.
For this, we introduce a relation $M = g[M_\star,Z(z)]$, where $Z(z)$ is the redshift dependent stellar metallicity.
In practice, to obtain the distribution as a function of BH masses we invert the relation to specify what stellar mass gave rise to a BH of a given mass, which we denote as $M_\star = g^{-1}[M,Z(z)]$.
For $g$, we adopt the delayed model of Ref.~\cite{Fryer:2011cx} and for $Z(z)$ we use the prescription of Ref.~\cite{Ma:2015ota}.
Although not considered in our work, we note that the star-to-BH mapping is more complex than the above treatment and in general could include dependencies on spins, initial magnetic fields, amongst other effects.

Of course, not all stars form BHs.
If, however, the remnant mass falls within the range $[5 \, M_{\odot},\, 50 \, M_{\odot}]$ we assume that a BH has formed.
This choice is consistent with conservative modeling of observed and theoretical stellar BH masses, although we relax this assumption as part of our systematics analyses.
In particular, the upper bound $50 \, M_\odot$ is a conservative estimate for the onset of the pair-instability mass gap~\cite{Belczynski:2016jno,Farmer:2019jed}.
However, the exact mass range of the pair instability gap could vary.
Recent GW observations suggest the existence of BHs with masses above $50\, M_\odot$~\cite{LIGOScientific:2025rsn}.
Other studies place the onset of the instability gap around $70\,M_\odot$~\cite{Woosley:2021xba,Farag:2022jcc} while models that include progenitor spin and magnetic fields can account for the recently observed BH masses above $70\,M_\odot$~\cite{Gottlieb:2025ugy}.
Motivated by these uncertainties, we vary the bound of the BH mass range as part of our systematics analysis.

Taken together, the BH formation rate $R_{\rm BH}$ can be written as,
\bea
R_{\rm BH} (z,M) &\equiv \frac{dN}{dt dV dM} (z,M) \\ &=  \int dM\,	\frac{dN_{\star}}{dt dV dM_{\star}}(t(z)-\tau(M_{\star}),M_{\star}) \\ &\times  \delta(M_{\star}-g^{-1}[M, Z(z)]).
\eea
Note that this expression encapsulates that the BH formation rate at a given moment is set by the star formation rate at an earlier time, with the offset controlled by the stellar lifetime $\tau(M_\star)$.

Finally, as was the case for galactic BHs, there is considerable uncertainty on the spin distribution of stellar BHs in the extragalactic population.
While we continue to adopt a uniform $\chi \in U[0,1]$ in our fiducial model, as chosen in the previous studies of Refs.~\cite{Brito:2017wnc, Brito:2017zvb, Tsukada:2018mbp}, we explore uncertainties on this choice as part of our systematics analysis.

\begin{table*}[!t]
\centering
\renewcommand{\arraystretch}{1.2} % for nicer spacing
\begin{tabular}{|p{4cm}|p{10cm}|}
\hline
\multicolumn{2}{|c|}{\textbf{Extragalactic BHs Systematic Configurations}} \\
\hline
\textbf{Name} & \textbf{Description} \\
\hline
Fiducial & The configuration for our fiducial analysis is defined as follows:
\newline \begin{minipage}[t]{\linewidth}
\begin{itemize}
  \item SFR $\psi(z)$ from Ref.~\cite{Vangioni:2014axa} (galaxy lum. function)
  \item IMF $\xi(M_{\star})$ from Ref.~\cite{Salpeter:1955it}
  \item Allowable BH mass range: $5 \, M_{\odot} \leq M \leq 50 \, M_{\odot}$
  \item Stellar lifetimes $\tau(M_{\star})$ from Ref.~\cite{Schaerer:2001jc}
  \item BH mass - stellar mass relation $g[M_{\star}, Z(z)]$ from Ref.~\cite{Fryer:2011cx}
  \item Metallicity evolution $Z(z)$ from Ref.~\cite{Ma:2015ota}
  \item Uniform initial spin distribution, $\chi \in [0,1]$
  \item Non-instantaneous superradiance emission $\dot{E}_h \neq E_h \delta(t-t_b)$
  \item Only the 211 superradiant mode is included
  \vspace{0.1cm}
\end{itemize}
\end{minipage} \\
\hline
Higher Modes & Inclusion of higher superradiant modes \\
\hline
Low BHs & Lower BH mass cutoff $5 \to 2\,M_{\odot}$ \\
\hline
High BHs & Upper BH mass cutoff $50 \to 100\,M_{\odot}$ \\
\hline
Cons. Spin & Conservative spin: initial spin distribution $\chi \in [0,0.5]$ \\
\hline
Opt. spin & Optimistic spin: initial spin distribution $\chi \in [0.5,1]$ \\
\hline
SFR GRB & Gamma-Ray Burst (GRB)-inferred alternate SFR $\psi(z)$ from Ref.~\cite{Vangioni:2014axa} \\
\hline
With BBH Mergers & Add the BBH merger population, which follows a different mass and spin distribution (see App.~\ref{app:ext_EG}), adopting the prescription in Ref.~\cite{Tsukada:2018mbp} \\
\hline
LIGO Mass Dist. & As for the galactic systematics in Tab.~\ref{tab:configs_GA} \\
\hline
LIGO Spin Dist. & \multicolumn{1}{|c|}{\textquotedbl} \\
\hline
LIGO Mass+Spin Dist.  & \multicolumn{1}{|c|}{\textquotedbl} \\
\hline
\end{tabular}
\caption{The list of systematic configurations we consider for the extragalactic search.
The results are shown in Fig.~\ref{fig:fig_1}.}
\label{tab:configs_EG}
\end{table*}

%%%%%%%%%%%%%%%%%%%%%%%%%%%%%%%%%
\subsection{Stochastic Signals and Systematics}
%%%%%%%%%%%%%%%%%%%%%%%%%%%%%%%%%

Having derived the rate of BH formation, $R_{\rm BH}$, we can now combine this with the axion superradiance calculation of Sec.~\ref{sec:superradiance} to obtain the local energy density in GWs, given by\footnote{Our extragalactic results are all computed using Eq.~\eqref{eq:EG_Omega}.
Nonetheless, to fully validate our procedure, we confirmed that a Monte Carlo simulation of the BH distribution produced matching results to what we show here.
The details of this cross check are provided in App.~\ref{Sec:MC_extragalactic}.}
\bea
\Omega_h(f) = \frac{f}{\rho_c} \int &dz dM d\chi dt_b\, \frac{dt}{dz} \frac{dN}{dt_b dV dM d\chi} \\ 
\times &\frac{dE_h}{dt}[t-t_b] \delta(f(1 + z) - f_s),
\label{eq:EG_Omega}
\eea
where $\rho_c = 3 H_0^2/8\pi G$ is the critical density and $f_s = \omega_R/\pi$ is the emitted GW linear frequency in the BH source frame (the $\delta$ function indicates we neglect any frequency drift, see Sec.~\ref{sec:superradiance}).
Examples of this distribution for different axion masses are shown in Fig.~\ref{fig:EG_Omegah} in the appendices.
This expression is derived in App.~\ref{app:Omegah}, although here we emphasize that it is derived by integrating over redshift $z$, BH mass $M$, BH birth time $t_b$, and spin $\chi$.
Here, $dE_h/dt$ is the time dependent GW emission from a BH given in Eq.~\eqref{eq:dotEh}, evaluated at the time since the BH birth, $t-t_b$.
This expression, of course, depends on the BH mass, spin, and the axion mass.
Further, as in the galactic case by default we consider only emission from the lowest 211 mode, exploring higher modes as a systematic.
We note Eq.~\eqref{eq:EG_Omega} differs from the expression commonly adopted in the literature which assumes all the GWs are emitted instantaneously at the moment of BH birth, see e.g. Refs.~\cite{Phinney:2001di,Brito:2017wnc, Brito:2017zvb, Tsukada:2018mbp}.
For our fiducial model where only the $m=1$ mode is included, Fig.~\ref{fig:Tg_Th} reveals that the relevant timescales are cosmologically short except for small $\mu$; therefore, the instantaneous approximation should generally be a good approximation for the lowest mode as we confirm in Fig.~\ref{fig:EG_Omegah}.
Nonetheless, for lower masses and higher modes there can be a significant discrepancy.

In summary, for a given axion mass $\mu$ and BH population model, we can compute $\Omega_h(f)$, which we can then combine with an experimental noise PSD, $S_n(f)$, to quantify whether this axion could be detectable.
For this, we compute the SNR of the signal as follows~\cite{Allen:1997ad}
\be
\SNR = \frac{3 H_0^2}{4 \pi^2} \left[ T_{\rm int} \int_0^\infty\! df\, \left( \frac{\Omega_h(f)}{f^3 S_n(f)} \right)^2 \right]^{1/2}\!,
\ee
where $T_{\rm int}$ is the total integration time, assumed to be 1\,yr for our fiducial analysis.
For the extragalactic analysis, longer observations can be of greater benefit than for the galactic analysis (cf. the discussion below Eq.~\eqref{eq:Tint-N}).
In particular, with a 10\,yr campaign, the fiducial LIGO O5 improves the lower edge of its mass reach by 10\% and the upper edge by just over 20\%.
Using this, we classify an axion as within reach when the expected signal predicts $\SNR \geq 8$, which we take as our sensitivity threshold following e.g. Refs.~\cite{Brito:2017zvb,Brito:2017wnc}.

Implementing the above prescription leads to the projected sensitivities as a function of axion mass for our fiducial astrophysical model shown in Fig.~\ref{fig:extragalactic}, where we find that LIGO O1 and O5 would have sensitivity from roughly $1 \times 10^{-13}\,\textrm{eV} \lesssim \mu \lesssim 4 \times 10^{-12}\,\textrm{eV}$, while future observatories like CE and ET will have extended sensitivities; CE in particular can extend from $7 \times 10^{-14}\,\textrm{eV} \lesssim \mu \lesssim 7 \times 10^{-12}\,\textrm{eV}$.
As shown, the fiducial LIGO sensitivity would be competitive with current constraints in just 1 year of observation time. 
In fact, LIGO has already performed searches for isotropic stochastic backgrounds using the O1 and O2 observing runs~\cite{LIGOScientific:2019vic}, and more recently with the O4 run~\cite{LIGOScientific:2025bgj}.
Although the results are not cast in a form we can immediately recast to a sensitivity, comparing their sensitivity to our templates for $\Omega_h(f)$, it appears clear that this dataset could already be used to constrain the axion parameter space.
Any such analysis would also need to account for the expected backgrounds arising from binary BHs and neutron stars~\cite{LIGOScientific:2017zlf}, which we have not considered here.

We next explore variations on our fiducial analysis for the extragalactic BH population search.
These involve the aforementioned astrophysical uncertainties on quantities such as the BH mass and spin distributions, and on the underlying SFR model.
We also explore the resulting sensitivity if one includes higher modes, and whether one also includes the BBH merger population (whose details are in App.~\ref{app:ext_EG}).
We list the ensemble of systematic configurations we examine in Tab.~\ref{tab:configs_EG}.
Then, we show how our axion mass sensitivities for LIGO O5 are affected from these systematic configurations in the right panel of Fig.~\ref{fig:fig_1}.
From these projections, we see that our extragalactic projections are generally quite robust.
The unique opportunity provided by extragalactic searches is to push to lower axion masses than generally accessible to galactic searches, and that remains true across the full set of systematic variations considered.
If one extends the BH population to higher masses, by increasing the upper end of the remnant mass window or using the LIGO inferred distribution of BH masses, the heavier BHs opens sensitivity to lighter axions still.
Further results exploring the effect of systematic uncertainties on the extragalactic BH axion search using other observatories are illustrated in App.~\ref{app:ext_EG}.

%%%%%%%%%%%%%%%%%%%%%%%%%%%%%%%%%
\section{Pushing to Higher Axion Masses}
\label{sec:highmasses}
%%%%%%%%%%%%%%%%%%%%%%%%%%%%%%%%%

Arguably the most tantalizing aspect of our results is that certain systematic configurations show that the resolvable galactic BH searches have the potential to probe axion masses substantially higher than existing constraints (see Fig.~\ref{fig:fig_1}).
In this section, we focus in on that exact possibility and study two particular systematic configurations meant to examine how high in axion mass superradiance might be able to achieve in the most optimistic scenario.

For both scenarios we adopt the optimistic log age distribution of Ref.~\cite{Sprague:2024lgq}.
In the first scenario, we combine our Gaussian mass distribution from Tab.~\ref{tab:configs_GA} \textit{with the addition} of higher modes to explore the maximal sensitivity to higher axion masses given a plausible galactic BH mass distribution; we reiterate that as outlined in Sec.~\ref{sec:superradiance} the higher modes are associated with considerable theoretical uncertainty (note that the existing X-ray spin-down constraints include both the $m=1,2$ modes~\cite{Baryakhtar:2020gao,Witte:2024drg}, with the latter also incorporating the effects of higher modes).
The second scenario invokes beyond-the-Standard-Model physics to source significantly smaller BH masses, in turn extending sensitivity to high axion masses.
In particular, we consider a scenario in which sub-stellar mass BHs can be formed through stellar capture of dark matter and subsequent collapse into a BH.
In this scenario, whose details are presented in App.~\ref{app:dm_collapse}, the BH mass distribution is extended with a sub-solar IMF (of slope $\alpha = 0.5$) reaching masses as small as $M = 0.075 \, M_{\odot}$, the limit for hydrogren core burning.
Note technically this scenario could allow more total BHs, although by default we keep the number fixed to $10^8$.
Another path to generating lighter BHs would be primordial black holes (PBHs), however we do not consider that here; as reviewed in App.~\ref{app:dm_collapse}, it is expected PBHs form with small spins, which presents a challenge to generating a large superradiance signal.

\begin{figure}[!tbh]
\centering
\includegraphics[width=0.49\textwidth]{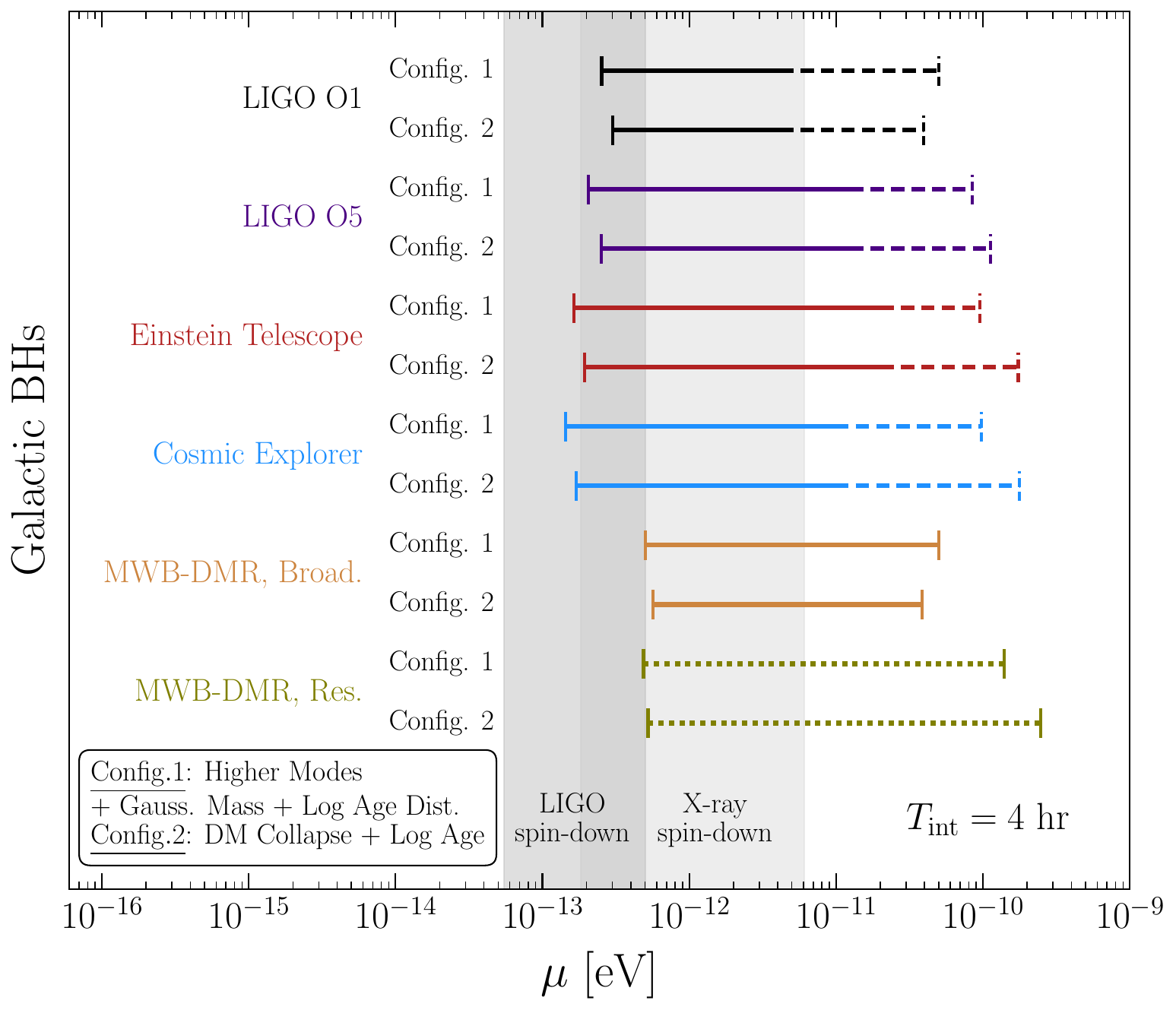}
\vspace{-0.4cm}
\caption{Sensitivity for two systematic configurations of our galactic BH ensemble particularly conducive to obtaining sensitivity to higher axion masses.
These two systematic configurations correspond to the inclusion of higher modes in addition to the Gaussian mass distribution (Config. 1), along with the DM Collapse BH mass distribution (Config. 2). For both configurations, we also adopt the log age distribution for BH ages.
We compare the sensitivities across all of our considered observatories using the benchmark $T_{\rm int} = 4$ hr.
Recall the dashed curves correspond to assuming the interferometers can extend their sensitivity range to higher frequencies.}
\label{fig:GA_high_mu}
\end{figure}

We illustrate, for all observatories considered in this work, the sensitivity to axion masses in these two scenarios in Fig.~\ref{fig:GA_high_mu}.
In doing so, we have implemented the same fixed $T_{\rm int} = 4\,$hrs and $T_{\rm obs} = 1\,$yr adopted in Figs.~\ref{fig:fig_1} and~\ref{fig:galactic} (see Eq.~\eqref{eq:Tint-N}).
For higher axion masses, the frequency drift of the signal grows larger and larger, reducing the period of time we can treat the signal as coherent.
However, as shown in App.~\ref{app:superradiance_extras}, for the 211 mode this drift is at most $\simeq 0.2\%$ of the signal frequency.
We do not expect it to be significantly larger for higher modes, although this is a point that remains to be substantiated to put the results for the first configuration in Fig.~\ref{fig:GA_high_mu} on firmer footing.
Regardless, the key takeaway from the results is that we see many instruments have the potential to push up to masses approaching $10^{-10}\,$eV.

It is worth reiterating a point discussed in Sec.~\ref{sec:gal}: the resonant MWB sensitivity must be interpreted carefully.
Due to the narrowband nature of the resonator, the device would have a lower duty cycle than the broadband detectors, and a scan strategy would have to be implemented.
In Fig.~\ref{fig:GA_high_mu}, we have made the simplifying assumption that at any given mass, and therefore GW frequency, the experiment would collect data for $T_{\rm int} =4\,$hr.
The sensitivity range shown is therefore indicative of the range of masses to which a resonant MWB is sensitive, not the range that could actually be tested in an integration time of a year.
However, it should be noted that the timescale on which such a device is likely to be built is long enough that LIGO might have narrowed the non-excluded mass range significantly, so that scanning the entire mass range is not necessary.
Finally, in the coming years, LIGO binary mergers might give us a better indication of the mass and spin distribution of galactic BHs, further narrowing down the parameter space that would need to be scanned.

If such a sensitivity could be achieved robustly, it would represent an exceptional discovery opportunity.
Indeed the mass range $\mu \sim 10^{-10}\,\mathrm{eV}$ is well-motivated from both experimental and theoretical perspectives.
Experimentally, axions with $\mu \lesssim 10^{-9} \,\mathrm{eV}$ remain largely below the reach of existing microwave-cavity haloscope experiments.
One of the most ambitious proposed axion-photon haloscopes is DMRadio-GUT and the lowest mass it is projected to reach the QCD axion prediction for is $4 \times 10^{-10}\,\textrm{eV}$.
CASPEr-Electric, an NMR based haloscope, could probe this mass range, but only if it reached its most optimistic design specifications~\cite{Budker:2013hfa}; cf. Ref.~\cite{Dror:2022xpi}.
Theoretically, the higher mass range is of interest due to arguments that UV completions involving string theory and grand unified theories (GUTs) can naturally populate this space.
In particular, requiring that the string and Kaluza–Klein scales lie above the grand-unification scale $M_{\rm GUT} \sim 10^{16} \,\mathrm{GeV}$ constrains the axion decay constant to be $f_a \sim 10^{15}$--$10^{17}\,\mathrm{GeV}$, corresponding to axion masses roughly in the range $10^{-11} \, \mathrm{eV} \lesssim \mu \lesssim10^{-8} \, \mathrm{eV}$~\cite{Benabou:2025kgx}.
Figure~\ref{fig:GA_high_mu} highlights that superradiance may be a path into this window.
Further, it is clear that unless interferometers can be modified to achieve their full high frequency potential, these searches could represent a unique discovery opportunity for devices like MWB and other high frequency GW detectors.

Of course, we emphasize once more that there are significant assumptions underpinning the analysis in this section.
At minimum, if the hints LIGO has obtained as to the existence of sub 5\,$M_{\odot}$ BHs is solidified~\cite{LIGOScientific:2020zkf,LIGOScientific:2024elc}, we take these results as motivating of further theoretical work to understand how higher superradiant modes can be included in our calculations in a controlled manner.
Further, it will be important to understand whether the optimistic MWB resonant sensitivity shown can be achieved with a realistic search strategy (see the discussion in Sec.~\ref{sec:gal}).
Nevertheless, the results in Fig.~\ref{fig:GA_high_mu} may be one of the most compelling science cases for high-frequency detectors, which is a notoriously challenging domain, see e.g. Refs.~\cite{TitoDAgnolo:2024res,Gasparotto:2025wok,Berlin:2026che}.

%%%%%%%%%%%%%%%%%%%%%%%%%%%%%%%%%
\section{Conclusion}
\label{sec:conclusion}
%%%%%%%%%%%%%%%%%%%%%%%%%%%%%%%%%

Superradiance is an exceptionally promising path to axion discovery.
If an ultralight boson simply exists in the spectrum of nature the startling consequence could be that BHs throughout the cosmos are shining GWs towards us, a testable hypothesis in the GW era.
Here we have taken steps to further quantify the axion reach of population-based studies. 
Already from our results in Fig.~\ref{fig:fig_1} it is clear that population-based studies can complement existing superradiance searches.
The ability to probe masses as low as $10^{-13}\,$eV appears relatively robust to systematics, whereas the high mass reach varies far more dramatically with systematic choices.
Of course, as we continue to learn the properties of BHs throughout the Milky Way and universe more broadly, these systematic uncertainties will be reduced.
However, even before that we argue superradiant searches are important: our results highlight that under optimistic assumptions axions even up to $\sim$$10^{-10}\,$eV could be discoverable, in which case axions could be the messenger through which we infer the BH population properties.

Although this work aims to provide a relatively comprehensive analysis of the population approach to superradiance, there remain a number of important open directions for future work.
For one, we have only considered a limited set of future GW detectors.
Our findings complement studies for the Levitated Sensor Detector~\cite{Sprague:2024lgq}; however, there are many other detectors one could consider, such as NEMO~\cite{Ackley:2020atn}, or lower frequency observatories such as AION~\cite{Badurina:2019hst} and MAGIS~\cite{MAGIS-100:2021etm}.
Secondly, there will undoubtedly be additional challenges in realizing the population level analyses presented in this work, especially in the case of searching for galactic monochromatic signals.
As noted already, our treatment involved a number of simplifying assumptions, and it is likely implementing an optimal analysis would require deploying techniques from machine learning.
Finally, the most significant omission in the present work is a study of the sensitivity of our results as a function of $f_a$.
Understanding both this and the reliability of pushing to higher modes is essential to quantifying how promising this search will for QCD axions.
Taken together, these improvements would sharpen the picture of superradiance as a discovery channel for new physics.

%%%%%%%%%%%%%%%%%%%%%%%%%%%%%%%%%
\section*{Acknowledgments}
%%%%%%%%%%%%%%%%%%%%%%%%%%%%%%%%%
%
We thank Valerie Domcke for collaboration in the early stages of this work.
Our work benefited from discussions with Asimina Arvanitaki, Christopher Dessert, Reed Essick, Michael Fedderke, Junwu Huang, Giuseppe Lucente, Giacomo Marocco, and Divya Singh.
We especially thank Richard Brito, Jacob Sprague, and Leo Tsukada for detailed feedback on their implementation of superradiance which helped us resolve several differences in the literature.
Lastly, we thank Masha Baryakhtar, Valerie Domcke, Reed Essick, Ben Safdi, and Sam Witte for feedback on a draft version of this work, and the anonymous referee for comments that further improved our work.
The work of ON is supported in part by the DOE award DESC0025293, as well as the NSF Graduate Research Fellowship Program under Grant DGE2146752.
The research of NLR was supported by the Office of High Energy Physics of the U.S. Department of Energy under contract DE-AC02-05CH11231.
JSE was supported by the National Science Foundation under cooperative agreement 202027 and by the by Japan Science and Technology Agency (JST) as part of Adopting Sustainable Partnerships for Innovative Research Ecosystem (ASPIRE), Grant Number JPMJAP2318.

\appendix

%%%%%%%%%%%%%%%%%%%%%%%%%%%%%%%%%
\section{Further Details on Superradiance}
\label{app:superradiance_extras}
%%%%%%%%%%%%%%%%%%%%%%%%%%%%%%%%%

In this appendix we provide several technical details of our superradiance implementation not included in the main text.

Firstly, for the computation of the cloud frequency $\omega = \omega_R + i \omega_I$ we use a combination of analytic and numeric methods.
For large $\alpha$ all results are computed numerically, although we ensure these match onto perturbative expressions; indeed our results in Fig.~\ref{fig:Tg_Th} use a combination of perturbative and numerical results.
The full perturbative expressions are given by~\cite{Baumann:2019eav},
\bea
\zeta_R &= 1 - \frac{\alpha^2}{2n^2} - \frac{\alpha^4}{8n^4} + \frac{f_{n\ell}}{n^3} \alpha^4 + \frac{h_{\ell}}{n^3} \chi m \alpha^5, \\
\zeta_I &= C_{n\ell} g_{\ell m} (m\chi - 2 \bar{r}_+ \alpha \xi_R) \alpha^{4\ell+4},
\label{eq:xiRI}
\eea
with $\zeta = \omega/\mu$.
The frequencies are defined using the following auxiliary quantities,
\bea
f_{n\ell} &= - \frac{6}{2\ell+1} + \frac{2}{n},\hspace{0.5cm}
h_{\ell} = \frac{16}{2\ell(2\ell+1)(2\ell+2)}, \\
C_{n\ell} &= \frac{2^{4\ell+1}(n+\ell)!}{n^{2\ell+4}(n-\ell-1)!} \left[ \frac{\ell!}{(2\ell)!(2\ell+1)!} \right]^2\!, \\
g_{\ell m} &= \prod_{k=1}^{\ell} \big[k^2 (1-\chi^2) + (\chi m - 2 \bar{r}_+ \alpha \zeta_R)^2 \big].
\eea
From these equations, $\zeta_I > 0$ requires $m\chi - 2 \bar{r}_+ \alpha \zeta_R > 0$, or equivalently $\Omega_H > \omega_R/m$, exactly as in Eq.~\eqref{eq:SR_condition}.
Further, in the limit of $\alpha \ll 1$, we can read off Eq.~\eqref{eq:tauc_sa} from the form of $\zeta_I$.

To obtain numerical results for $\omega$ we follow the approach laid out in Ref.~\cite{Dolan:2007mj} and further make use of the public code provided in Ref.~\cite{Superradiance_code}.
The starting point is the Kerr metric in Boyer--Lindquist coordinates,
\begin{align}
ds^2 &= \left(1-\frac{2 M r}{\rho^2}\right) dt^2+\frac{4 \chi M^2 r \sin^2\theta}{\rho^2} dt d\phi \nonumber\\ 
&- \frac{\rho^2}{\Delta} dr^2 - \rho^2 d\theta^2 \\ 
&-\left[(r^2+\chi^2M^2)+\frac{2 M r}{\rho^2} \chi^2 M^2 \sin^2\theta\right] \sin^2\theta\, d\phi^2, \nonumber
\end{align}
where we introduce $\Delta = r^2 -2 M r + \chi^2 M^2$ and $\rho^2 = r^2 + \chi^2 M^2 \cos^2\theta$.
As noted in the main text, the solutions for a scalar field in this background are separable and can be written as $\Phi = e^{-i\omega t} e^{i m \phi}\, R(r)\,S(\theta)$.
Substituting this into the scalar field equation, Eq.~\eqref{eq:KG-Kerr} we obtain an equation for the radial function,
\begin{align}
\partial_r (\Delta \partial_r R)
+ &\Bigg[ \frac{\omega^2(r^2+\chi^2 M^2)^2-4\chi M^2 m \omega r + m^2 \chi^2 M^2}{\Delta} \nonumber\\
&- \omega^2 \chi^2 M^2 - \mu^2 r^2 - \Lambda \Bigg] R = 0,
\end{align}
and spheroidal function,
\bea
\frac{1}{\sin \theta} \partial_{\theta}(\sin \theta\, &\partial_{\theta} S) + \Big[  \Lambda - \frac{m^2}{\sin^2 \theta} \\
&+ \chi^2 M^2 (\omega^2-\mu^2) \cos^2 \theta\Big] S = 0.
\eea
Here $\Lambda$ is the separation constant.

\begin{figure*}
\centering

\begin{minipage}{0.42\textwidth}
\includegraphics[width=\linewidth]{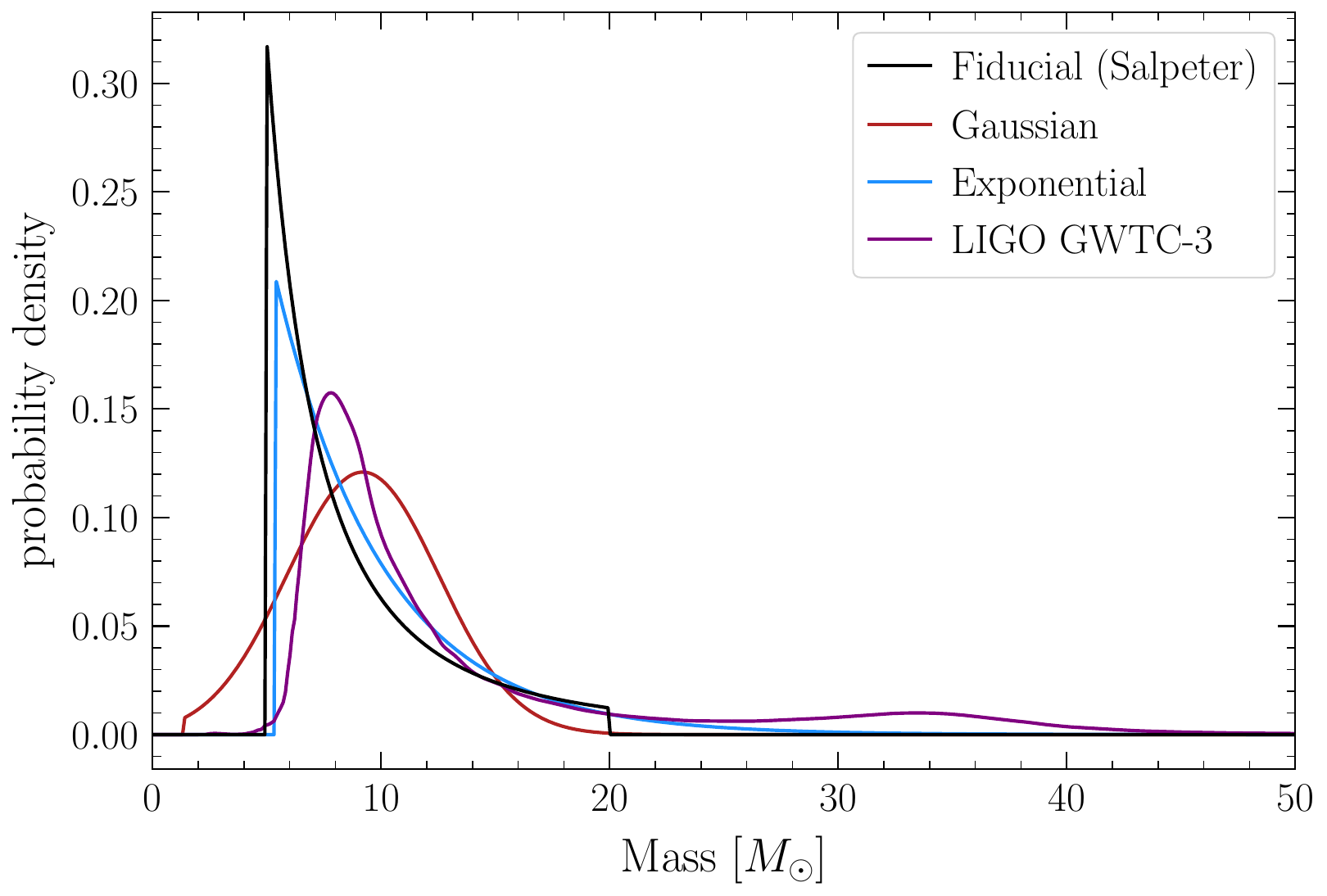}
\end{minipage}
\hspace{0.3cm}
\begin{minipage}{0.42\textwidth}
\includegraphics[width=\linewidth]{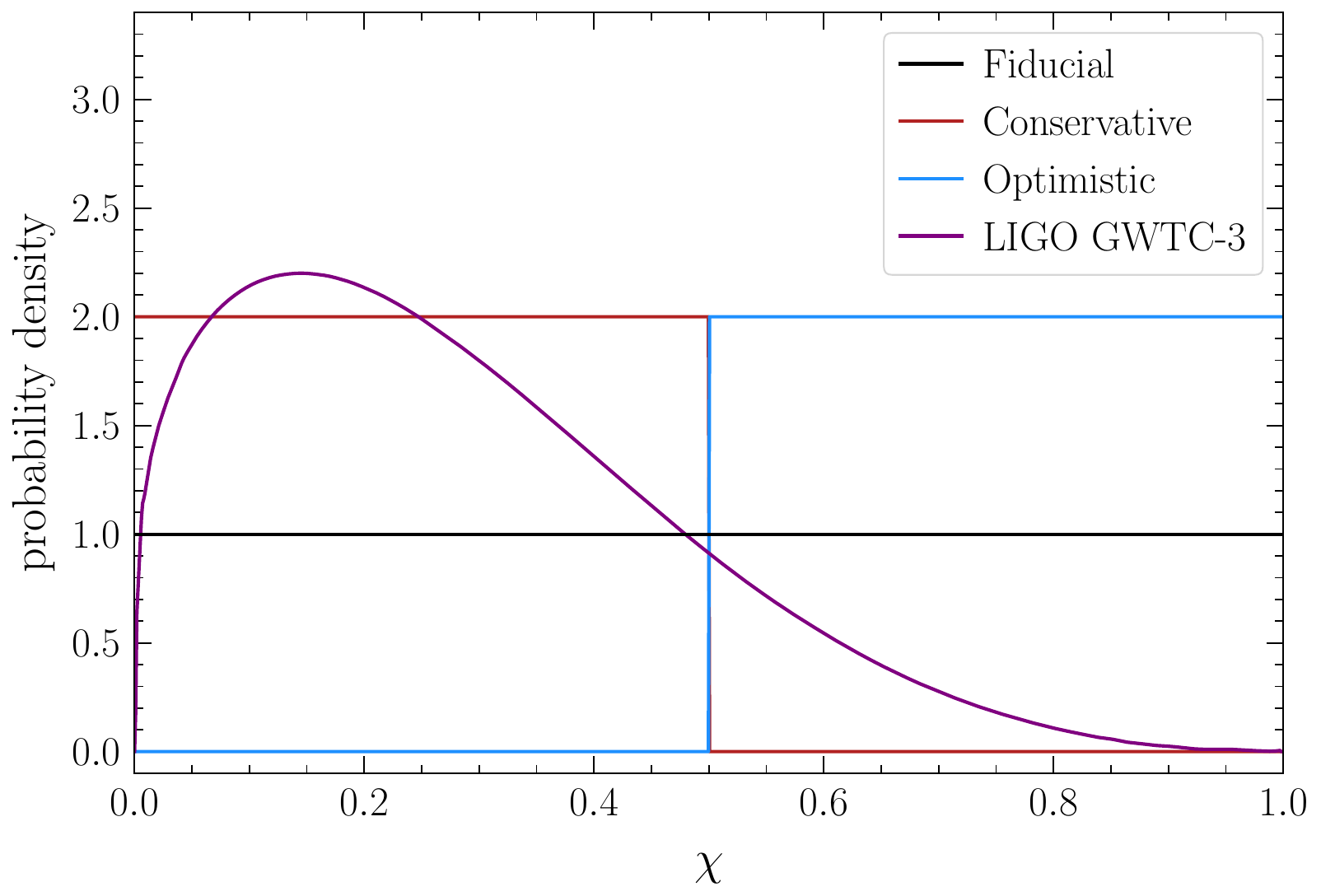}
\end{minipage}

\vspace{0.3cm}

\begin{minipage}{0.42\textwidth}
\includegraphics[width=\linewidth]{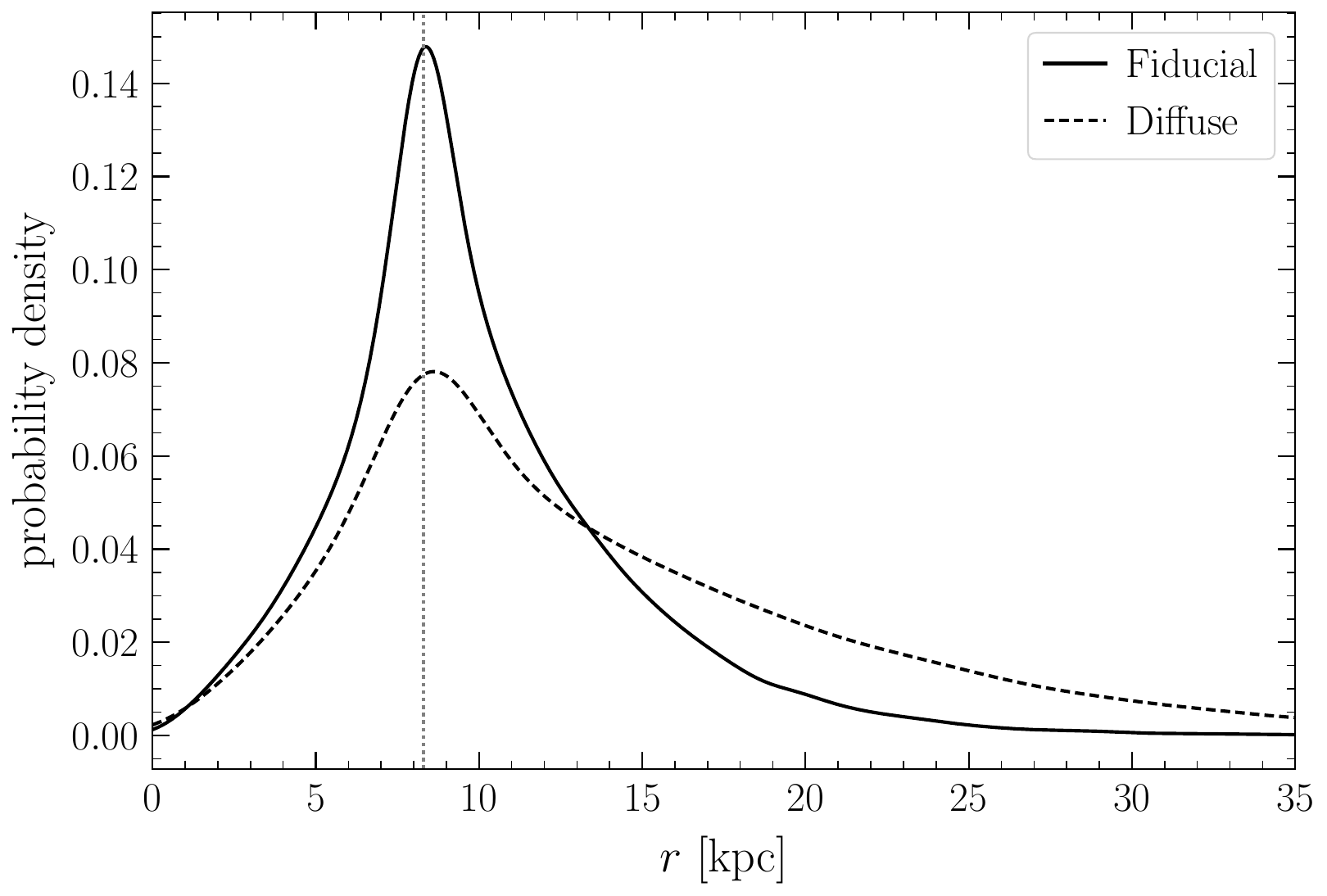}
\end{minipage}
\hspace{0.3cm}
\begin{minipage}{0.42\textwidth}
\includegraphics[width=\linewidth]{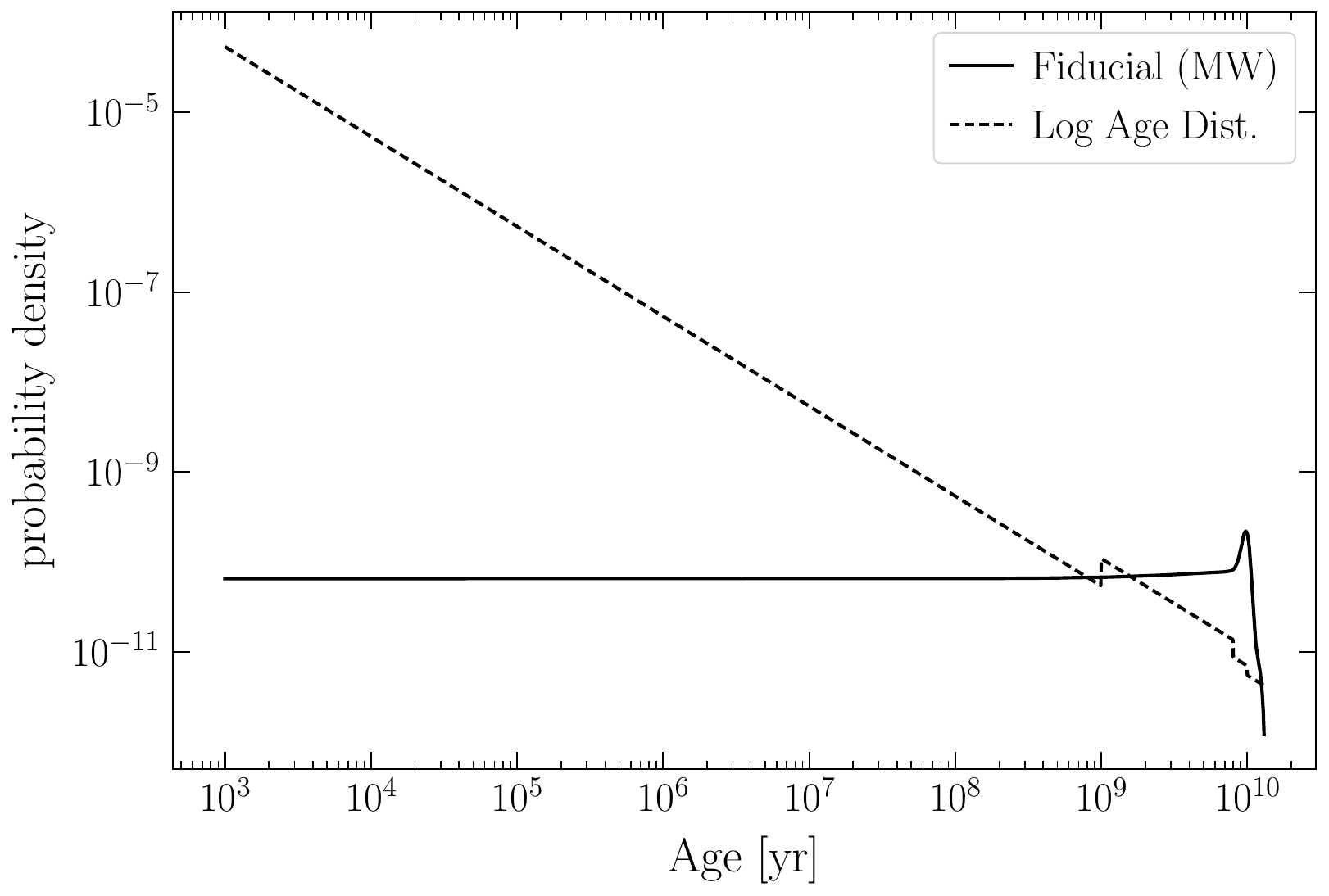}
\end{minipage}
\vspace{-0.2cm}
\caption{Comparisons of the astrophysical systematic configurations we explore for our galactic BH analysis.
We show the distributions for mass (top left), spin (top right), distance (bottom left), and age (bottom right).
For the distance distribution, the solar distance with respect to the galactic center is shown as a vertical dotted line.}
\label{fig:GA_show_configs}
\end{figure*}

\begin{figure*}
\centering
\includegraphics[width=0.49\textwidth]{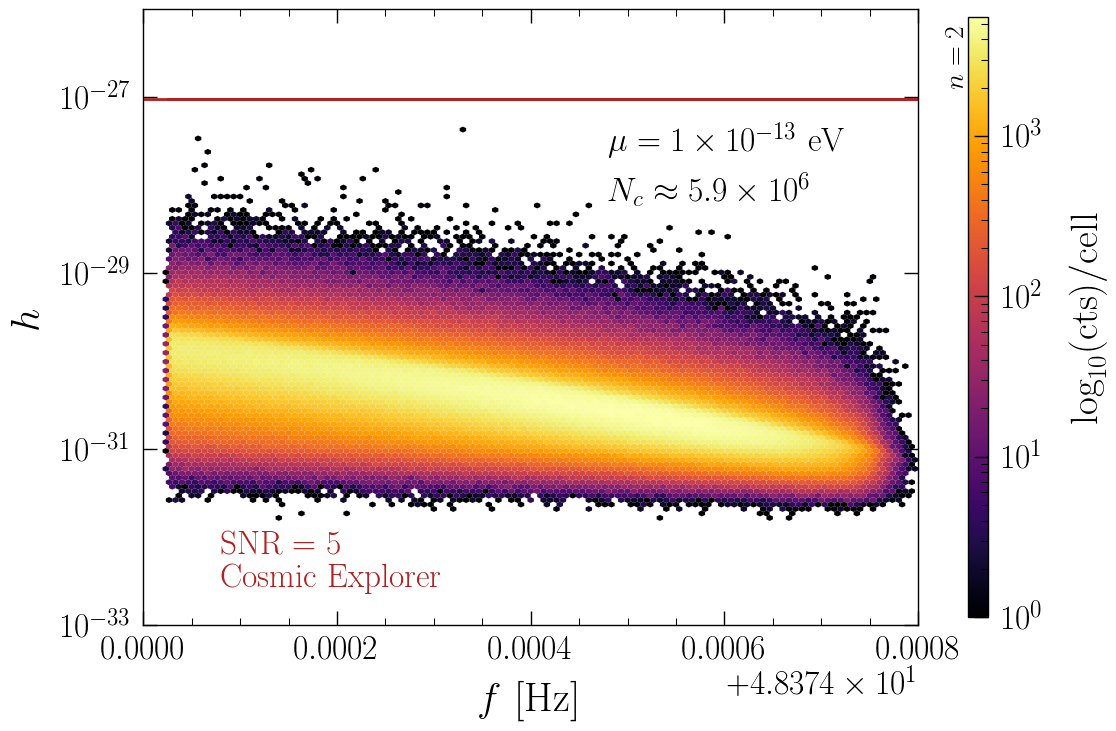}
\includegraphics[width=0.49\textwidth]{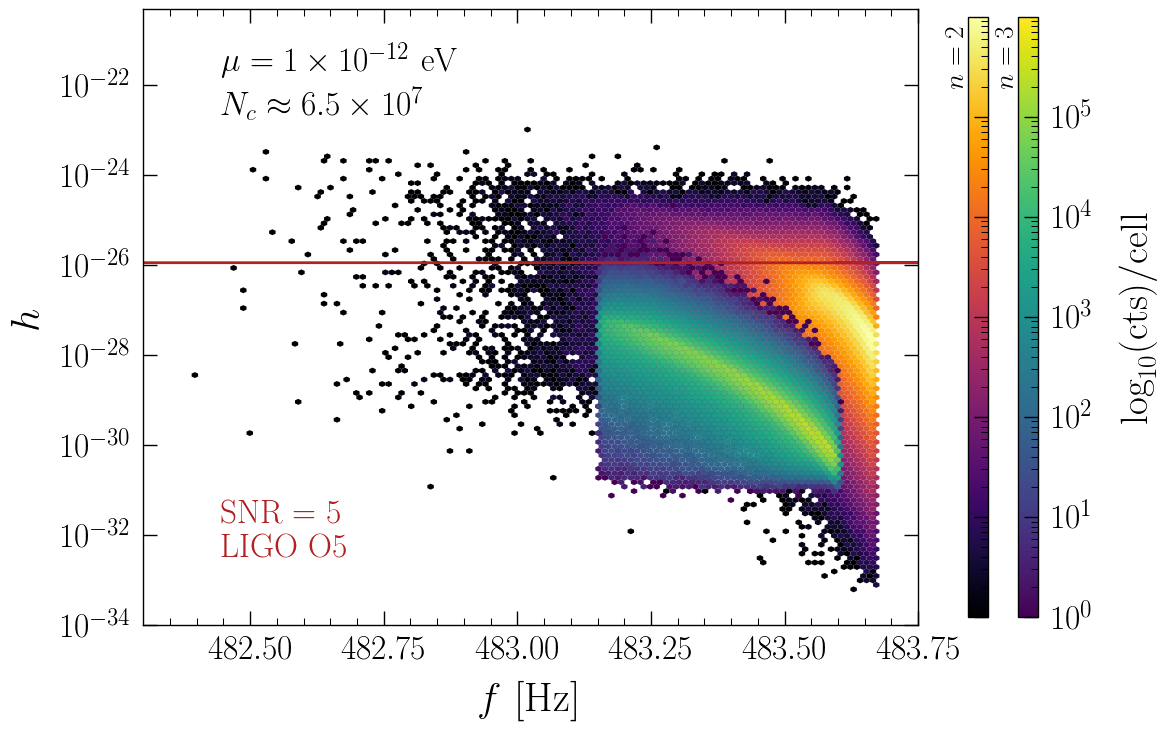}
\vspace{-0.4cm}

\caption{Illustrations of the final distribution of strains from resolvable axion clouds around extant BHs which emit GW signals reached by Earth today, used in our Galactic BH superradiance scenario, for axion masses $\mu = 10^{-13}$\,eV and $10^{-12}$\,eV.
The distribution is plotted as a density map, with separate colorbars partitioned by final mode $n$.
$N_c$ denotes the total number of extant clouds.
The SNR$=5$ threshold under the given chosen observatory for $T_{\rm obs} = 1$\,yr is also shown for each case.
Note that for $\mu = 10^{-13}$\,eV we show the Cosmic Explorer sensitivity to highlight that even for a future device sensitivity to these lower masses is challenging.}
\label{fig:GA_strains_compare_1}
\end{figure*}

\begin{figure*}
\centering
\includegraphics[width=0.52\textwidth]{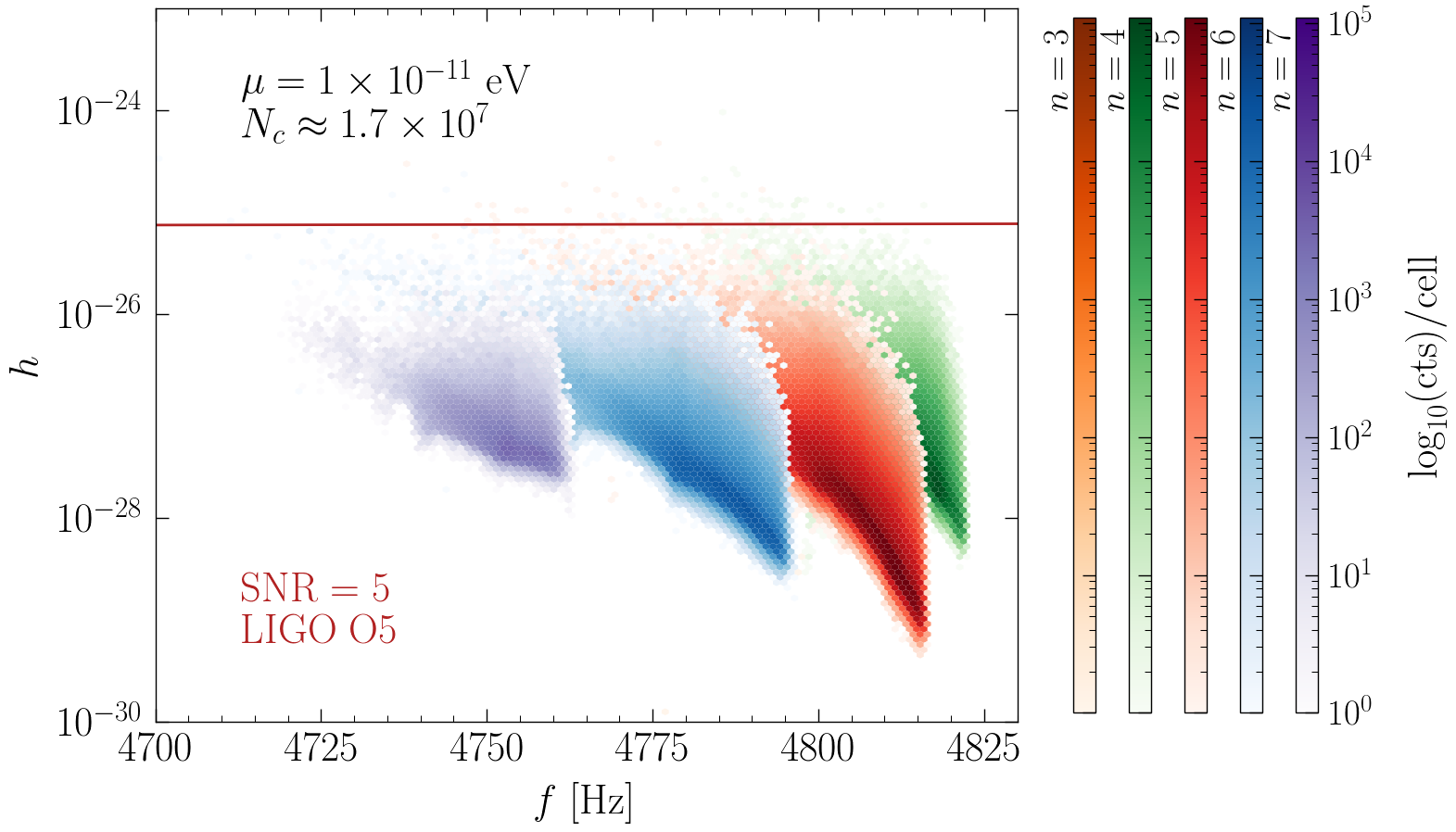}
\includegraphics[width=0.46\textwidth]{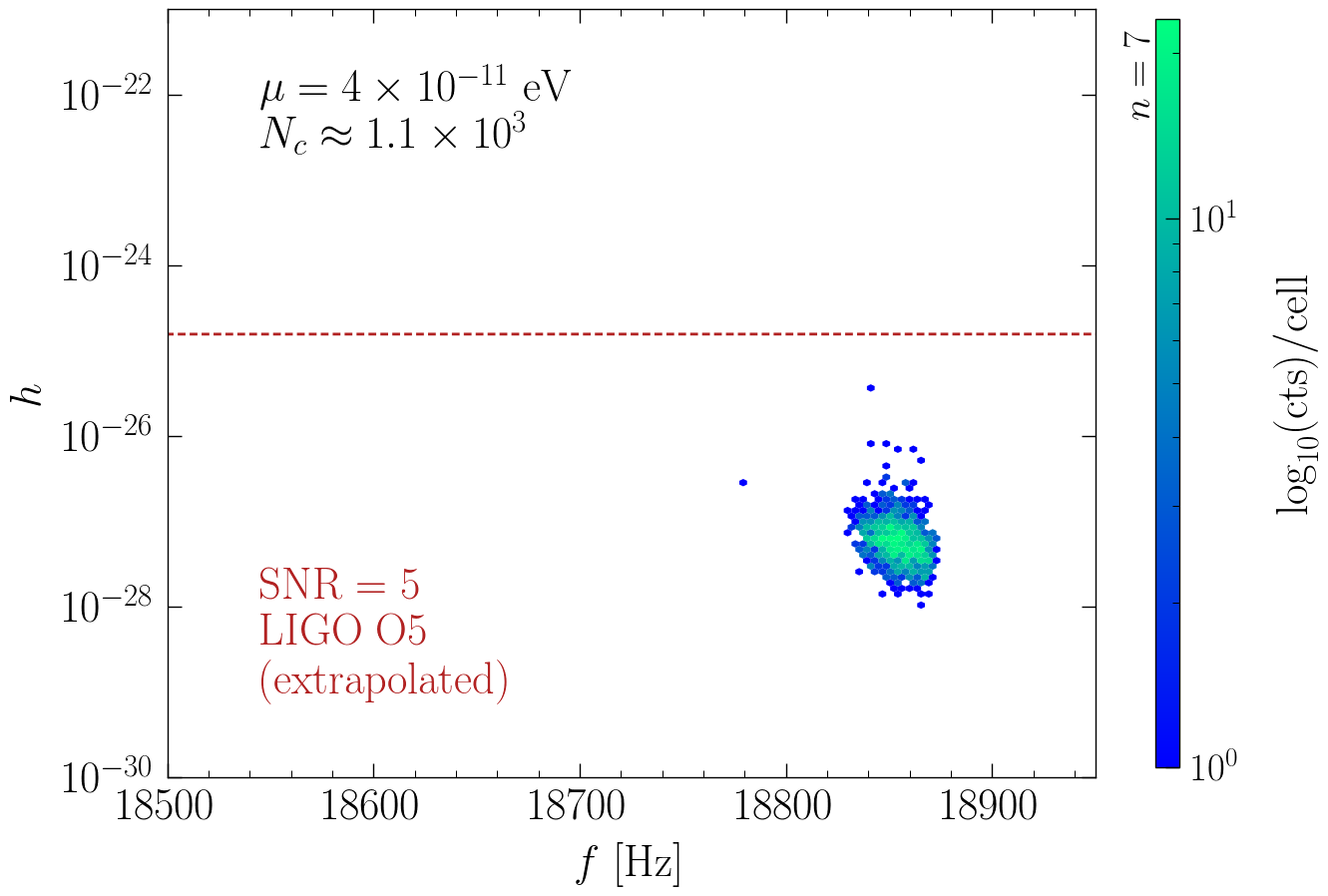}
\vspace{-0.4cm}

\caption{As in Fig.~\ref{fig:GA_strains_compare_1} but for the case of $\mu=10^{-11}$\,eV, which has 5 active final modes $n$, and $\mu=4\times 10^{-11}$\,eV, which has only one active final mode $n$.
We note that for the latter, in this frequency range we extrapolate LIGO O5's sensitivity (dashed line), as remarked in the main text.}
\label{fig:GA_strains_compare_2}
\end{figure*}

Reference~\cite{Dolan:2007mj} outlined a procedure for solving the equations for $R$ and $S$ to determine $\omega$ for each mode, which reduces the problem to numerically finding the root of a continued fraction relation.
We implement this algorithm, partly drawing on the code in Ref.~\cite{Superradiance_code}.
Taking $n-1 = \ell=m$, for each mode we compute the frequencies over a grid $\alpha \in [10^{-3},5]$ and $\chi \in [0.09,0.99]$, with steps of $\Delta \alpha = 2 \times 10^{-3}$ and $\Delta \chi = 0.05$.
We then interpolate between and extrapolate from these values.\footnote{This procedure can occasionally generate numerical artifacts in the form of downward fluctuations in the Monte Carlo strain distributions (typically at higher $\mu$), making them occasional low-$h$ outliers.
For illustration purposes, in showing our strain distributions at high $\mu$ we apply a mild local smoothing procedure removing a small fraction of these outliers which lie significantly below typical strain amplitudes within narrow frequency bins.}

For the cloud evolution, the key parameters are the final mass and spin of the BH after the system has moved below the death line.
If we define the ratio of the cloud to initial BH mass as $\kappa_f = M_c^f/M$, then solving the system in Eq.~\eqref{eq:MJ_DE} yields,
\be
\kappa_f = 1 - \frac{m^3 - m \sqrt{m^4-16 \alpha^2 \zeta_R^2(m - \chi \alpha \zeta_R)^2}}{8 \alpha^2 \zeta_R^2 (m - \chi \alpha \zeta_R)}.
\label{eq:kappa_f}
\ee
Here $M$, $\chi$, and $\alpha$ are all initial BH parameters.
For $\alpha \ll 1$, we have $\kappa_f \simeq \alpha \chi/m$, consistent with Eq.~\eqref{eq:Mcf}.
Using this we can write the final BH spin as
\be
\chi_f = \frac{4m \alpha \zeta_R(1-\kappa_f)}{m^2 + 4 \alpha^2 \zeta_R^2 (1-\kappa_f)^2},
\label{eq:chi_f}
\ee
which for small gravitational coupling becomes $\chi_f \simeq 4\alpha/m$.

To describe the decay of the cloud into GWs we need to specify the numerical coefficient in Eq.~\eqref{eq:dotEh},
\be
D_{n\ell} \!=\! \frac{2^{4(\ell+1)} \ell(2\ell\!-\!1)}{n^{4(\ell+2)}(\ell\!+\!1)(4\ell\!+\!2)!} \!\left( \frac{(2\ell\!-\!2)!(n\!+\!\ell)!}{(\ell!)^2(n\!-\!\ell\!-\!1)!} \right)^2\!.
\ee

Finally, the GW emission from the superradiant cloud experiences frequency drift due to self-gravity of the cloud and self-interactions between axions.
In the limit of small self-interactions, the rate of drift of the GW annihilation frequency for the $211$ mode ($\dot f_{\rm ann} = 2 \dot f_{211}$) is~\cite{Baryakhtar:2020gao}
\be
\dot{f}_{\rm ann} = - \frac{\mu \alpha^2}{\pi}\left[\frac{1}{4}\frac{\dot{\alpha}}{\alpha}+ 0.19\, \alpha\, \dot{\varepsilon}_{211} + 0.57 \,\dot{\alpha}\, \varepsilon_{211} \right]\!,
\ee
where $\varepsilon_{211} = N_{211}/M^2 = M_c/\mu M^2$ is the normalised occupation number of the mode and $\alpha$ and $M$ are evaluated for the final BH mass.
The GW signature is dominant once the cloud has fully formed, at which time we can take $\dot{\alpha}=0$, so that the only relevant contribution to the frequency shift arises from $\dot{\varepsilon}_{211} = - \dot{E}_{h,211}/\mu M^2$, with $\dot{E}_{h,211}$ given in Eq.~\eqref{eq:dotEh}.
Combining this with Eq.~\eqref{eq:Mct} we then find
\be
\dot f_{\rm ann}(t) = \frac{0.19\,D_{21}}{\pi}\,
\frac{\alpha^{15}\mu^2\kappa_f^2(1-\kappa_f)^{13}}{(1+t/T_h)^2},
\ee
with $\kappa_f$ given by the full expression in Eq.~\eqref{eq:kappa_f}.
This result requires tracking the difference between the initial and final BH mass; in particular, $\alpha$ in this expression is evaluated for the initial BH mass.
The drift rate is largest at $t=0$, when the cloud is most massive.
Integrating over the cloud lifetime with Eq.~\eqref{eq:Th}, the total fractional shift is
\be
\frac{\Delta f_{\rm ann}}{f_{\rm ann}} = 0.19\,\alpha^2\kappa_f (1- \kappa_f).
\ee
which never exceeds 0.2\% for the 211 mode across all $\alpha$ and $\chi$.
This is consistent with the intuition that the drift is a gravitational redshift controlled by the cloud mass, which is bounded by $\kappa_f \lesssim 0.1$ for all modes.

Whilst the total frequency drift is small, the relevant question for our galactic semi-coherent search is whether the superradiance signal remains coherent over the search.
Quantitatively, over a search of length $T_{\rm int}$ we require it to remain within a single frequency bin of width $1/T_{\rm int}$.
Assuming over this time the frequency shift is constant, we require $\Delta f \simeq \dot{f}_{\rm ann}(t) T_{\rm int} < 1/T_{\rm int}$, or using the maximal ($t=0$) drift rate,
\be
T_{\rm int} < \sqrt{\frac{\pi}{0.19\,D_{21}}}\,\frac{1}{\alpha^{7.5}\mu\,\kappa_f(1 - \kappa_f)^{6.5}}.
\ee
In the most extreme case, with $\chi =1$ and $\alpha$ close to 0.5, this condition is violated for $\mu = 10^{-11}\,$eV and our benchmark $T_{\rm int} = 4\,$hr.
Nonetheless, clouds with such large spin and coupling grow and dissipate extremely rapidly, and are not objects driving our sensitivity.
Instead, if we consider the signals in Tab.~\ref{tab:GA_BH_list_12}, the bound becomes $T_{\rm int} \lesssim 10^4\,$hr for the loudest event, or $3 \times 10^4$ for the second, both very safe.
For the higher modes relevant for the signals in Tab.~\ref{tab:GA_BH_list_11}, the generalisation of the non-relativistic self-gravity correction is discussed in Ref.~\cite{Siemonsen:2022yyf}.
Yet as the frequency drift arises from the change in mass of the cloud, we suspect it is unlikely the higher modes would significantly change the conclusion.

%%%%%%%%%%%%%%%%%%%%%%%%%%%%%%%%%
\section{Extended Results and Systematics for the Galactic BH Search}
\label{app:ext_GA}
%%%%%%%%%%%%%%%%%%%%%%%%%%%%%%%%%

Here we show extended results for our galactic BH search, and give further details regarding systematic variations around our fiducial analysis discussed in the main text.

In Fig.~\ref{fig:GA_show_configs}, we show the underlying distributions for each of the relevant BH properties.
In particular, we show the different choices considered for the BH distributions in mass, spin, distance, and age.
All of these contribute discernible differences in our overall axion sensitivity, as shown in Fig.~\ref{fig:fig_1}.
In Tab.~\ref{tab:spatial_params} we list the parameters used to define our fiducial BH spatial distribution model, i.e. the parameters required for Eq.~\eqref{eq:gal_mass_dens}.

\begin{figure*}[!htb]
\centering

\begin{minipage}{0.42\textwidth}
\includegraphics[width=\linewidth]{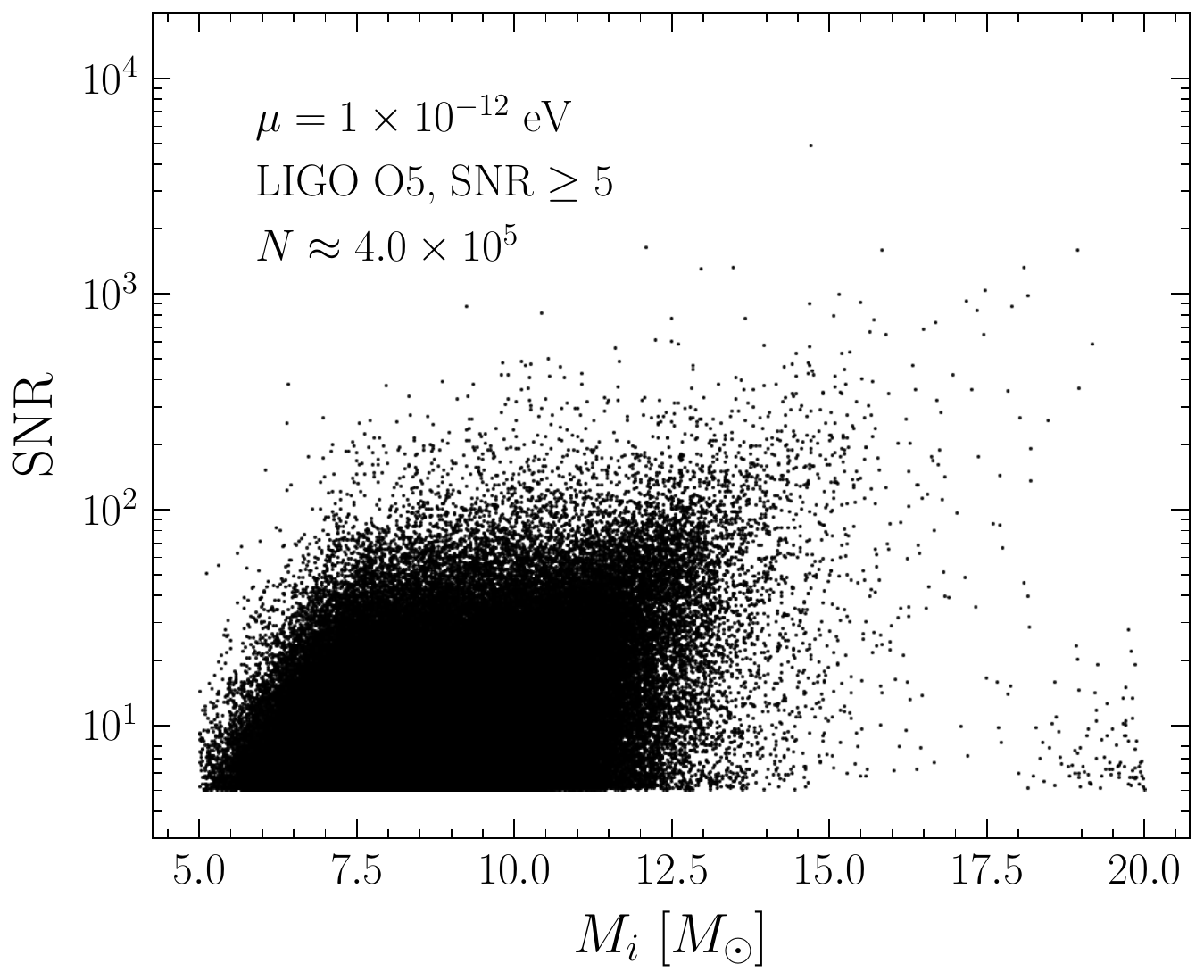}
\end{minipage}
\hspace{0.3cm}
\begin{minipage}{0.42\textwidth}
\includegraphics[width=\linewidth]{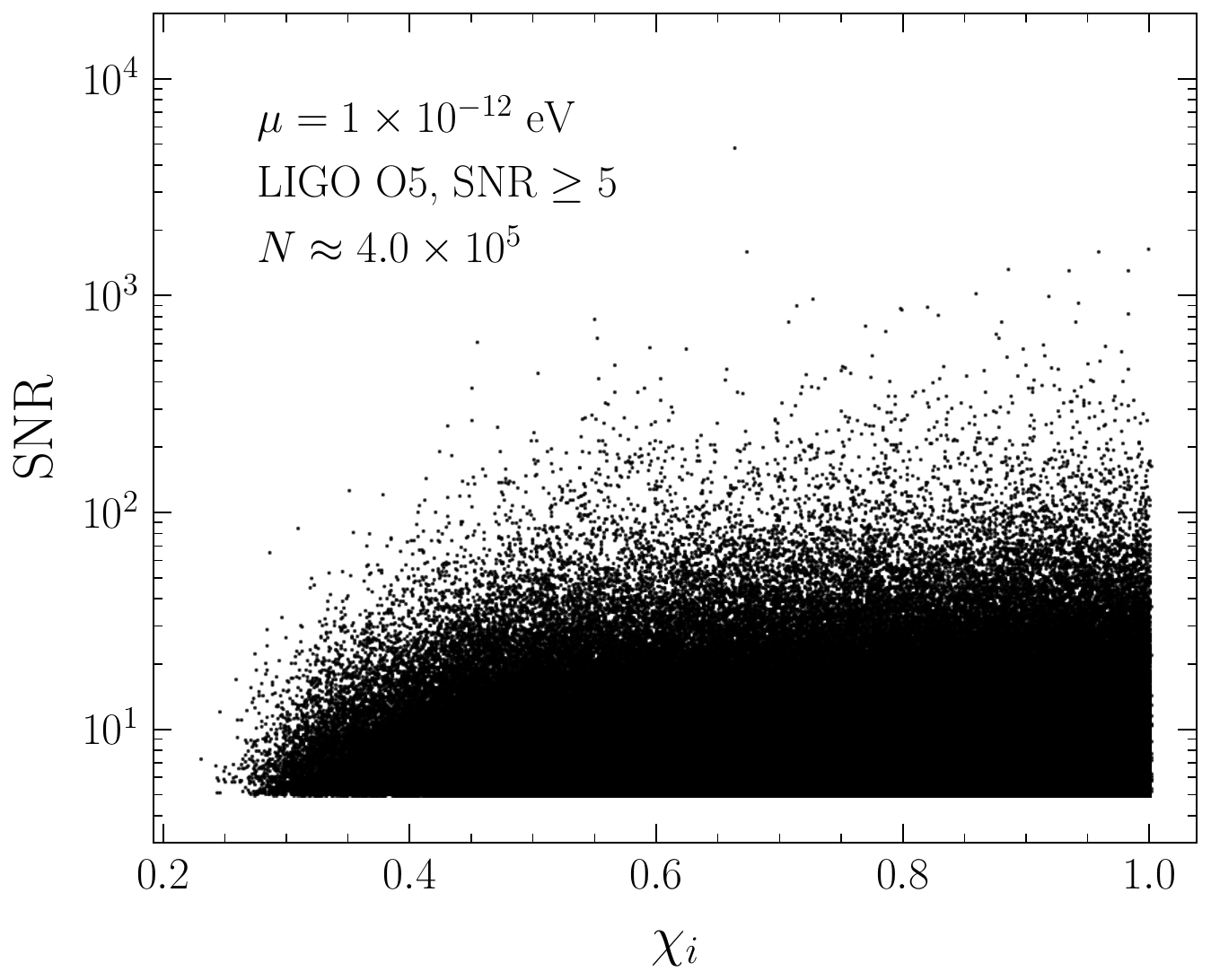}
\end{minipage}

\vspace{0.3cm}

\begin{minipage}{0.42\textwidth}
\includegraphics[width=\linewidth]{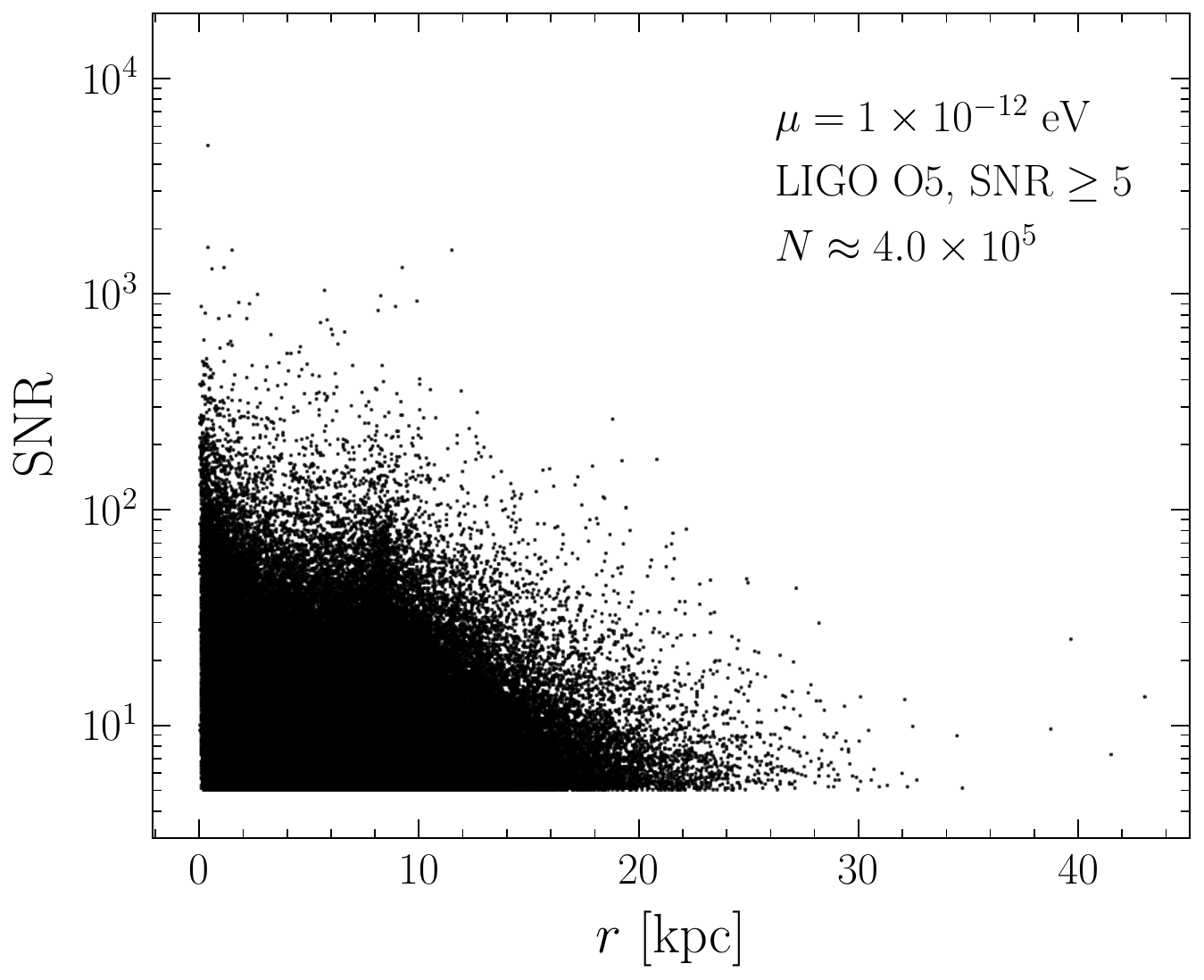}
\end{minipage}
\hspace{0.3cm}
\begin{minipage}{0.42\textwidth}
\includegraphics[width=\linewidth]{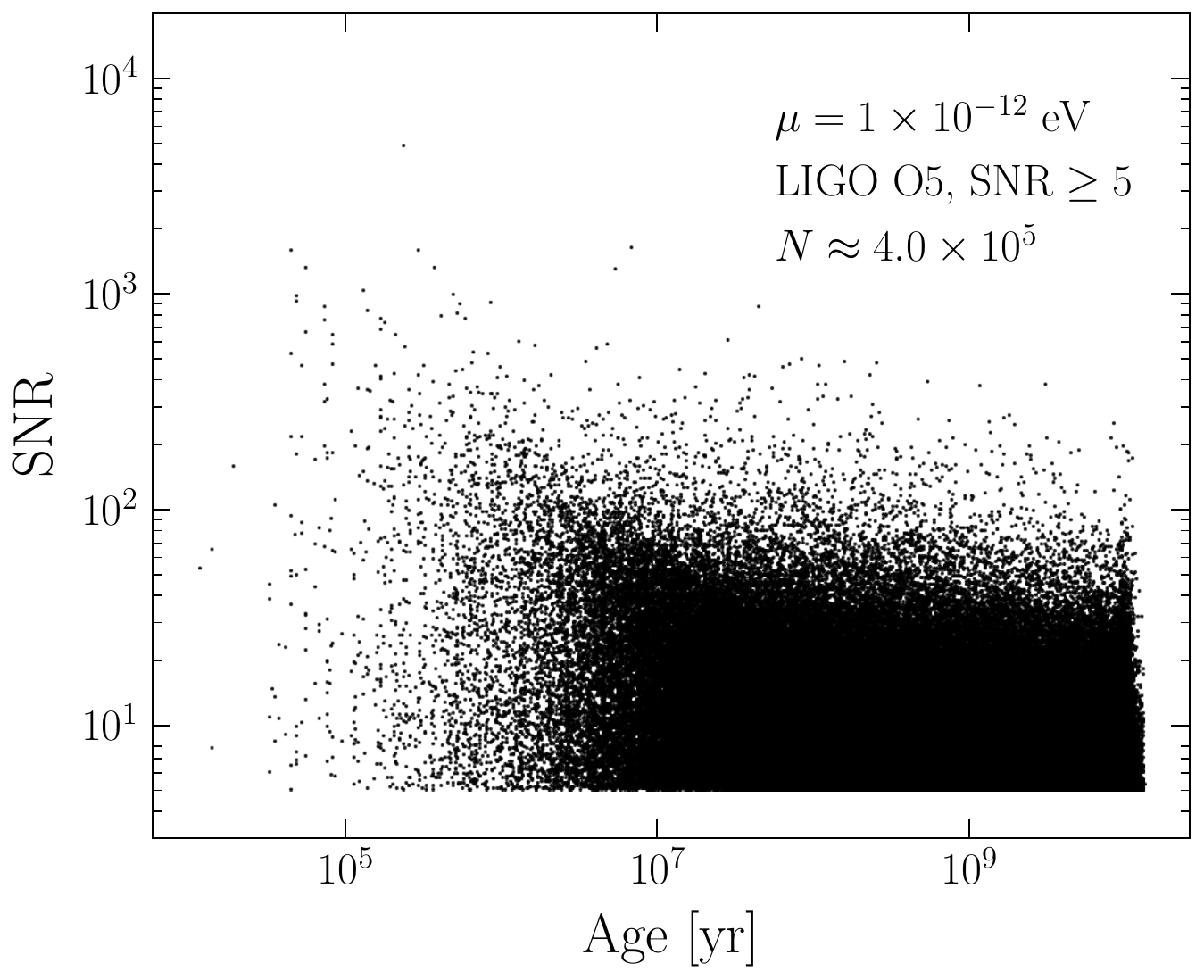}
\end{minipage}

\caption{Distribution of the SNRs for LIGO O5 (where $\rm{SNR} \geq 5$) over BH masses, spins, distances, and ages in our fiducial galactic BH ensemble, assuming axion mass $\mu = 10^{-12}$\,eV.
$N$ denotes the number of signals above the SNR threshold.
In the context of Tab.~\ref{tab:GA_BH_list_12}, we note here that our strongest individual BHs with respect to SNR are well within the overall distributions of BH properties, such that even if one were to discard them, there would still exist enough signals to confidently rule out the purported axion mass.}
\label{fig:SNR_by_property}
\end{figure*}

In Figs.~\ref{fig:GA_strains_compare_1} and~\ref{fig:GA_strains_compare_2}, we illustrate the final distribution of resolvable strains arising from superradiance occurring in the axion clouds surrounding our fiducial distribution of galactic BHs, for several axion masses $\mu$.
We additionally illustrate the ${\rm SNR}=5$ threshold strain from LIGO O5 (over $T_{\rm obs} = 1$\,yr) overlaid on these figures to show the expected number of detectable signals used in our statistical analyses, for each $\mu$ (for $\mu = 1\times 10^{-13}$ eV, we show the threshold strain from Cosmic Explorer).
For $10^{-11}\,$eV and $10^{-12}$\,eV, we also show, in Tabs.~\ref{tab:GA_BH_list_12} and~\ref{tab:GA_BH_list_11} the BH properties corresponding to the top 10 strongest and weakest final strains $h$ (above ${\rm SNR} =5$ for LIGO O5) from these distributions.
We note that, as expected, for the strongest signal BHs, reasonably high spins, short distances, and young ages dominate, although, as we point out in the SNR distributions by BH properties in Fig.~\ref{fig:SNR_by_property}, these values are not anomalous outliers.
From Fig.~\ref{fig:SNR_by_property}, we see that even if we were to rule out these values, there would still exist enough signals to confidently rule out axion masses.

\begin{figure}[!htb]
\centering
\includegraphics[width=0.49\textwidth]{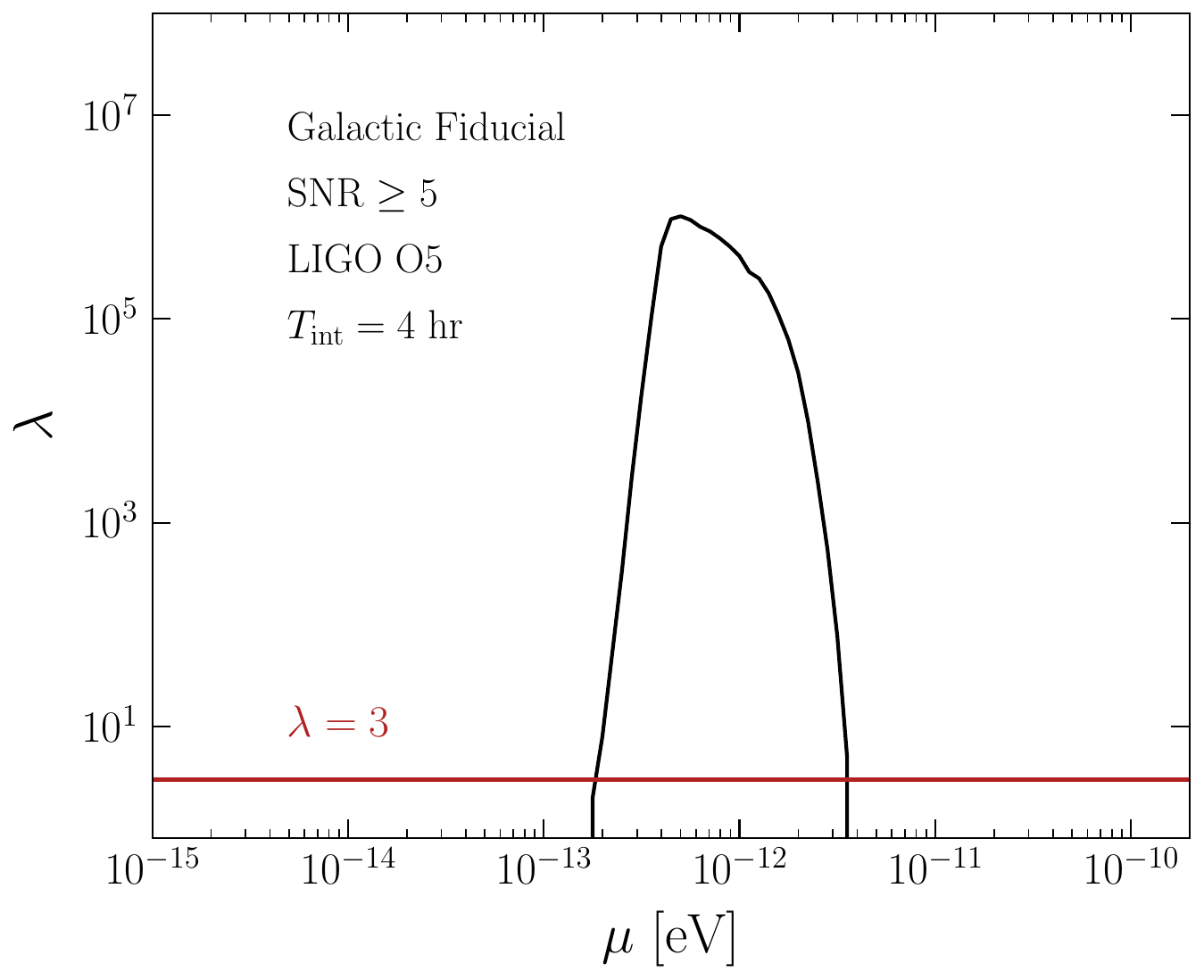}
\vspace{-0.4cm}
\caption{An illustration of the distribution of $\lambda$ over axion mass $\mu$, as defined in the statistical framework described in the main text for the galactic BH population search.
We have sensitivity to axion masses, at 95\% confidence, for which $\lambda \geq 3$.}
\label{fig:GA_stat}
\end{figure}

\begin{figure*}%[!htb]
\centering

\begin{minipage}{0.45\textwidth}
    \includegraphics[width=\linewidth]{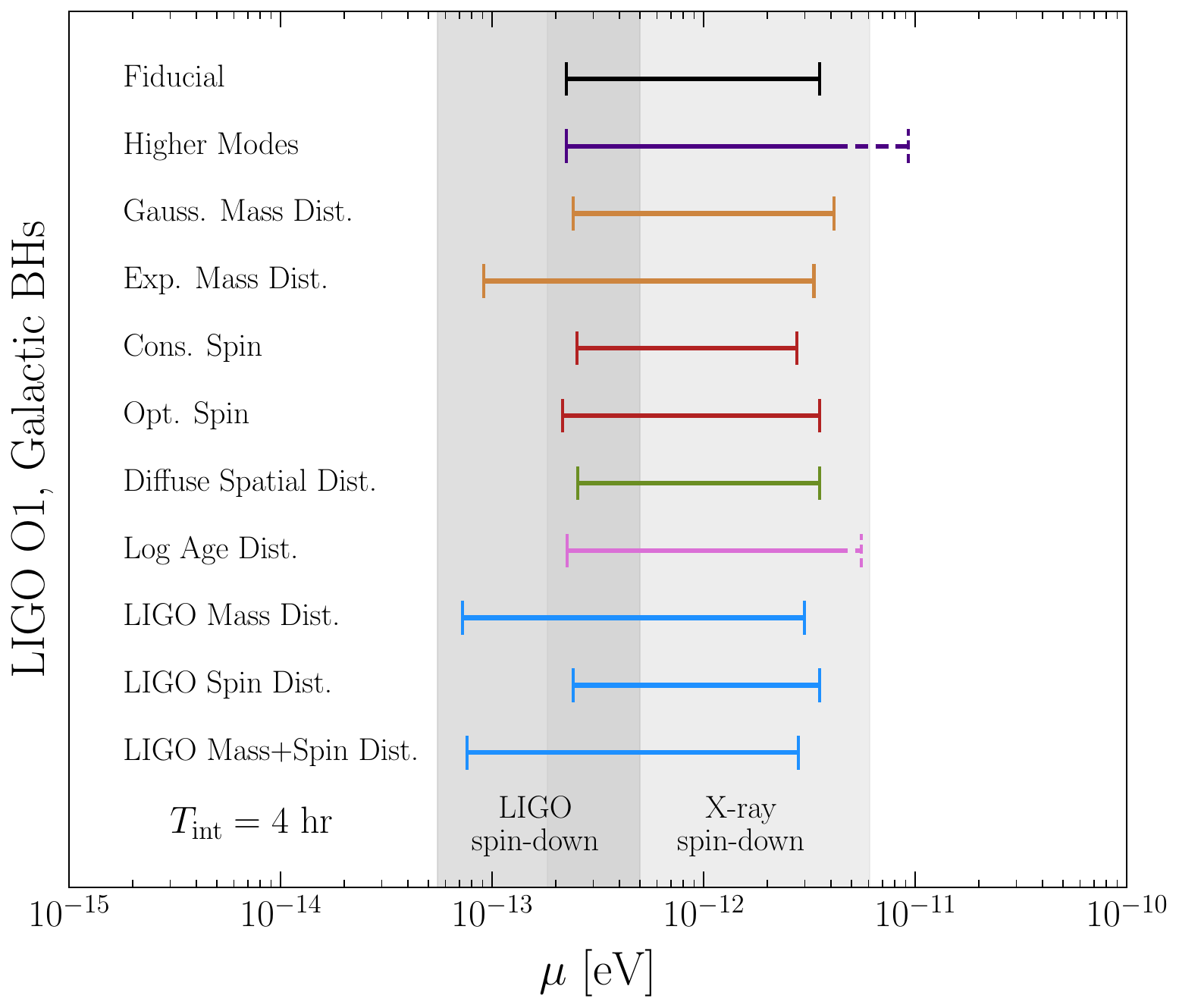}
\end{minipage}

\vspace{0.3cm}

\begin{minipage}{0.45\textwidth}
    \includegraphics[width=\linewidth]{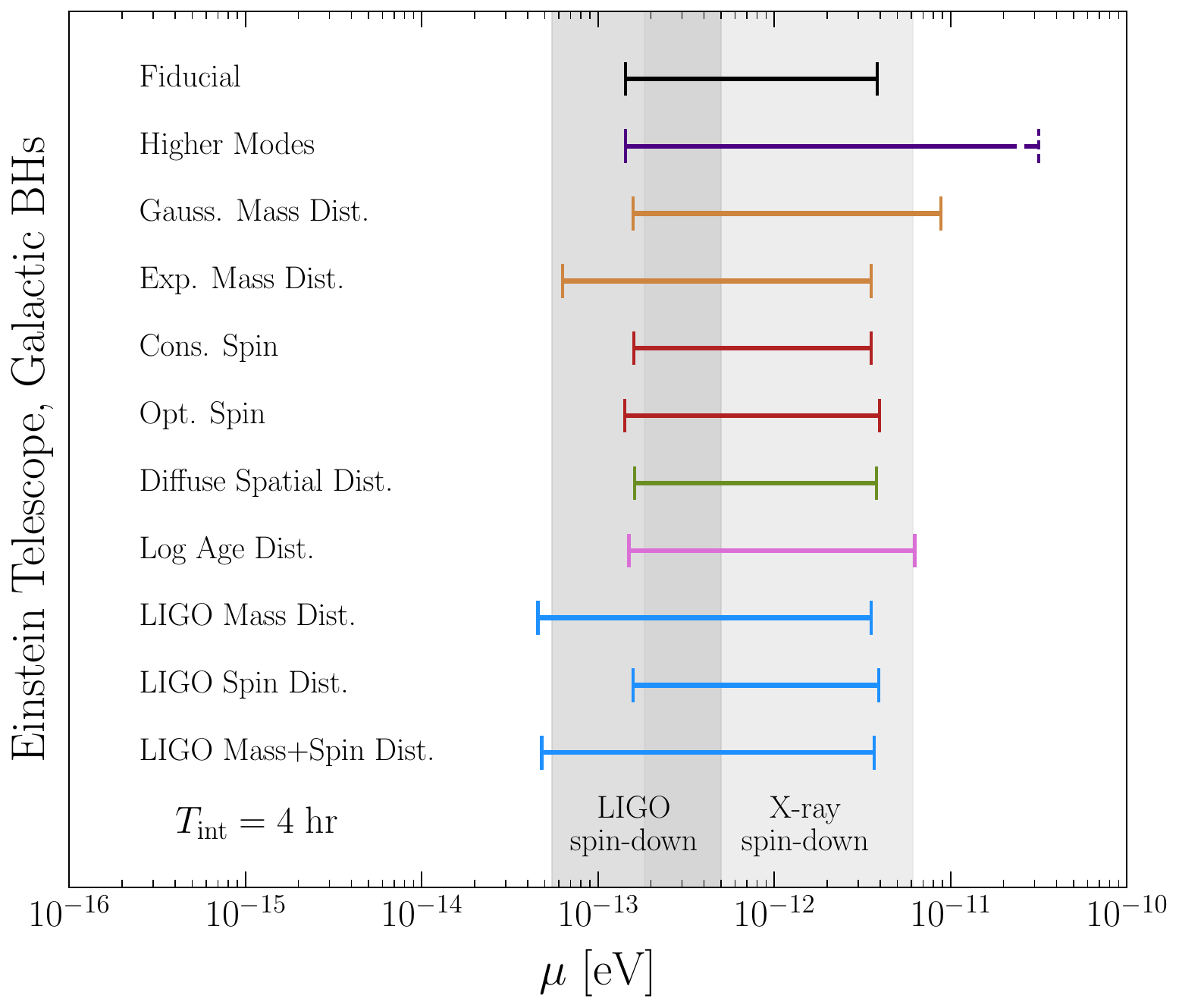}
\end{minipage}
\hspace{0.3cm}
\begin{minipage}{0.45\textwidth}
    \includegraphics[width=\linewidth]{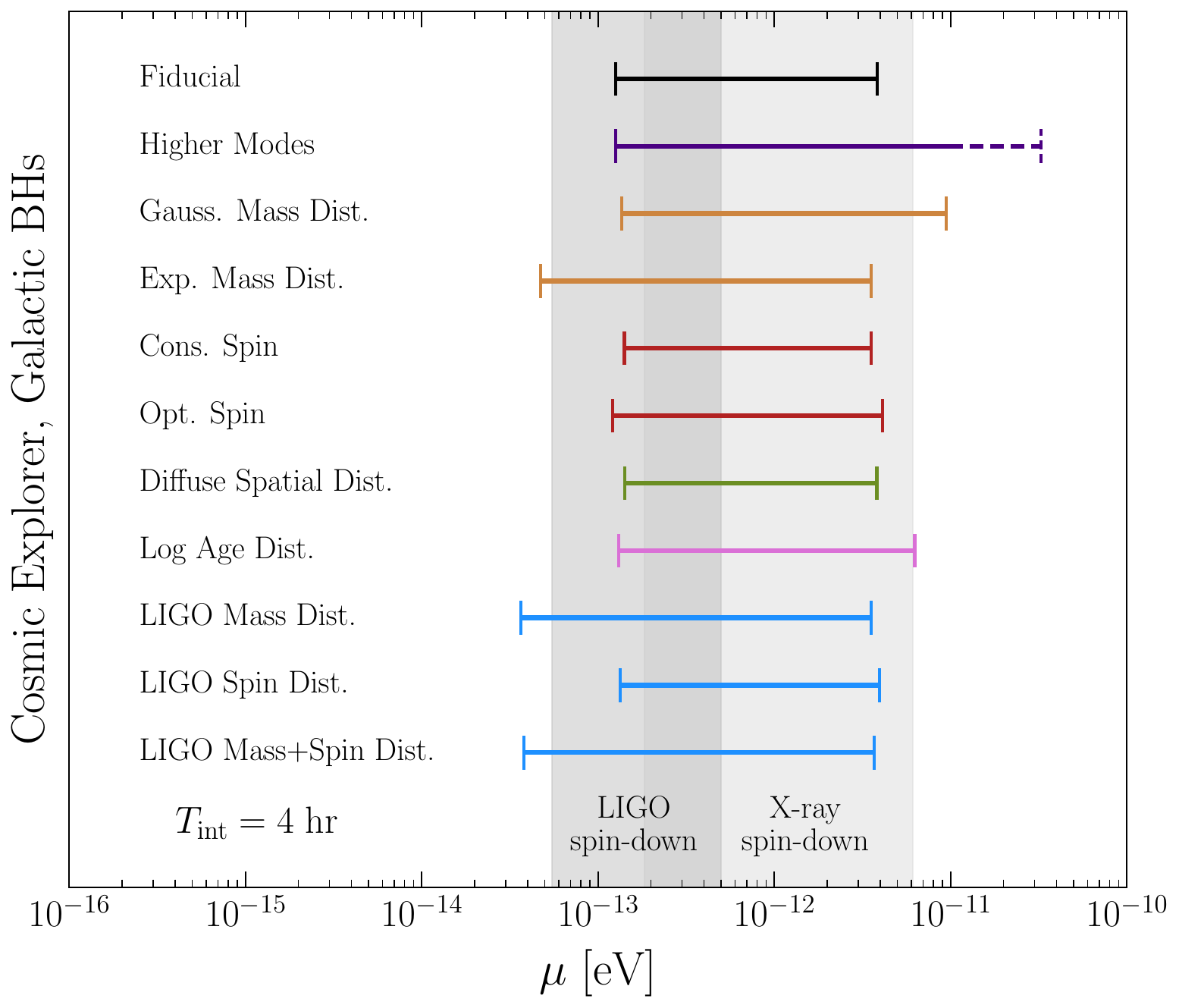}
\end{minipage}

\vspace{0.3cm}

\begin{minipage}{0.45\textwidth}
    \includegraphics[width=\linewidth]{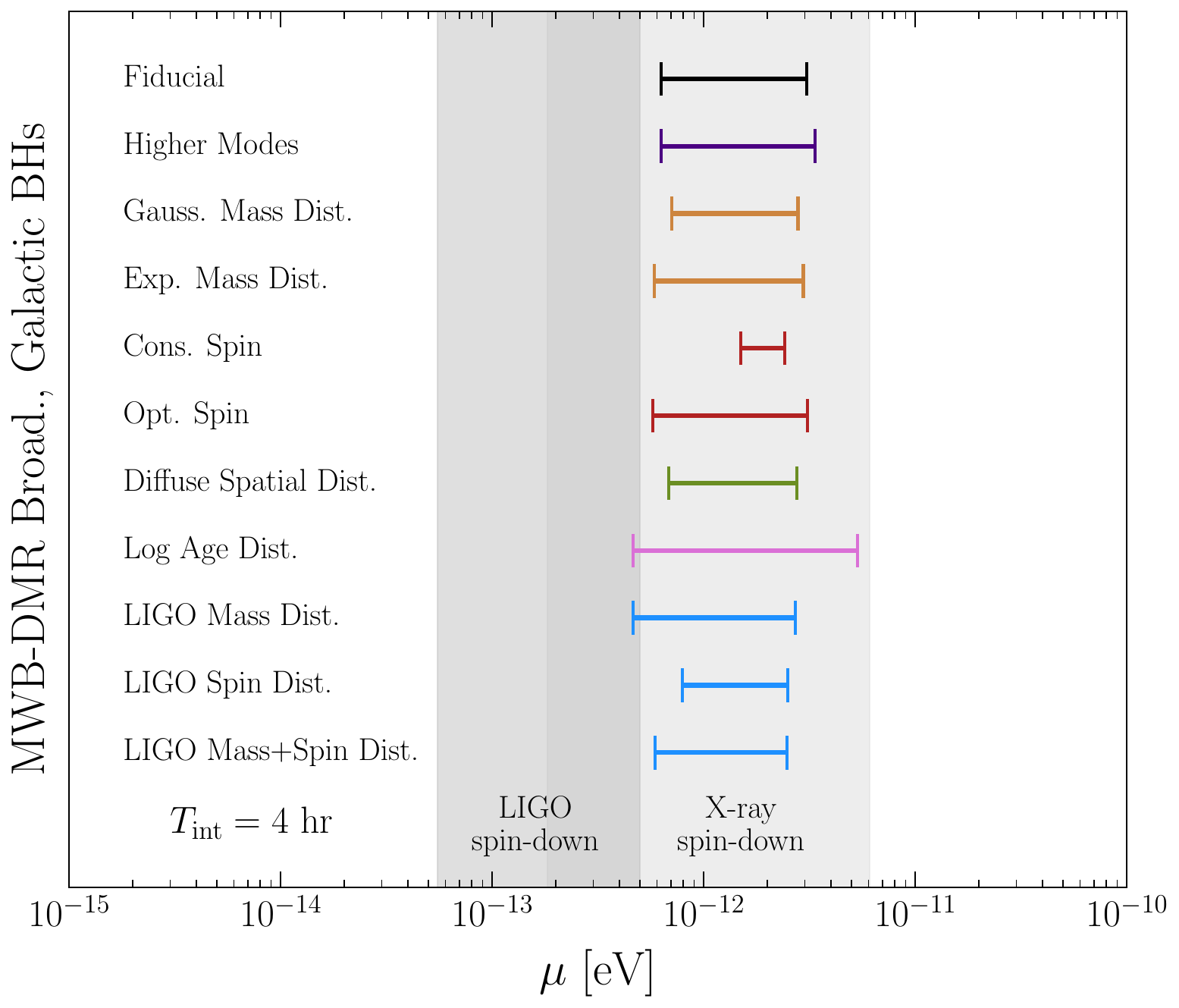}
\end{minipage}
\hspace{0.3cm}
\begin{minipage}{0.45\textwidth}
    \includegraphics[width=\linewidth]{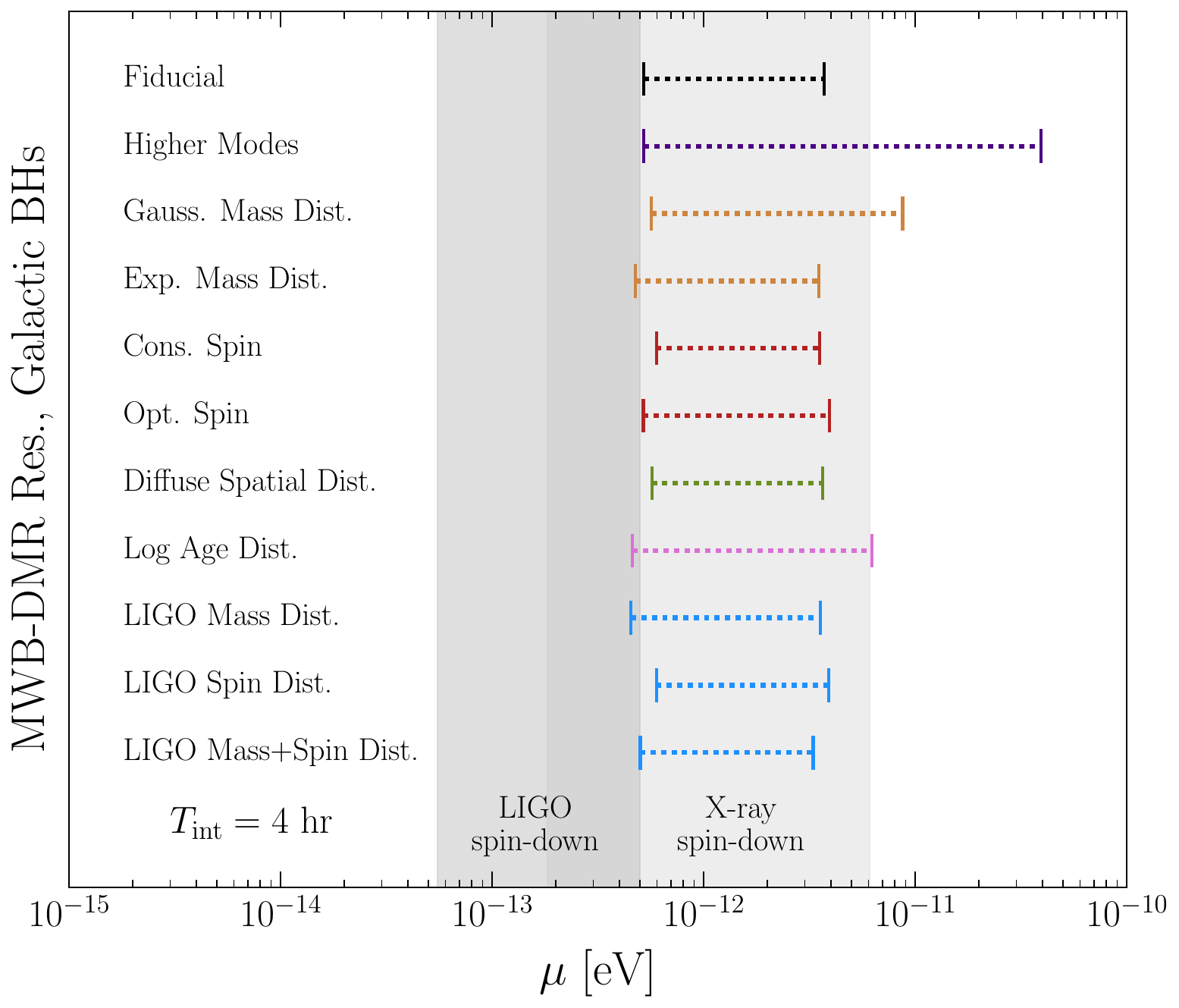}
\end{minipage}
\vspace{-0.2cm}
\caption{As in the left of Fig.~\ref{fig:fig_1}, but for the additional instruments considered in our work here: LIGO O1, Einstein Telescope, Cosmic Explorer, and MWBs.}
\label{fig:GA_observatory_sys}
\end{figure*}

We show an illustration of our statistical analysis for the resolvable signals in Fig.~\ref{fig:GA_stat} for our fiducial galactic model, for LIGO O5, as described in the main text.
Finally, we extend the analyses of astrophysical systematics for LIGO O5 shown in Fig.~\ref{fig:fig_1} to the other observatories considered in our work.
We illustrate analogous results in Fig.~\ref{fig:GA_observatory_sys}.

\begin{table}[!htb]
\centering

\textbf{Galactic Spatial Distribution}
\vspace{2pt}

\begin{tabular}{p{4cm} p{2cm}}
\hline
\multicolumn{2}{c}{\textbf{Disk Parameters}} \\
\hline
Parameter & Value \\
\hline
$R_{d,{\rm thin}}$ & 3.00\,kpc \\
$R_{d,{\rm thick}}$ & 3.29\,kpc \\
$z_{d,{\rm thin}}$ & 0.3\,kpc \\
$z_{d,{\rm thick}}$ & 0.9\,kpc \\
$R_\odot$ & 8.29\,kpc \\
\hline
\multicolumn{2}{c}{\textbf{Bulge Parameters}} \\
\hline
Parameter & Value \\
\hline
$\alpha_{\rm b}$ & 1.8 \\
$r_0$ & 0.075\,kpc \\
$r_{\mathrm{cut}}$ & 2.1\,kpc \\
$q$ & 0.5 \\
\hline
\end{tabular}
\caption{Parameters used in the construction of our galactic BH spatial distribution as defined in Eq.~\eqref{eq:gal_mass_dens}.
These choices are adopted from the prescription in Ref.~\cite{Sprague:2024lgq}, which in turn were taken from Refs.~\cite{McMillan:2011wd, SDSS:2005kst, Bissantz:2001wx}.}
\label{tab:spatial_params}
\end{table}

\begin{table*}[htb!]
\centering
\renewcommand{\arraystretch}{1.3}
\setlength{\tabcolsep}{11pt}
\begin{tabular}{c c c c c c c c}
\hline\hline
\multicolumn{8}{c}{Fiducial Galactic BHs, $\mu = 10^{-12}\,\mathrm{eV}$} \\
\hline
Rank & SNR & $h$ & $M\,(M_\odot)$ & $\chi$ & $r$ (kpc) & Age (yr) & Final $m$ \\
\hline\hline
1  & $4.771\times10^{3}$ & $1.090\times10^{-23}$ & 14.69 & 0.6628 & 0.3490 & $2.348\times10^{5}$ & 1 \\
\hline
2  & $1.627\times10^{3}$ & $3.718\times10^{-24}$ & 12.08 & 0.9980 & 0.3596 & $6.701\times10^{6}$ & 1 \\
\hline
3  & $1.582\times10^{3}$ & $3.613\times10^{-24}$ & 15.81 & 0.6724 & 1.4510 & $2.883\times10^{5}$ & 1 \\
\hline
4  & $1.577\times10^{3}$ & $3.600\times10^{-24}$ & 18.93 & 0.9577 & 11.45  & $4.407\times10^{4}$ & 1 \\
\hline
5  & $1.309\times10^{3}$ & $2.990\times10^{-24}$ & 13.47 & 0.8846 & 1.1090 & $3.717\times10^{5}$ & 1 \\
\hline
6  & $1.296\times10^{3}$ & $2.960\times10^{-24}$ & 18.07 & 0.9820 & 9.1880 & $5.470\times10^{4}$ & 1 \\
\hline
7  & $1.295\times10^{3}$ & $2.958\times10^{-24}$ & 12.95 & 0.9341 & 0.5677 & $5.339\times10^{6}$ & 1 \\
\hline
8  & $1.024\times10^{3}$ & $2.339\times10^{-24}$ & 17.45 & 0.8586 & 5.6850 & $1.301\times10^{5}$ & 1 \\
\hline
9  & $9.818\times10^{2}$ & $2.243\times10^{-24}$ & 15.15 & 0.9177 & 2.6460 & $4.875\times10^{5}$ & 1 \\
\hline
10 & $9.650\times10^{2}$ & $2.204\times10^{-24}$ & 18.12 & 0.7266 & 8.2630 & $4.772\times10^{4}$ & 1 \\
\hline
\multicolumn{8}{c}{$\vdots$} \\
\hline
405809 & 5.00 & $1.143\times10^{-26}$ & 8.532 & 0.8689 & 7.078 & $9.239\times10^{8}$ & 1 \\
\hline
405810 & 5.00 & $1.143\times10^{-26}$ & 7.767 & 0.7963 & 4.062 & $1.066\times10^{9}$ & 1 \\
\hline
405811 & 5.00 & $1.143\times10^{-26}$ & 9.019 & 0.7908 & 9.096 & $5.697\times10^{8}$ & 1 \\
\hline
405812 & 5.00 & $1.143\times10^{-26}$ & 7.397 & 0.8927 & 2.954 & $2.986\times10^{9}$ & 1 \\
\hline
405813 & 5.00 & $1.143\times10^{-26}$ & 7.891 & 0.7123 & 3.18 & $2.982\times10^{9}$ & 1 \\
\hline
405814 & 5.00 & $1.143\times10^{-26}$ & 8.459 & 0.8713 & 7.846 & $5.363\times10^{8}$ & 1 \\
\hline
405815 & 5.00 & $1.143\times10^{-26}$ & 7.427 & 0.6205 & 2.00 & $3.918\times10^{9}$ & 1 \\
\hline
405816 & 5.00 & $1.143\times10^{-26}$ & 9.273 & 0.7403 & 10.45 & $4.044\times10^{8}$ & 1 \\
\hline
405817 & 5.00 & $1.143\times10^{-26}$ & 7.501 & 0.7387 & 3.185 & $3.694\times10^{8}$ & 1 \\
\hline
405818 & 5.00 & $1.143\times10^{-26}$ & 9.261 & 0.6241 & 9.106 & $3.540\times10^{8}$ & 1 \\
\hline
\end{tabular}
\caption{The ten strongest and weakest final strains $h$ with SNR $\geq 5$ for LIGO O5 from our fiducial resolvable galactic BH population, as well as the initial BH properties from which the strain was sourced, for $\mu = 10^{-12}$\,eV.
We additionally show, in the rightmost column, the final mode $m$ corresponding to this strain.
(Recall we only consider clouds with $n-1=\ell=m$; see Sec.~\ref{sec:superradiance}.)
We note that the masses and distances to the BHs yielding the strongest signals are similar to those of recently-measured BHs~\cite{El-Badry:2022zih,Gaia:2024ggk}.}
\label{tab:GA_BH_list_12}
\end{table*}

\begin{table*}[htb!]
\centering
\renewcommand{\arraystretch}{1.3}
\setlength{\tabcolsep}{11pt}
\begin{tabular}{c c c c c c c c}
\hline\hline
\multicolumn{8}{c}{Fiducial Galactic BHs, $\mu =10^{-11}\,\mathrm{eV}$} \\
\hline
Rank & SNR & $h$ & $M\,(M_\odot)$ & $\chi$ & $r$ (kpc) & Age (yr) & Final $m$ \\
\hline\hline
1  & $6.448\times10^{1}$ & $9.916\times10^{-25}$ & 8.116 & 0.9505 & 0.6189 & $1.352\times10^{5}$ & 3 \\
\hline
2  & $2.921\times10^{1}$ & $4.483\times10^{-25}$ & 10.43 & 0.8277 & 0.2706 & $5.176\times10^{5}$ & 4 \\
\hline
3  & $2.232\times10^{1}$ & $3.370\times10^{-25}$ & 18.13 & 0.9210 & 6.2430 & $1.128\times10^{5}$ & 5 \\
\hline
4  & $2.176\times10^{1}$ & $3.335\times10^{-25}$ & 12.08 & 0.9980 & 0.3596 & $6.701\times10^{6}$ & 4 \\
\hline
5  & $1.926\times10^{1}$ & $2.931\times10^{-25}$ & 13.50 & 0.9458 & 5.2880 & $7.546\times10^{4}$ & 4 \\
\hline
6  & $1.783\times10^{1}$ & $2.743\times10^{-25}$ & 8.344 & 0.9992 & 1.2710 & $6.698\times10^{5}$ & 3 \\
\hline
7  & $1.484\times10^{1}$ & $2.283\times10^{-25}$ & 7.528 & 0.8484 & 1.6760 & $5.469\times10^{5}$ & 3 \\
\hline
8  & $1.495\times10^{1}$ & $2.279\times10^{-25}$ & 15.02 & 0.7716 & 0.2006 & $2.479\times10^{7}$ & 5 \\
\hline
9  & $1.358\times10^{1}$ & $2.088\times10^{-25}$ & 7.866 & 0.7534 & 3.0430 & $1.373\times10^{4}$ & 3 \\
\hline
10 & $1.225\times10^{1}$ & $1.875\times10^{-25}$ & 11.48 & 0.8885 & 1.4550 & $1.880\times10^{6}$ & 4 \\
\hline
\multicolumn{8}{c}{$\vdots$} \\
\hline
75 & 5.222 & $8.043\times10^{-26}$ & 9.209 & 0.8236 & 0.08483 & $1.213\times10^{9}$ & 4 \\
\hline
76 & 5.211 & $8.034\times10^{-26}$ & 6.778 & 0.6825 & 0.6476 & $9.684\times10^{6}$ & 3 \\
\hline
77 & 5.206 & $8.016\times10^{-26}$ & 7.966 & 0.9957 & 4.884 & $5.470\times10^{4}$ & 3 \\
\hline
78 & 5.177 & $7.972\times10^{-26}$ & 7.253 & 0.8092 & 2.46 & $1.741\times10^{6}$ & 3 \\
\hline
79 & 5.171 & $7.928\times10^{-26}$ & 10.73 & 0.7471 & 0.9328 & $7.237\times10^{6}$ & 4 \\
\hline
80 & 5.155 & $7.922\times10^{-26}$ & 8.427 & 0.9611 & 12.23 & $8.571\times10^{4}$ & 3 \\
\hline
81 & 5.119 & $7.896\times10^{-26}$ & 6.639 & 0.8005 & 1.621 & $5.498\times10^{5}$ & 3 \\
\hline
82 & 5.101 & $7.860\times10^{-26}$ & 6.998 & 0.7609 & 2.096 & $1.488\times10^{6}$ & 3 \\
\hline
83 & 5.095 & $7.777\times10^{-26}$ & 12.62 & 0.9236 & 9.191 & $2.410\times10^{5}$ & 4 \\
\hline
84 & 5.068 & $7.769\times10^{-26}$ & 10.82 & 0.7300 & 1.050 & $1.824\times10^{6}$ & 4 \\
\hline
\end{tabular}
\caption{As in Tab.~\ref{tab:GA_BH_list_12} but for $\mu = 10^{-11}\,\textrm{eV}$.
We remark on the importance of contributions from modes higher than the first for this axion mass.}
\label{tab:GA_BH_list_11}
\end{table*}

%%%%%%%%%%%%%%%%%%%%%%%%%%%%%%%%%
\section{Extended Results and Systematics for the Extragalactic BH Search}
\label{app:ext_EG}
%%%%%%%%%%%%%%%%%%%%%%%%%%%%%%%%%

In this section we show extended results illustrating aspects of our extragalactic BH search and analysis, as well as a description of our modeling for the remnant BBH merger population, which is one of the systematic tests shown in Fig.~\ref{fig:fig_1}.

For our fiducial extragalactic model, we show the resulting $\Omega_h$ over frequencies in Fig.~\ref{fig:EG_Omegah} for various axion masses $\mu$.
We remark that in this fiducial model, the GW energy density tends to peak around $\mu\sim 10^{-12}$ eV, suggesting where our stochastic GW background search should be the most sensitive.

\begin{figure}[!htp]
\centering
\includegraphics[width=0.49\textwidth]{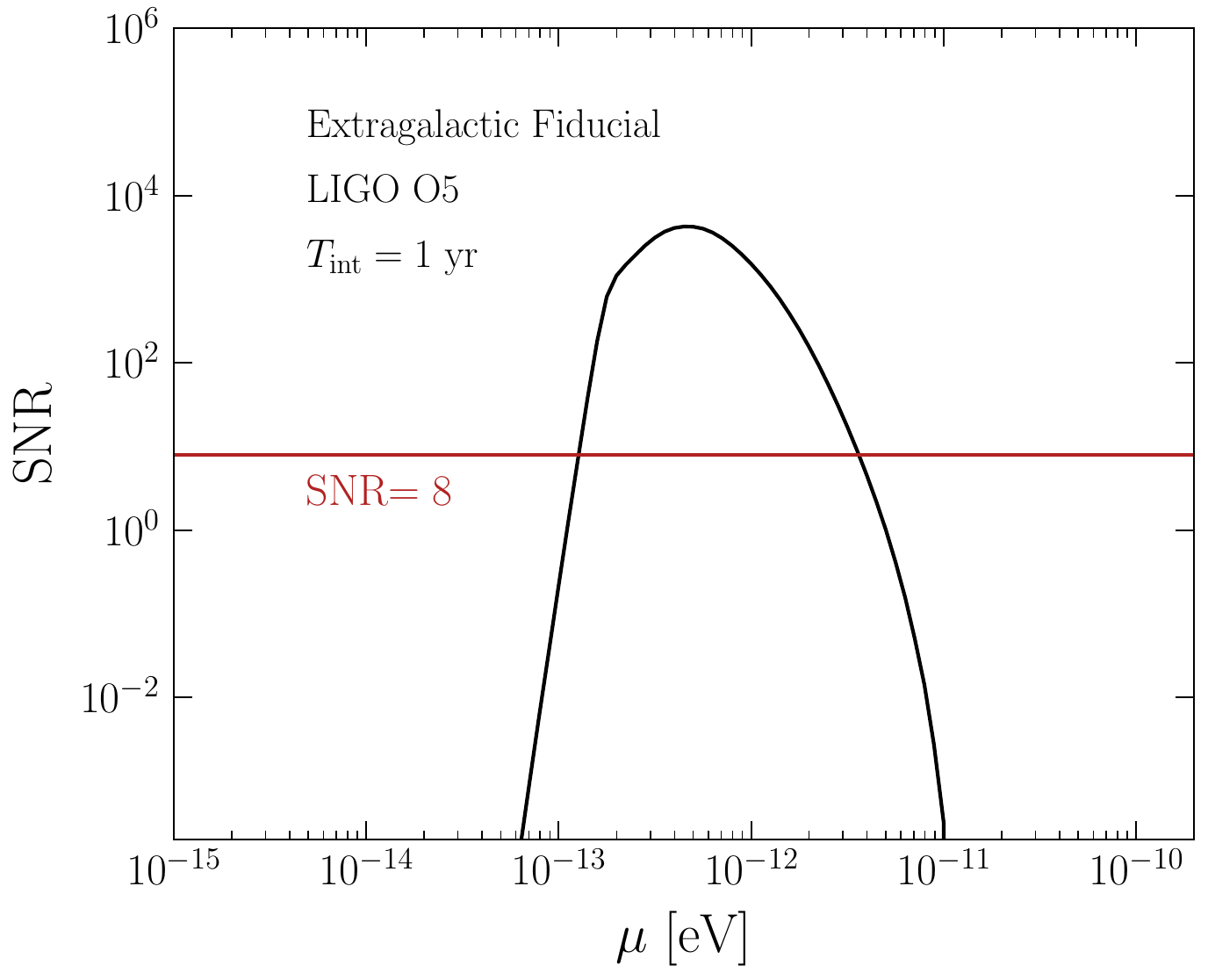}
\vspace{-0.4cm}
\caption{An illustration of the distribution of SNR over axion mass $\mu$, as defined in the statistical framework described in the main text for the extragalactic BH population search.
We have sensitivity to axion masses, at 95\% confidence, for which SNR $\geq 8$.}
\label{fig:EG_stat}
\end{figure}

\begin{figure*}[!htp]
\centering

\begin{minipage}{0.45\textwidth}
\includegraphics[width=\linewidth]{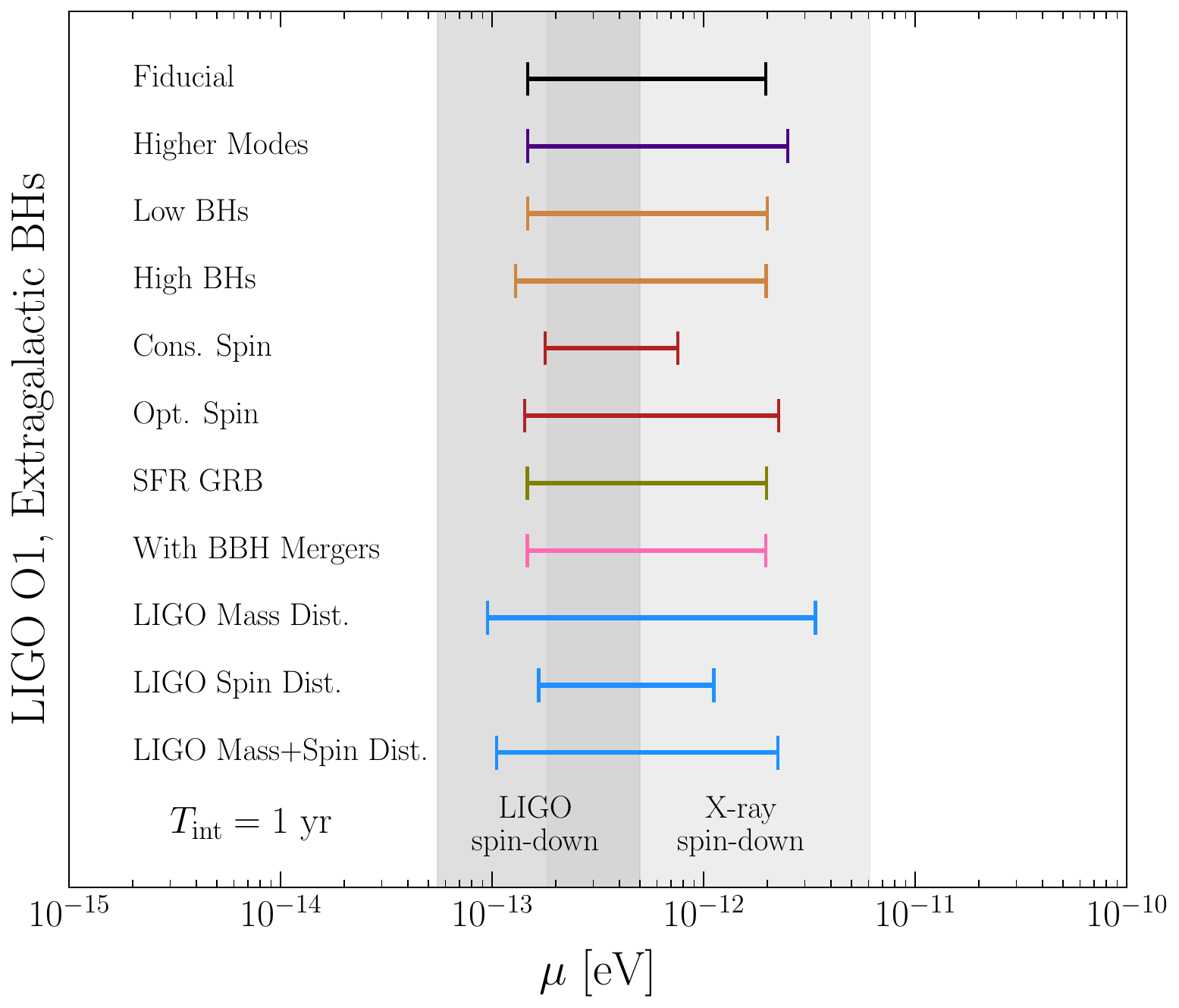}
\end{minipage}

\vspace{0.3cm}

\begin{minipage}{0.45\textwidth}
\includegraphics[width=\linewidth]{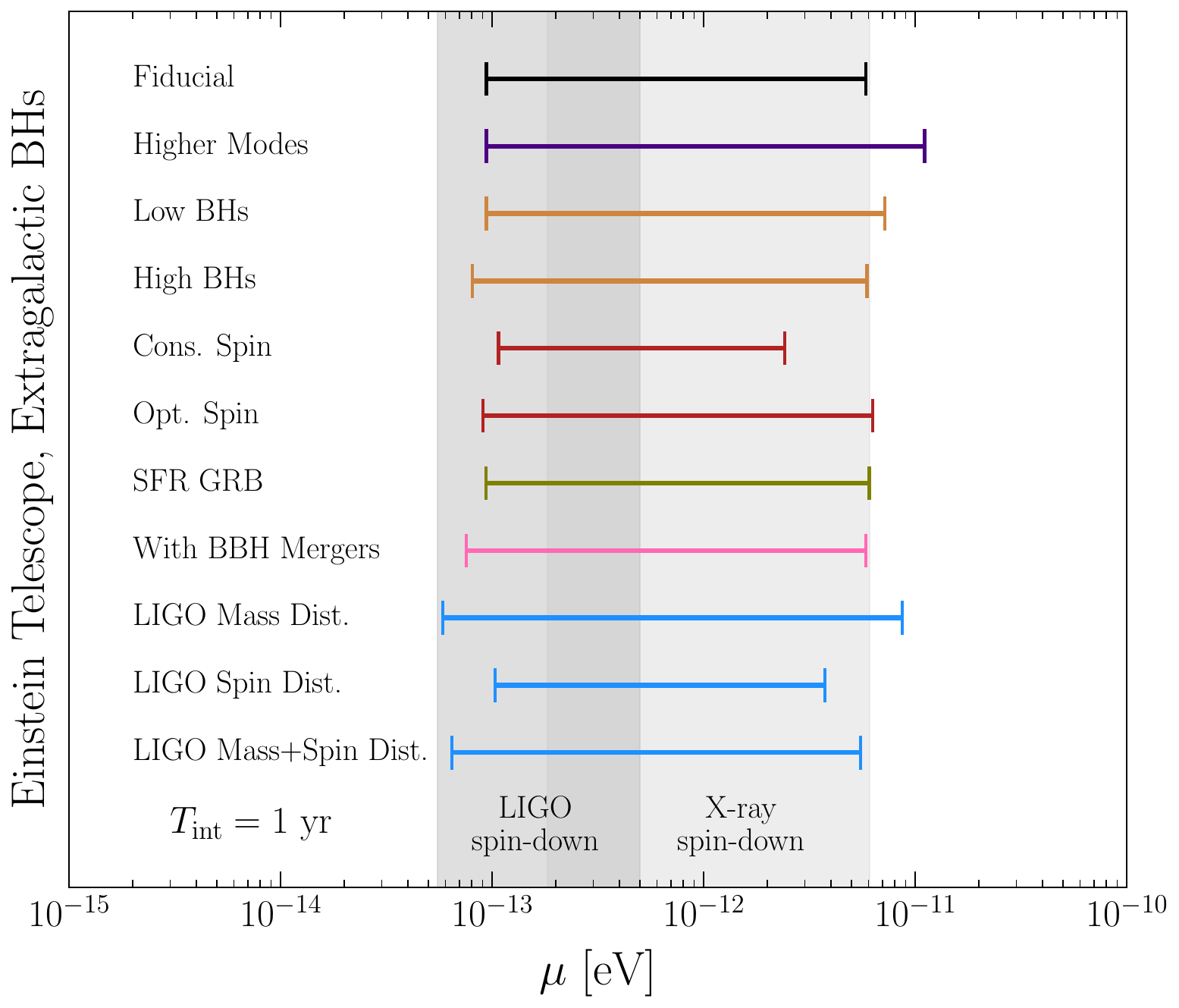}
\end{minipage}
\hspace{0.3cm}
\begin{minipage}{0.45\textwidth}
\includegraphics[width=\linewidth]{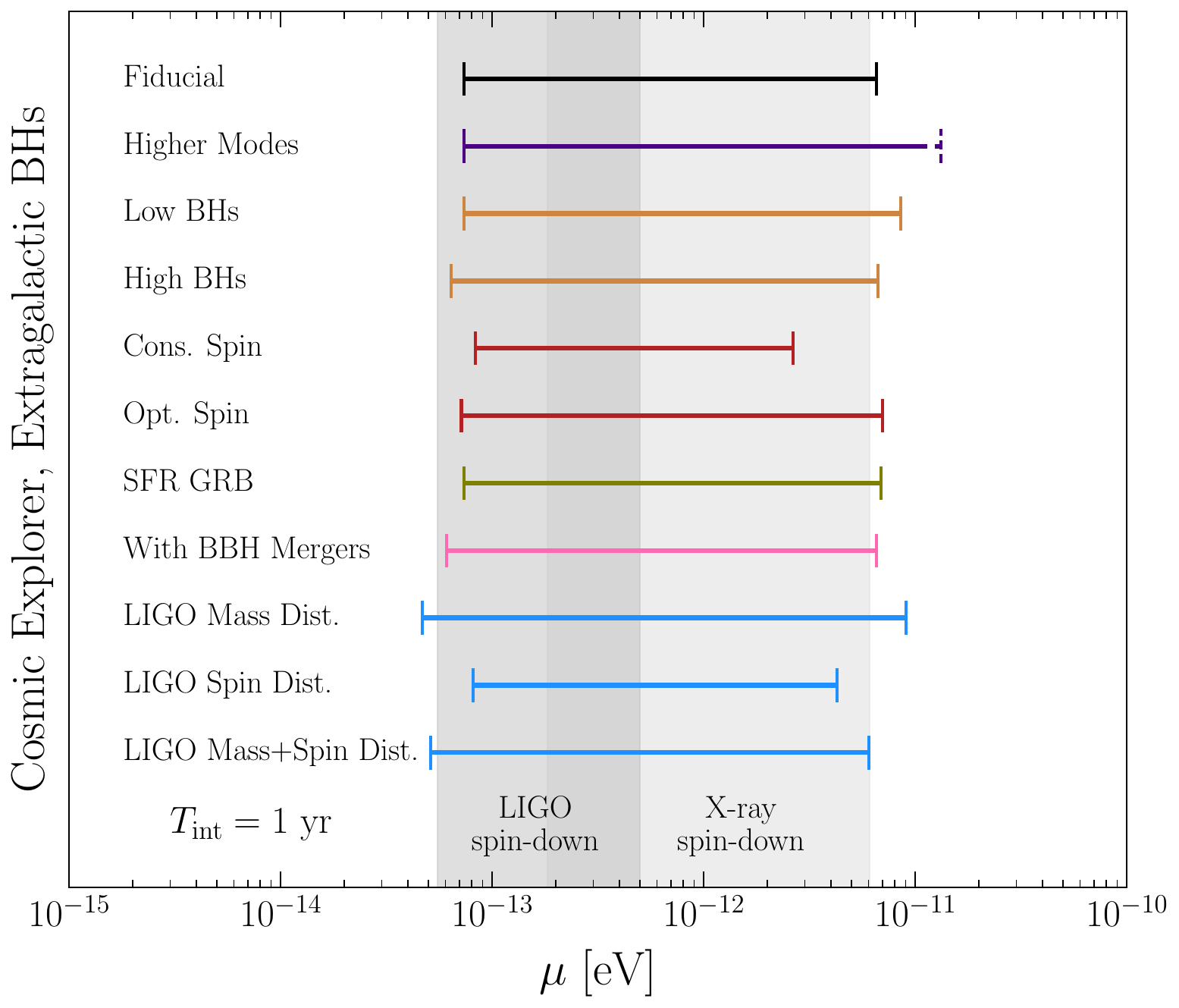}
\end{minipage}
\vspace{-0.2cm}
\caption{Extensions of the right panel of Fig.~\ref{fig:fig_1} toward other observatories considered in our work here: LIGO O1, Einstein Telescope, Cosmic Explorer.
We note that the Magnetic Weber Bars are not shown here as they posses no appreciable sensitivity for nearly all of the astrophysical systematic configurations considered, under the assumed $T_{\rm int}$ and SNR $\geq 8$ threshold.}
\label{fig:EG_observatory_sys}
\end{figure*}

In Fig.~\ref{fig:EG_stat} we show an illustration of our SNR over axion mass $\mu$ used in our statistical analysis for stochastic signals discussed in the main text, for LIGO O5 with threshold ${\rm SNR}=8$.
The results of the systematic analyses for all observatories (note LIGO O5 is presented in Fig.~\ref{fig:fig_1} of the main text) are shown in Fig.~\ref{fig:EG_observatory_sys}.

Finally, we discuss our modeling for the remnant BBH merger population, which we include as a systematic astrophysical configuration (`With BBH Mergers' in Tab.~\ref{tab:configs_EG}).
We broadly adopt the model in Ref.~\cite{Tsukada:2018mbp}, which calls for a BH primary and secondary mass distribution from Refs.~\cite{LIGOScientific:2017zlf,LIGOScientific:2016dsl,LIGOScientific:2016jlg}, described by
\be
p(m_1, m_2) \propto \frac{m_1^{-2.35}}{m_1 - 5 \, M_{\odot}},
\ee
where we constrain $5 \, M_{\odot} < m_{1,2} < 95 \, M_{\odot}$ and $m_1 + m_2 < 100 \, M_{\odot}$.
The mass of the remnant BH after subtracting the binding energy of the innermost stable orbit~\cite{2009GReGr..41.1667H} is
\be
M \simeq m_1 + m_2 - 5.7 \times 10^{-2} \frac{m_1 m_2}{m_1 + m_2} \,.
\ee

Next, the merger rate is taken as~\cite{LIGOScientific:2016fpe,LIGOScientific:2016jlg, Dominik:2013tma}
\be
R_m(z) = \int_{t_{\rm min}}^{t_{\rm max}}\!dt_d\, \psi(z_f) \, p(t_d),
\ee
which integrates over the time delay $t_d$, whose distribution $p(t_d)$ weights the SFR $\psi$.
Note that $z_f$ is the redshift taken at the time $t_f = t(z) - t_d$, which represents the binary formation epoch ($t(z)$ is the cosmic time at the merger event).
We take the distribution $p(t_d) \propto 1/t_d$ for $t_{\rm min} < t_d < t_{\rm max}$ for $t_{\rm min} = 50$\,Myr and $t_{\rm max} = t_H$ (the Hubble time), following Refs.~\cite{LIGOScientific:2016fpe, Dominik:2013tma}.
Implicit in this formulation is the assumption that, broadly, the formation rate of BBHs follows the isolated extragalactic BH SFR through redshift dependence only, though in reality the situation is likely more nuanced (for example, see Ref.~\cite{Madau:2014bja} for possible metallicity dependence).

The merger rate is then combined with our integration over redshift, mass, spin, and merger birth time $t_b$ to infer the BBH merger background GW density
\begin{align}
\Omega_{\rm BBH}(f) &= \frac{f}{\rho_c} \int \!dz dm_1 dm_2 d\chi dt_b \, \frac{dt}{dz} R_m(z) \\ &\times p(m_1, m_2) p(\chi) \frac{dE_h}{dt}[t-t_b] \delta(f(1 + z) - f_s), \nonumber
\end{align}
where, similar to the formulation of $\Omega_h$ in Sec.~\ref{sec:extragal}, $\rho_c = 3 H_0^2/8\pi G$ is the critical density and $f_s = \omega_R/\pi$ is the emitted GW linear frequency in the BBH source frame.
Following Ref.~\cite{Tsukada:2018mbp}, and motivated by LIGO/Virgo observations~\cite{LIGOScientific:2017bnn,LIGOScientific:2017ycc,LIGOScientific:2017vox,LIGOScientific:2016dsl} as well as numerical simulations~\cite{Gonzalez:2006md,Berti:2007fi,Rezzolla:2007xa}, we simply take $p(\chi) = \delta(\chi - 0.7)$ for all BBH remnants.

Finally, we note that following estimates in Ref.~\cite{LIGOScientific:2017bnn}, we normalize the local merger rate at $z = 0$ as
\be
\!\int \!dm_1 dm_2\, p(m_1, m_2) R_m(z{=}0) = 103 \, {\rm Gpc}^{-3} \, {\rm yr}^{-1},
\ee
with the understanding that in reality this rate is only suggestive, as other estimates place credible bounds between $\sim$$10$--$100$\,Gpc$^{-3}$\,yr$^{-1}$~\cite{Tsukada:2018mbp}.

As can be seen in Figs.~\ref{fig:fig_1} and~\ref{fig:EG_observatory_sys}, the addition of the BBH merger population under these assumptions ultimately makes a marginal impact on the overall axion mass sensitivity.

%%%%%%%%%%%%%%%%%%%%%%%%%%%%%%%%%
\section{Sub-Solar Mass Black Holes \\ and Dark Matter}
\label{app:dm_collapse}
%%%%%%%%%%%%%%%%%%%%%%%%%%%%%%%%%

Standard astrophysics does not yield BHs with masses below the Tolman-Oppenheimer-Volkoff (TOV) bound.
Standard astrophysical objects with masses below the TOV bound can withstand gravitational collapse due to Fermi degeneracy pressure of neutrons.
Therefore, for there to be a population of sub-stellar mass BHs, ingredients beyond the standard astrophysical paradigm are required.

A natural candidate for the necessary additional ingredient is dark matter.
Below we discuss how two dark-matter scenarios could lead to sub-solar mass BHs: primordial-origin BHs with small masses, and stellar capture of dark matter leading to BH formation in astrophysical objects.

It is well-known that PBHs formed in the early universe are a viable dark matter candidate~\cite{Carr:2016drx,Sasaki:2018dmp}.
However, in the mass range $0.1 \lesssim M/M_\odot \lesssim 5$, PBHs can only comprise less than 10\% of the observed dark matter~\cite{Tisserand_2007,Wyrzykowski_2011}.
As a result, at most $0.1 M_{\rm halo}$ can be in the form of PBHs, leading to an expected number of PBHs that ranges from $N_{\rm PBH} \sim 10^{11} -10^{12}$, which would be distributed according to the dark-matter profile in the Milky Way.
Thus, at first glance it would seem that even in this mass range where PBHs cannot be all of the dark matter, they might nevertheless be sufficiently plentiful to lead to measurable superradiance signals. 
However, two complications arise that prevent us from studying this case in detail.
The first is that due to the formation mechanism of PBHs as the spherical collapse of a matter overdensity in the early universe, the spin of typical PBHs at birth is very small.
The expectation is that PBHs could have slightly non-zero spin, with $\chi_{_{\rm PBH}} \sim 0.01$ expected~\cite{DeLuca:2019buf,Mirbabayi:2019uph}.
While the PBH could in principle spin up further through the formation of an accretion disk, accretion is inefficient for light BHs.
If the spin remains small, only axions with $\alpha \ll 1$ can experience superradiant instabilities (see Eq.~\eqref{eq:SR_condition}).
As a result, cloud growth, controlled by Eq.~\eqref{eq:Tc}, and GW emission, controlled by Eq.~\eqref{eq:dotEh}, are slow.
The second complication is that the mass function of PBHs is unknown.
Therefore, ad hoc assumptions would have to be made in order to study a population of sub-solar mass PBHs and their associated superradiance signals.
For these two reasons primarily, we do not study the scenario of sub-solar mass PBH superradiance (although see e.g. Ref.~\cite{Jia:2025vqn}), although it could be examined carefully in future work.

The second dark-matter scenario that can lead to sub-solar mass BHs is dark-matter capture and collapse.
This process has been extensively studied, particularly for neutron stars~\cite{Goldman:1989nd,Gould:1989gw,deLavallaz:2010wp,Kouvaris:2010jy,Kouvaris:2011fi,McDermott:2011jp,Kouvaris:2018wnh,Garani:2021gvc,Bhattacharya:2023stq}, but also for sun-like and other stellar populations~\cite{Acevedo:2020gro,Dasgupta:2020mqg,Ellis:2021ztw,Bramante:2024idl}.

For this scenario to be viable, certain ingredients are required.
The DM capture rate should be efficient, in which case it saturates at the geometric capture rate $C \sim \pi R_\star^2 n_{_{\rm DM}}(r) v_{\rm esc}^2/v_0 $, where $n_{_{\rm DM}}(r)$ is the dark-matter number density at the stellar location $r$.
Further, the star has radius $R_\star$, an escape velocity $v_{\rm esc}$ and the dark matter has a dispersion velocity $v_0$.
The total captured dark-matter mass should accumulate in the core of the star and exceed the self-gravitation mass
\be
M_{\rm sg} \gtrsim \sqrt{\frac{3}{\pi \rho_\star^c}} \left(\frac{T_\star^c}{G m_{_{\rm DM}}}\right)^{3/2},
\ee
where $T_\star^c$ and $\rho_\star^c$ are the core temperature and density of the star respectively.
The captured dark-matter mass should also exceed the Chandrasekhar limit for the dark matter, $M_{\rm Ch} \sim M_\textrm{pl}^3/m_{_{\rm DM}}^2$, or else the dark matter will collapse to a compact object sustained by degeneracy pressure. The nascent BH should accrete more dark matter and stellar material faster than the Hawking evaporation rate, and sufficiently fast so as to convert the star within a reasonable time.
As the accretion rate scales as $\dot{M}_{\rm BH} \propto M_{\rm BH}^2 \rho_\star$ and evaporation as $\dot{M}_{\rm BH} \propto - 1/M_{\rm BH}^2$, this requirement translates into a requirement that the BH at formation should be sufficiently massive.
This in turn requires that the dark matter be sufficiently light. Despite all these requirements, it is fairly straightforward to find dark matter parameter space where stellar collapse to BHs is viable.

If we assume that this process is viable for sub-solar mass stars, e.g. M-dwarfs and brown dwarf stars, then BHs with masses as low as $M_{\star} \sim 0.08\,M_\odot$ could exist in the galaxy. The mass distribution would follow the IMF of the corresponding stellar population.
Studies have shown that the IMF for such stars follows a broken power law, where the stars with $M_{\star} \gtrsim 0.4\,M_\odot$ follow the Salpeter IMF, while stars with $0.02\, M_\odot \lesssim M_{\star} \lesssim 0.4\,M_\odot$ have an IMF $\xi(M_\star) \sim M_\star^{-\alpha}$ with $0.5 \lesssim \alpha \lesssim 1$~\cite{B_jar_2011, Mu_i__2019, Almendros_Abad_2023, Mu_ic__2025, Rom_2026}.
The result is obtained from observations of young clusters in our galaxy.
In our analysis, we fix the slope at the shallowest edge, i.e. $\alpha = 0.5$.
The total number of BHs will depend on whether the dark-matter capture and conversion process is sufficiently efficient at that position in the galaxy.
Without considering specific dark-matter model parameters, we cannot precisely model the number and distribution of BHs.
Such an analysis is well beyond the scope of this work. However, given that there are thought to be up to one brown dwarf for every two main sequence stars~\cite{scholz2012substellar}, there could be as many as $N_{\rm BH} \sim 10^{11}$ sub-solar BHs in the galaxy.
In our analysis, we fix a total of $N_{\rm BH} = 10^8$, distributed according to the aforementioned IMF.

%%%%%%%%%%%%%%%%%%%%%%%%%%%%%%%%%
\section{Stochastic Backgrounds from Non-Instantaneous Emission}
\label{app:Omegah}
%%%%%%%%%%%%%%%%%%%%%%%%%%%%%%%%%

In Sec.~\ref{sec:extragal} we studied the stochastic background that can be generated from the GW emission from superradiant clouds throughout the entire universe.
Our key expression for computing this is given in Eq.~\eqref{eq:EG_Omega}, which specifies the local energy density these GW generate as a function of frequency.
A detail we emphasized is that this result accounts for the fact that the energy emitted during the superradiance process need not be instantaneous; indeed, from Fig.~\ref{fig:Tg_Th} we can see that the timescales can be cosmological.
In this appendix we derive that expression and demonstrate that under the assumption of instantaneous emission it reduces to the result from Ref.~\cite{Phinney:2001di}, widely adopted in the literature.

Our starting point is to compute the differential flux of gravitons being emitted from a differential unit of time and volume in the cosmos.
To obtain this, we need the birth rate of BHs.
We let $dN/dt_b dV$ be the distribution of BHs born per unit time and comoving volume; we work exclusively in comoving coordinates throughout this calculation.
Once the BH is born, we wish to determine whether it forms a superradiant cloud, which also depends on the BH mass and spin.
Therefore, we consider the BH distribution fully differential in these variables,
\be
\frac{dN}{dt_b dV dM d\chi}.
\label{eq:dNdVdtbdMdchi}
\ee
This distribution will vary with redshift, however by cosmological isotropy should not depend on a specific direction.
Our model for this distribution, which is derived from the expected star formation rate, is detailed in Sec.~\ref{sec:extragal}.

Next we need the rate at which these BHs produce gravitons.
From the discussion of cloud decay in Sec.~\ref{sec:superradiance}, for a given initial BH mass and spin $M$ and $\chi$ we can determine the energy being emitted into GW using Eq.~\eqref{eq:dotEh}, which specifies $\dot{E}_h[M,\chi,\mu,t-t_b]$.
Importantly, the size of the power emitted depends upon the time since the birth of the BH, $t-t_b$.
When $T_c+T_h$ is a cosmologically short timescale, we can approximate all the energy as being emitted instantaneously, which corresponds to setting $\dot{E}_h = E_h \delta(t-t_b)$, with $E_h$ integrated total energy the cloud emits over its lifetime.\footnote{From Eqs.~\eqref{eq:dotEh} and \eqref{eq:Mct} we have $E_h = D_{n\ell} ( M_c^f/M_f )^2 T_h \alpha^{4\ell+10}$.}
Regardless, as the energy is emitted approximately monochromatically into gravitons with a source energy $\omega_s = 2\omega_R$, we can compute the differential number of gravitons being emitted as
\be
\frac{dN_h}{d\omega dt} = \frac{dE_h}{dt}[t-t_b]\, \frac{\delta(\omega-\omega_s)}{\omega}.
\ee
This specifies the rate gravitons are produced per BH so that when combined with the BH distribution from Eq.~\eqref{eq:dNdVdtbdMdchi} we can compute the total differential flux of gravitons as,
\be
\!\frac{dN_h}{dtd\omega dV} =\! \int \! dMd\chi dt_b\frac{dN}{dt_b dV dM d\chi} \frac{dE_h}{dt} \frac{\delta(\omega{-}\omega_s)}{\omega},
\label{eq:Rateh}
\ee
i.e. we simply combine the distributions marginalizing over the additional parameters.

So far we have computed the rate of gravitons being produced per unit time.
We next use this to compute the energy density of gravitons arriving at our detector.
If our detector has a differential area for detecting gravitational radiation of $dA$, then it will capture a fraction $dA/4\pi r^2$ of the flux emitted from a volume element a comoving distance $r$ away.
Accounting for the full extragalactic contribution requires integrating over the comoving volume, $dV = r^2 dr d\Omega$.
With the assumed isotropy, the solid angle integral can be performed immediately, and the graviton flux on our detector is
\bea
\frac{dN_h}{dt d\omega dA} = \int dzdMd\chi dt_b\,&(1+z) \frac{dt}{dz} \frac{dN}{dt_b dV dM d\chi} \\
\times &\frac{dE_h}{dt} \frac{\delta(\omega(1+z)-\omega_s)}{\omega(1+z)}.
\label{eq:grav-penultimate}
\eea
In writing this expression we accounted for the redshifting of the graviton energy and further changed integration variables from $r$ to $z$ using $dr/dz = (1+z)\, dt/dz$.
This expression specifies the number of gravitons incident on our detector per unit time, energy, and area. To convert this to an energy density, we divide by the graviton speed $c=1$ and multiply by energy $\omega$; in detail, $d\rho_h/d\omega = \omega \times dN_h/dtd\omega dA$.
We can then rewrite these with respect to the critical energy density, $\rho_c$, using,
\be
\Omega_h(\omega) = \frac{1}{\rho_c} \frac{d\rho_h}{d\ln \omega} = \frac{\omega}{\rho_c} \frac{d\rho_h}{d\omega}.
\ee
Combining this with Eq.~\eqref{eq:grav-penultimate} and rewriting the energy in terms of the linear frequency $f$, we arrive at
\bea
\Omega_h(f) = \frac{f}{\rho_c} \int dzdMd\chi dt_b\,&\frac{dt}{dz} \frac{dN}{dt_b dV dM d\chi} \\
\times &\frac{dE_h}{dt} \delta(f(1+z)-f_s),
\eea
as given in Eq.~\eqref{eq:EG_Omega}.

\begin{figure}[!t]
\centering
\includegraphics[width=0.47\textwidth]{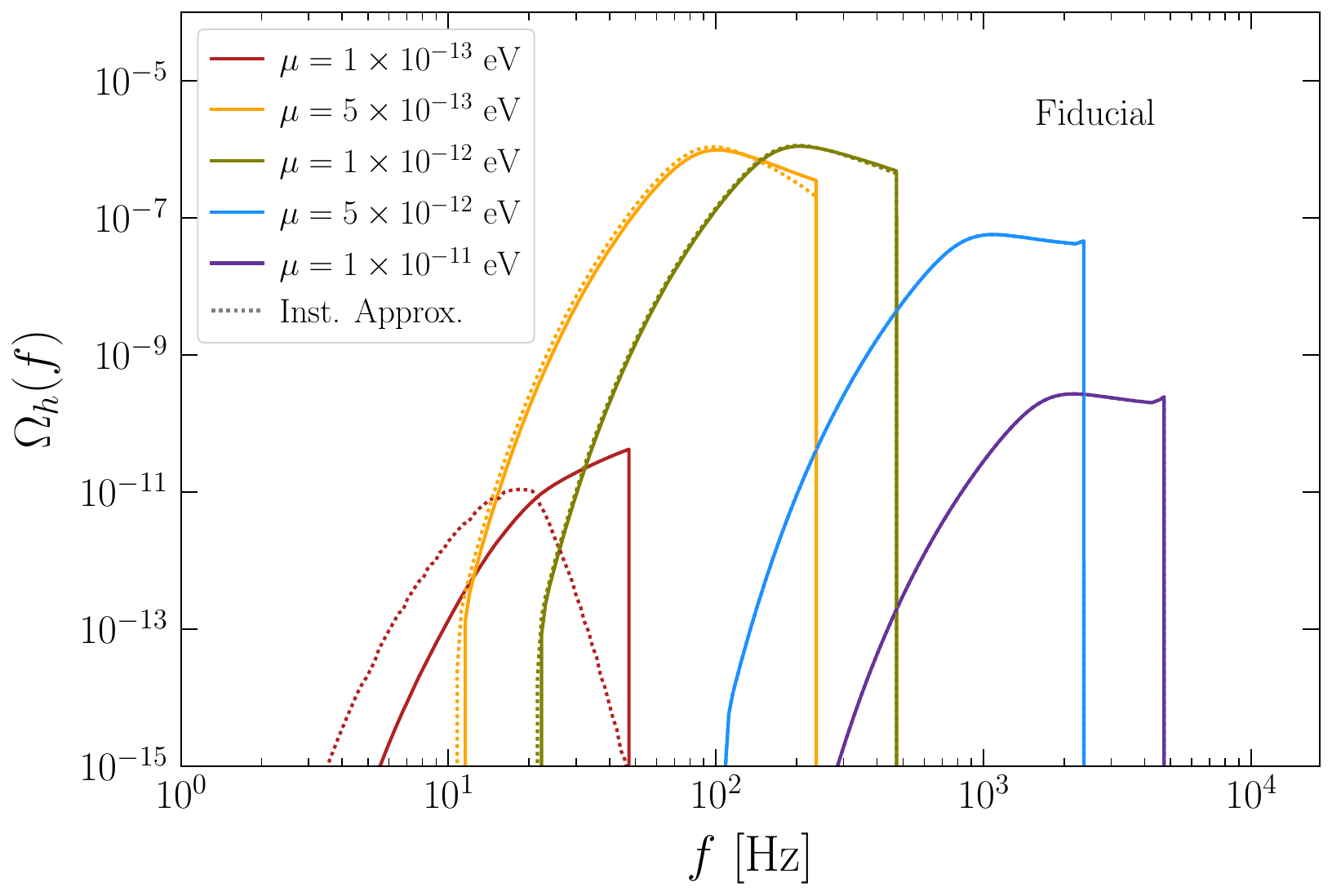}
\vspace{-0.2cm}
\caption{The stochastic GW energy density $\Omega_h(f)$ from our fiducial distribution of extragalactic BHs across selected axion masses $\mu$.
This signal would need to be detected over not only instrumental backgrounds, but also a contribution from the stochastic signal of binary BHs and neutron stars~\cite{LIGOScientific:2017zlf}.
We also compare results to the commonly adopted instantaneous emission approximation, showing that for our fiducial model that includes only the $m=1$ mode the difference is only pronounced for lower masses where the timescales become cosmological, see Fig.~\ref{fig:Tg_Th}.
When higher modes are included the discrepancy is enhanced.}
\label{fig:EG_Omegah}
\end{figure}

As emphasized, the above result distinguishes between emission time of the GWs, encoded in $t$ or equivalently $z$, and the birth time of the BH, $t_b$.
If we assume the emission is instantaneous, $\dot{E}_h = E_h \delta(t-t_b)$, the result becomes
\bea
\Omega^\textrm{inst.}_h(f) = \frac{f}{\rho_c} \int dzdMd\chi\,&\frac{dt}{dz} \frac{dN}{dt dV dM d\chi} \\
\times &E_h \delta(f(1+z)-f_s),
\eea
which is the form widely adopted in the literature, see Refs.~\cite{Phinney:2001di,Brito:2017wnc, Brito:2017zvb, Tsukada:2018mbp}.
By default, we use the non-instantaneous formalism, although we show both results for our fiducial model in Fig.~\ref{fig:EG_Omegah}.
There we can see that only lower masses lead to an appreciable difference.
The reason is, for $m=1$, only lower $\mu$, and hence $\alpha$, can have a sufficiently long timescale for the cloud decay that the instantaneous approximation begins to break down.
When higher modes are included, the difference between the two approaches is more pronounced across the mass range, cf. Fig.~\ref{fig:Tg_Th}.

%%%%%%%%%%%%%%%%%%%%%%%%%%%%%%%%%
\section{Extragalactic Monte Carlo Simulation}
\label{Sec:MC_extragalactic}
%%%%%%%%%%%%%%%%%%%%%%%%%%%%%%%%%

In the following, we describe in detail our Monte Carlo simulation for the GW signal from the extragalactic BH population that we performed as a cross check on the approach in Eq.~\eqref{eq:EG_Omega}.
The key complication is that we cannot just simulate all BHs in our universe because there are simply too many.
Instead we simulate only a fraction and then implement a rescaling procedure to compute the full GW signal.
The technique is fast and allows for the inclusion of all physical effects.
It ultimately confirms the validity of the prescription used in Sec.~\ref{sec:extragal}.

For the simulation, we use a co-moving volume of size $\Delta V = (1\,{\rm Mpc})^3$.
For comparison, the co-moving volume of our universe is $V_u^{\rm cm} \sim (10^4\,{\rm Mpc})^3$.
We also use time steps of size $\Delta t=10^6\,$yrs.

We begin the simulation at $t_s = 10^8$\,yrs, and continue the evolution until today.
Within each time step the SFR tells us how much mass, $\Delta M_\star$, is used to form stars:
\be
\psi(t)=\frac{\Delta M_\star}{\Delta t \Delta V}.
\ee
where the SFR is parameterized as in Eq.~\eqref{eq:SFR}.
The number of stars $\Delta N_\star$ that were formed in the mass interval $\Delta M_\star$, volume $\Delta V$ and time interval $\Delta t$ is determined by
\be
\frac{\Delta N_\star}{\Delta t \Delta V \Delta M_\star}=\frac{\psi(t)\xi(M_\star)}{\int_{M^\textrm{min}_\star}^{M_\star^\textrm{max}}\!dM_{\star}\, M_{\star} \,\xi(M_{\star})},
\label{eq:dNdtdVdM}
\ee
and we adopt the same prescriptions as described around Eq.~\eqref{eq:dNdtdVdMs_main} in the main text.

We then loop over all time-step intervals, in each of which we calculate the number of stars formed according to Eq.~\eqref{eq:dNdtdVdM}.
Since we know the stellar lifetimes (cf. Ref.~\cite{Schaerer:2001jc}), we can also determine the time at which the stars born in each interval collapse into BHs.
This fixes the BH birth time.
Further, we track the relation between metallicity and cosmic time, which is needed to calculate the BH mass from the stellar mass.
We adopt the metallicity-mass relation as in the main text.
Other parameters such as the minimum and maximum remnant mass taken to be associated with BHs and the spin distribution match the fiducial extragalactic model defined in Tab.~\ref{tab:configs_EG}.

The above procedure yields a list of BHs, which in the next step we analyze one-by-one.
Given an axion mass we check for each BH if the superradiance condition is fulfilled, starting from the 211 mode.
If the superradiance condition is not satisfied, we proceed to the next mode.
If the superradiance condition is fulfilled for a specific mode $m$, we compute the cloud growth time $T_c$ and the time it takes to emit GWs $T_h$ after the cloud has stopped growing.
Using this we generate a list of GW emission events.

Next we account for the spatial distribution of BHs.
We distribute the events randomly only within the volume of the universe for which the GW radiation is guaranteed to still be present on Earth today.
To account for this restricted volume, we down-weight the GW events by the ratio of the volume in which the GW events can be placed to the total comoving volume of the universe.
Once a position for a GW event has been selected, we redshift the GW frequency to its value as observed today on Earth.

Finally, the calculation of the stochastic GW spectrum proceeds as follows.
We loop over all GW emission events and calculate the amplitude $h$ with Eq.~\eqref{eq:h}.
We then calculate the GW energy density for each individual GW emission event, using $\rho_h^i = \omega^2\, h^2/8\pi$, with $i$ indexing the event.
From here, the GW spectrum is typically specified as 
\be
\Omega_h = \frac{1}{\rho_c} \frac{\Delta\rho_{\rm h}}{\Delta f}\,f.
\ee
Because we do not simulate the entire GW population we scale the GW spectrum in the end by the volume ratio $V^{\rm cm}_u/\Delta V$.

\bibliographystyle{apsrev4-1}
\bibliography{refs}

\end{document}